\documentclass[lettersize,journal]{IEEEtran}
\usepackage{hyperref}
\usepackage{amsmath,amsfonts}
\usepackage{algorithmic}
\usepackage{algorithm}
\usepackage{array}
\usepackage{booktabs}
\usepackage[caption=false,font=normalsize,labelfont=sf,textfont=sf]{subfig}
\usepackage{textcomp}
\usepackage{stfloats}
\usepackage{url}
\usepackage{verbatim}
\usepackage{graphicx}
\usepackage{cite}
\usepackage{tabularx}
\usepackage{multirow}   
\usepackage{tikz}
\usepackage{adjustbox}
\usepackage[edges]{forest}

\definecolor{framework-blue}{RGB}{47, 85, 151}
\tikzset{%
    parent/.style = {align=center,text width=2.5cm,rounded corners=3pt, line width=0.3mm, fill=gray!10,draw=gray!80},
    child/.style = {align=center,text width=2.3cm,rounded corners=3pt, fill=blue!10,draw=blue!80,line width=0.3mm},
    grandchild/.style = {align=center,text width=2cm,rounded corners=3pt},
    greatgrandchild/.style = {align=center,text width=1.5cm,rounded corners=3pt},
    greatgrandchild2/.style = {align=center,text width=1.5cm,rounded corners=3pt},    
    referenceblock/.style =  {align=center,text width=1.5cm,rounded corners=2pt},
    dimension_based/.style= {align=center,text width=2.2cm,rounded corners=3pt, fill=white,draw=framework-blue,line width=0.3mm},
    dimension_based_work/.style= {align=center, text width=4.5cm,rounded corners=3pt, fill=white,draw=framework-blue,line width=0.3mm},
    perception_based/.style= {align=center,text width=2.2cm,rounded corners=3pt, fill=white,draw=framework-blue,line width=0.3mm},
    perception_based_work/.style= {align=center, text width=4.5cm,rounded corners=3pt, fill=white,draw=framework-blue,line width=0.3mm},
    logic_based/.style= {align=center,text width=2.2cm,rounded corners=3pt, fill=white,draw=framework-blue,line width=0.3mm},
    logic_based_work/.style= {align=center, text width=4.5cm,rounded corners=3pt, fill=white,draw=framework-blue,line width=0.3mm},
    interaction_based/.style= {align=center,text width=2.2cm,rounded corners=3pt, fill=white,draw=framework-blue,line width=0.3mm},
    interaction_based_work/.style= {align=center, text width=4.5cm,rounded corners=3pt, fill=white,draw=framework-blue,line width=0.3mm},
}

\newcolumntype{P}[1]{>{\centering\arraybackslash}p{#1}}
\begin{document}

\title{Nature's Insight: A Novel Framework and Comprehensive Analysis of Agentic Reasoning Through the Lens of Neuroscience}
% \title{From Brain to Machine: A Neural Definition of AI Reasoning}


\author{Zinan Liu$^{*}$, Haoran Li$^{*}$, Jingyi Lu, Gaoyuan Ma, Xu Hong, Giovanni Iacca~\IEEEmembership{Senior Member,~IEEE}, Arvind Kumar,\\ Shaojun Tang, and Lin Wang$^{\dag}$~\IEEEmembership{Member,~IEEE}
\thanks{$^{*}$ Equal contribution. $^{\dag}$ Corresponding author.}
\thanks{Zinan Liu, Gaoyuan Ma and Lin Wang are with the School of EEE, Nanyang Technological University (NTU), Singapore (email: zinan001@e.ntu.edu.sg; C230096@e.ntu.edu.sg; linwang@ntu.edu.sg). }
\thanks{Haoran Li is a visiting student at the School of EEE, NTU, Singapore  and is also with the School of Cyber Science and Technology, University of Science and Technology of China, China (e-mail: n2409944c@e.ntu.edu.sg).}
\thanks{Jingyi Lu is currently a research intern at the School of EEE, NTU, Singapore, and is with the School of Integrated Circuits and Electronics, Beijing Institute of Technology, China (email: 1120223707@bit.edu.cn).}
\thanks{Hong Xu is with Psychology in the School of Social Sciences, Nanyang Technological University, Singapore. Email: xuhong@ntu.edu.sg}
\thanks{Giovanni Iacca is with the Department of Information Engineering and Computer Science, University of Trento, Italy. Email: giovanni.iacca@unitn.it}
\thanks{Arvind Kumar is with the Division of Computational Science and Technology, KTH Royal Institute of Technology, Sweden, Email:arvkumar@kth.se }
\thanks{Shaojun Tang is with the Bioscience and Biomedical Engineering, Hong Kong University of Science \& Technology, Email: shaojuntang@ust.hk }}

\markboth{Journal of \LaTeX\ Class Files,~Vol.~14, No.~8, August~2021}%
{Shell \MakeLowercase{\textit{et al.}}: A Sample Article Using IEEEtran.cls for IEEE Journals}


\maketitle

\begin{abstract}
Autonomous AI is no longer a hard-to-reach concept--it enables the machines (agents) to move beyond executing tasks to independently addressing complex problems, adapting to change while handling the uncertainty of the environment. However, what makes the agents truly autonomous? It is agentic reasoning, that is crucial for foundation models to develop symbolic logic, statistical correlations, or large-scale pattern recognition to process information, draw inferences, and make decisions. However, it remains unclear why and how existing agentic reasoning approaches work, in comparison to biological reasoning, which instead is deeply rooted in neural mechanisms involving hierarchical cognition, multimodal integration, and dynamic interactions. In this work, we propose a novel neuroscience-inspired framework for agentic reasoning. Grounded in three cognitive neuroscience-based definitions of reasoning, supported by corresponding mathematical formulations and biological reasoning pathways, we develop a unified framework that models the full reasoning process from sensory input to action. Within this framework, we identify four core reasoning types--perceptual, dimensional, logical, and interactive--inspired by distinct functional roles observed in the human brain. We apply this framework to systematically classify and analyze existing AI reasoning methods, evaluating their theoretical foundations, computational designs, and practical limitations. We further explore the implications for developing more generalizable and cognitively aligned agents in both physical and virtual settings. Finally, based on our framework, we outline future directions for AI reasoning and introduce new reasoning methods inspired by neural models, analogous to chain-of-thought prompting. By bridging cognitive neuroscience and AI, this work offers a theoretical foundation and practical roadmap for advancing agentic reasoning in intelligent systems. \textbf{\textit{The associated project can be found at: }}\url{https://github.com/BioRAILab/Awesome-Neuroscience-Agent-Reasoning}.
\end{abstract}


\section{Introduction}
\begin{figure}[t!]
    \centering
    \includegraphics[width=0.95\linewidth]{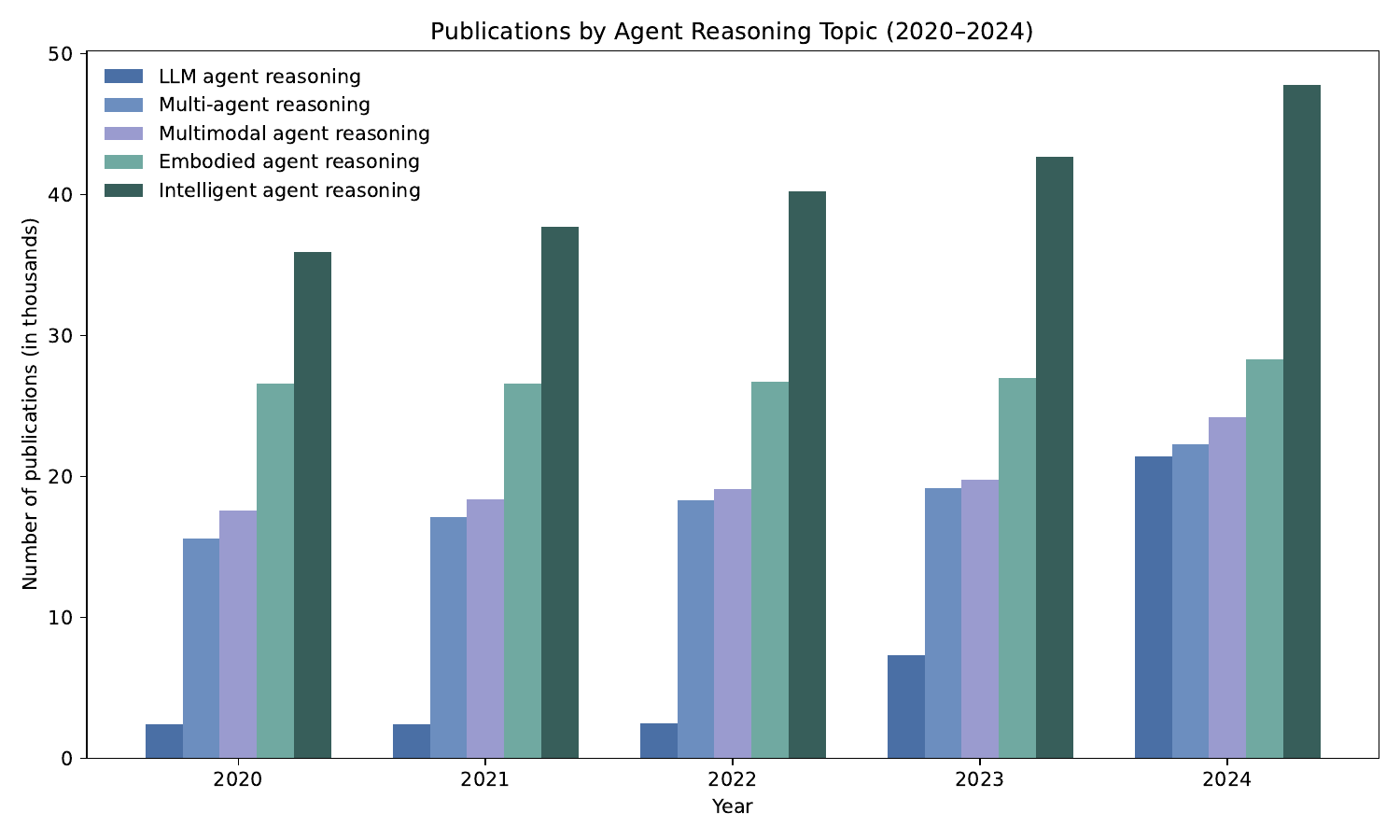}
    \caption{Google Scholar results for research topics related to agentic reasoning. The vertical axis represents the number of publications (in thousands), while the horizontal axis denotes the publication year. The figure highlights a significant rise in "\textbf{LLM agentic reasoning}" publications since 2023, reflecting the impact of large language models on the field. }
    \label{fig:reasoning_trend}
    \vspace{-10pt}
\end{figure}

\IEEEPARstart{R}EASONING is the process of drawing conclusions from premises~\cite{proudfoot2009routledge}. It forms a cornerstone of human intelligence~\cite{russell2016artificial,khorasani2008artificial,poole1998computational,nilsson1998artificial} and enables individuals to interpret the world, anticipate future events, and solve complex problems across a wide range of domains. Similarly, for artificial agents, reasoning is fundamental to adaptive decision-making, generalization, and problem-solving in dynamic environments. As shown in Fig.~\ref{fig:reasoning_trend}, recent years have witnessed a surge in research interest surrounding agentic reasoning, particularly in Large Language Model (LLM)-based reasoning, highlighting the growing impact of large language models in this field. In the development of autonomous artificial intelligence (AI)—systems capable of independently perceiving, reasoning, and acting in complex, uncertain environments, reasoning stands as a critical prerequisite. Unlike narrow AI systems that excel in specialized tasks but struggle with abstraction and transfer learning, autonomous AI requires robust reasoning mechanisms to synthesize information, infer hidden relationships, and adaptively navigate novel situations without explicit human intervention. Therefore, advancing the reasoning capabilities of AI agents is not merely an incremental improvement--it is a necessary step toward building more intelligent, self-directed agents that can move beyond pattern recognition and reactive behavior.

\par
\begin{figure*}[t!]
    \centering
    \includegraphics[width=1.0\linewidth]{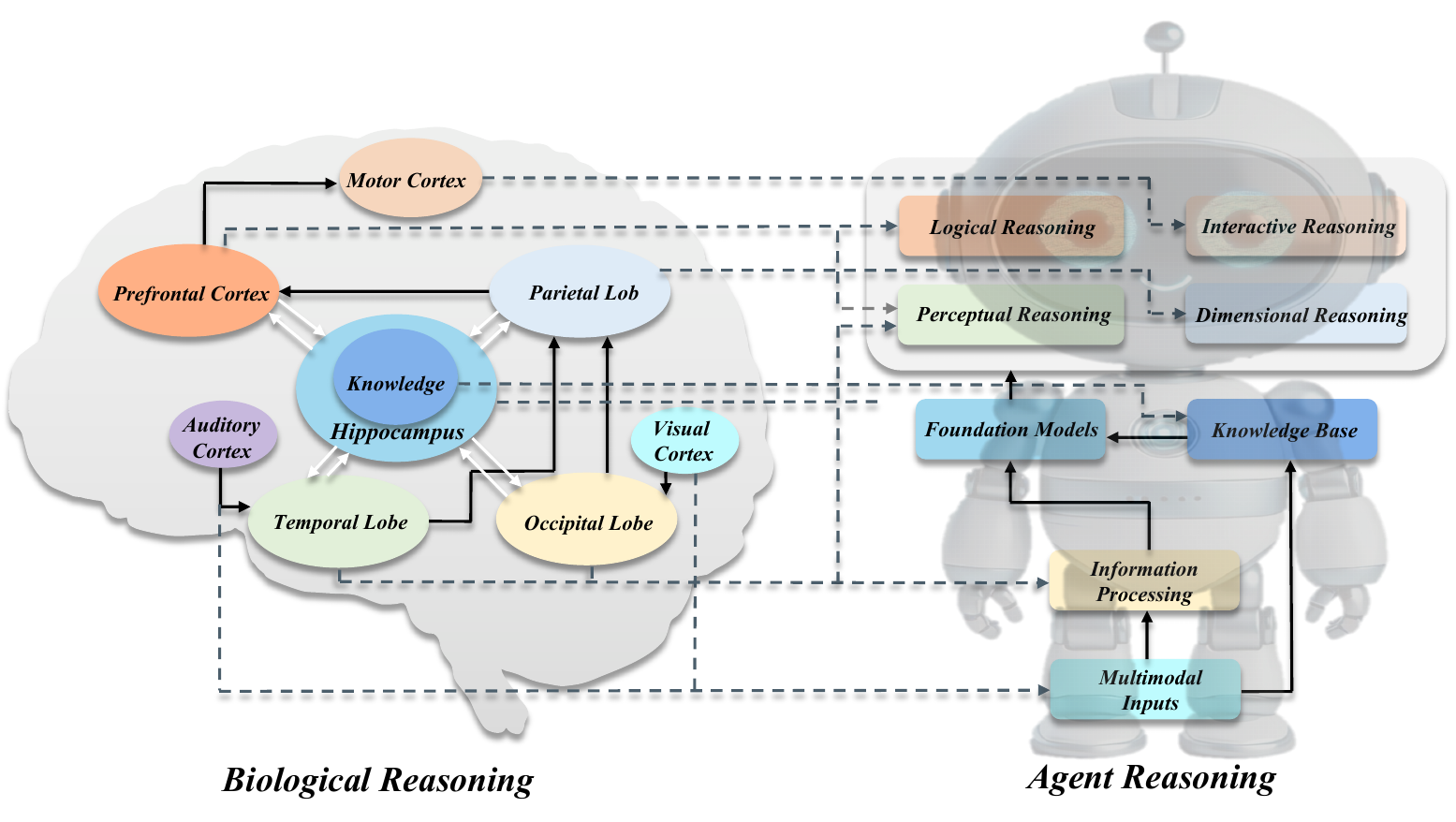}
    \vspace{-20pt}
    \caption{ The proposed neuroscience-inspired framework for agentic reasoning. The left panel illustrates the human brain’s reasoning process, where sensory inputs are processed through modality-specific cortices and integrated in higher association areas such as the parietal and prefrontal cortices. This enables abstract reasoning and decision-making, supported by predictive coding mechanisms and memory retrieval from the hippocampus. Inspired by this cognitive flow, the right panel presents a corresponding architecture for AI agents, consisting of sensory input, multi-level information processing, foundational understanding (via foundation models), factual memory storage (knowledge base), and a centralized reasoning module for adaptive and context-aware decision-making. White arrows denote top-down predictive signals based on predictive coding; black arrows represent the forward reasoning process; and dashed lines indicate the conceptual mapping between human brain functions and agent modules.}
    \label{fig:neuro_reasoning}
\end{figure*}

 Human reasoning is a \textbf{continuous} and 
\textbf{dynamic cycle} that enables individuals to process information, generate inferences, take actions, and refine knowledge over time as shown in Fig.~\ref{fig:neuro_reasoning} (left). This process begins with multi-modal perception, where external stimuli--such as visual, auditory, and textual inputs~\cite{ficsek2023cortico, warrier2009relating}--are integrated with prior knowledge and lived experience. While this may resemble Bayesian inference processes used in artificial agents, human reasoning exhibits distinct capabilities: it operates in highly uncertain, open-ended environments, leverages abstract analogies, and adapts flexibly in real-time based on minimal cues. For instance, a human can intuitively infer a person’s emotional state from subtle shifts in tone, gesture, or eye movement and adjust behavior accordingly, something current AI agents still struggle to do reliably~\cite{shank2019feeling}. This kind of nuanced, socially grounded inference arises not just from data-driven computation but from embodied experiences, neural priors, and a deep contextual understanding of the world. Once information is processed, the human mind engages in inference mechanisms, evaluating possibilities, predicting outcomes, and formulating decisions~\cite{fuster2008prefrontal}. These inferences are not static; they evolve in response to feedback, continuously updating internal cognitive models~\cite{rao1999predictive}. The reasoning outputs manifest as actions~\cite{ebbesen2017motor} that interact with the environment, and critically, the results of these actions are internalized as structured memory and knowledge. This recursive reasoning-action loop enables continual learning, robust generalization, and effective decision-making in dynamic, ambiguous scenarios.

\par \textbf{Human reasoning mechanisms provide valuable insights for enhancing AI agents' ability to handle complex tasks.} One notable example is how the brain tackles high-complexity reasoning problems under limited attentional resources. Due to the bottleneck in attentional capacity~\cite{otermans2022working,tombu2011unified}, humans cannot process large amounts of information simultaneously. Instead, they rely on \textbf{serial reasoning}, where problems are broken down into manageable steps and solved iteratively as shown in Tab.~\ref{tab:multistep-reasoning} (ACR-T~\cite{anderson2014atomic}). This principle directly aligns with Chain of Thought (CoT)~\cite{wei2022chain} reasoning for LLMs, which structures problem-solving as a sequence of intermediate inference steps to enhance accuracy and coherence. By mirroring this stepwise approach, AI agents can better manage computational complexity and improve reasoning performance.

\par Despite these insights, \textbf{AI agentic reasoning mechanisms still fall short of human cognition}, particularly in autonomous agents navigating dynamic and unpredictable environments.  Most AI models, including LLMs and reinforcement learning agents, rely on static architectures and feedforward processing, \textbf{lacking the iterative refinement and feedback mechanism}~\cite{lamme2000distinct,semedo2022feedforward}. Unlike human cognition, which continuously integrates new information to refine understanding, AI systems typically cannot adjust their reasoning in real time. Another key limitation is \textbf{long-term adaptability}. Humans not only adjust their immediate reasoning steps but also update their internal mental models~\cite{rodrigo1992updating} when exposed to new experiences. In contrast, AI agents typically operate within fixed training paradigms, restricting their ability to incorporate novel knowledge into existing frameworks. This rigidity leads to poor generalization in novel or complex scenarios. Furthermore, AI agents struggle with \textbf{multi-modal integration}. Human cognition seamlessly combines sensory inputs—such as vision, sound, and touch—into a coherent understanding. For example, we can easily relate derivatives to slopes in mathematics, drawing analogies across different domains~\cite{holyoak2025thinking}. In contrast, AI models process each modality separately, limiting their ability to perform cross-modal reasoning and effectively interpret ambiguous situations. Finally, agentic reasoning \textbf{lacks global comprehension and causal inference}. Many models, especially LLMs, rely on autoregressive predictions based on local context rather than a holistic understanding. This results in strong pattern recognition but weak causal reasoning, long-term dependencies, and counterfactual thinking—key elements of human intelligence essential for complex decision-making.

\par \textbf{To address these challenges, this paper explores agentic reasoning inspired by neuroscience, examining how current agentic reasoning mechanisms compare to human cognitive processes.} By analyzing how agents process information, adapt to new knowledge, integrate memory, and perform cross-modal reasoning, we aim to highlight both strengths and limitations in existing agent systems. Our goal is to provide insights that guide the development of more flexible, adaptive, and robust reasoning models, ultimately advancing AI agents toward greater autonomy and generalization capabilities.
% \begin{figure}
%     \centering
%     \includegraphics[width=0.95\linewidth]{figures/reasoning_trend.pdf}
%     \caption{Google Scholar results for research topics related to agent reasoning. The vertical axis represents the number of publications (in thousands), while the horizontal axis denotes the publication year. The figure highlights a significant rise in "\textbf{LLM agent reasoning}" publications since 2023, reflecting the impact of large language models on the field. }
%     \label{fig:reasoning_trend}
% \end{figure}
\par Our paper is the \textbf{\textit{first}} to systematically examine agentic reasoning from a \textbf{\textit{neuroscience}} perspective, distinguishing itself from prior works that primarily focus on foundation models such as LLMs~\cite{sun2023survey,sui2025stop,li2025system} and multimodal models~\cite{wang2025multimodal,bi2025reasoning} \textbf{as they are more akin to relatively static and passive knowledge repositories in the human brain rather than complete reasoning systems.} In contrast, ~\cite{sun2023survey} primarily explores reasoning mechanisms within LLMs, while ~\cite{li2025system} differentiates between heuristic and deliberate reasoning, but neither provides a systematic discussion on how an AI agent, as a whole, performs reasoning. \textbf{Our key contribution lies in establishing a comprehensive agentic reasoning framework that spans from sensory to motor action, grounded in neuroscience principles as shown in Fig.~\ref{fig:neuro_reasoning} (right).} This structured definition lays the foundation for future research on enhancing AI agents' reasoning capabilities.
\par In Sec.~\ref{sec:definition}, we establish the conceptual and theoretical basis for agentic reasoning by integrating insights from cognitive neuroscience. We begin by introducing three formal definitions of reasoning derived from neuroscientific perspectives, which are complemented by corresponding mathematical formulations and grounded in biological reasoning processes observed in the brain. Building on these foundations, we develop a unified framework that captures the full reasoning cycle—from sensory perception to decision-making and action. Central to this framework is the identification of four core reasoning modalities: perceptual, dimensional, logical, and interactive. These categories reflect distinct functional subsystems within the human brain and serve as the organizational backbone for our subsequent analysis.  In Sec.~\ref{sec:taxonomy}, \textbf{rather than merely categorizing existing reasoning methods, we systematically reinterpret and introduce them through the lens of our neuroscience-inspired agentic reasoning framework.} By situating current approaches within this structured paradigm, we examine their underlying technical mechanisms, assess the extent to which they align with human cognitive processes, and identify key limitations that hinder their generalization. This analytical perspective not only clarifies the current landscape of AI reasoning but also reveals critical gaps and opportunities for future development. In Sec.~\ref{sec:benchmark}, we systematically categorize existing reasoning tasks and datasets based on our proposed agentic reasoning framework. Rather than providing a general overview, we align benchmarks with specific reasoning types—perceptual, dimensional, logical, and interactive reasoning—allowing for a structured analysis of current evaluation methods. \textbf{Furthermore, we identify key gaps in existing benchmarks and propose new challenge tasks that better capture the complexities of real-world AI agentic reasoning, paving the way for more comprehensive and rigorous evaluation standards.} Sec.~\ref{sec:application} examines the applications of current reasoning methods. Here, we review practical implementations of AI agentic reasoning techniques across diverse domains, including autonomous navigation, visual question answering, robotics, and human-agent interaction. The discussion not only illustrates the strengths and limitations of existing approaches but also underscores the importance of multi-modal integration and dynamic decision-making in real-world scenarios. Finally, Sec.~\ref{sec:future_work} explores future directions for agentic reasoning by identifying key limitations of current AI agents and drawing insights from neuroscience. Rather than merely outlining general trends, we propose new research directions inspired by cognitive models. \textbf{We examine additional cognitive architectures and mechanisms that could inspire more advanced AI agentic reasoning paradigms.} By leveraging established neuroscience models, we highlight potential pathways for improving AI agents' adaptability, sequential inference, and knowledge integration, providing a biologically motivated perspective on the future of agentic reasoning.

In summary, our major contributions are as follows:
\begin{itemize}

 \item \textbf{Establishing a Neuroscience-Based AI Agentic Reasoning Framework.} Unlike prior surveys that primarily focus on reasoning in foundation models, we are the first to systematically examine agentic reasoning from a neuroscience perspective. We construct a comprehensive framework spanning from perception to action, providing a structured foundation for future research.

\item \textbf{A Framework-Based Analysis with Systematic Analysis of Reasoning Methods.} Unlike conventional task-based surveys, our work adopts a novel framework-based approach. We systematically categorize and analyze existing reasoning methods within our neuro-inspired framework, evaluating their technical characteristics, alignment with human cognition, and key challenges.

\item \textbf{Identifying Limitations and Proposing Future Directions.} We systematically identify key limitations in current agentic reasoning models, including challenges in adaptability, generalization, and multistep reasoning. Based on these insights, we propose future research directions to enhance agentic reasoning capabilities.

\item \textbf{Developing an Open-Source Repository for Agentic Reasoning Research.} To facilitate future studies, we curate and release a structured repository that organizes benchmark tasks, datasets, and reasoning-related papers based on our proposed framework, serving as a valuable resource for advancing AI agentic reasoning. We will continuously update the repository to enhance its utility.

\end{itemize}
\section{Neuroscience-inspired Agentic Reasoning}
\label{sec:definition}

Understanding the nature of reasoning requires an interdisciplinary approach, drawing insights from cognitive science, psychology, and neuroscience. From a neuro-scientific perspective, reasoning is not a singular or isolated cognitive function but rather a dynamic and multi-faceted process that enables individuals to derive conclusions, solve problems, and make decisions. It involves the interaction of memory, perception, and executive functions, orchestrated across various specialized 
 neural circuits. Reasoning allows for adaptive responses to novel situations by leveraging prior experiences while continuously incorporating new information ~\cite{bichindaritz1995case, mansouri2011ontology}.  

The underlying mechanisms of reasoning can be characterized by three fundamental principles~\cite{krawczyk2017reasoning}. First, reasoning operates as a hybrid process, integrating prior knowledge with newly acquired information to support both familiar and innovative outcomes across varied contexts. Second, it functions as an integrative and recursive system that synthesizes multiple diverse inputs into a coherent output, whether a mental representation or a physical action. This output can, in turn, serve as a new input for subsequent reasoning, enabling continuous refinement and dynamic adaptation. Third, reasoning follows a structured, multistep progression, ensuring that mental processes are systematically navigated toward a conclusion. These principles collectively define reasoning as a core cognitive function with a structured yet flexible nature.

\begin{figure}
    \centering
     
    \includegraphics[width=0.95\linewidth]{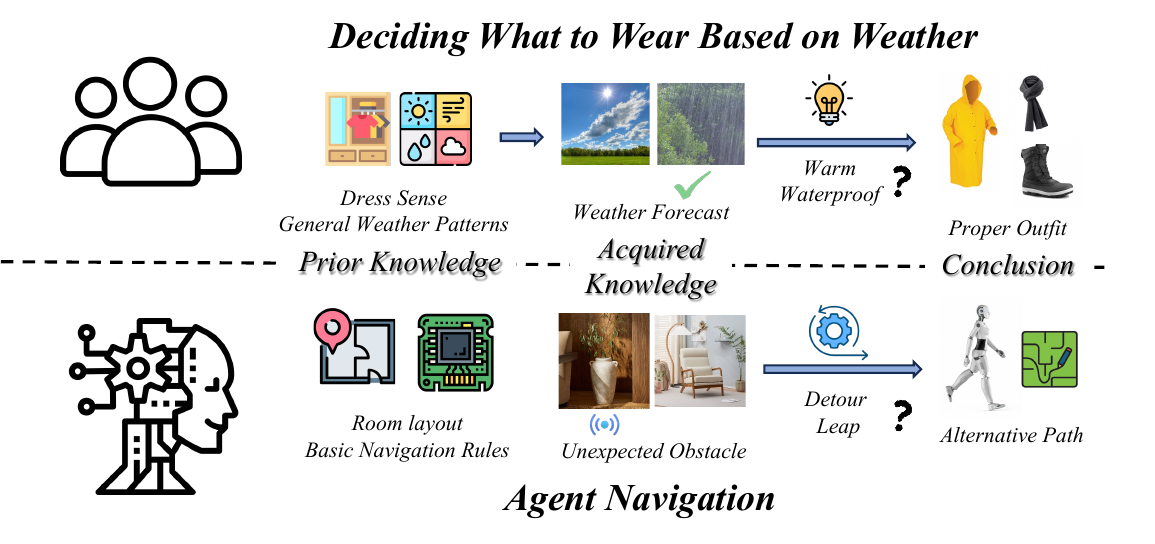}
    \vspace{-10pt}
    \caption{\textbf{The hybrid nature of reasoning in humans and AI agents}. Reasoning is a fusion of prior knowledge and new information, forming a hybrid process. This section provides examples: 1) Human Reasoning, deciding what to wear based on past knowledge and weather forecasts, and 2) Agentic Reasoning, adjusting navigation in response to unexpected obstacles.}
    \label{fig:reasoning_comparison}
\end{figure}

\noindent \textbf{Hybrid Nature of Reasoning.} Reasoning is inherently a complex hybrid process that synthesizes prior knowledge with new information, as shown in Fig. \ref{fig:reasoning_comparison}. Some outcomes arise from novel recombination of past experiences, while others depend on the integration of entirely new inputs. This dual mechanism highlights the balance between learned patterns and the generation of original solutions, which makes reasoning both adaptive and deeply creative. 

\noindent \textbf{Recursive Input-Output Integration.} As an essential cognitive mechanism, reasoning processes diverse inputs to produce meaningful outputs. These outputs can manifest as internally generated thoughts or externally executed actions, both of which result from complex neural computations. Specifically, it can be considered as a recursive cognitive process in which outputs often serve as new inputs, enabling continuous refinement of thought and behavior.  The ability to combine information from different sources is fundamental for logical deduction, problem solving, and decision making.  

\noindent \textbf{Multistep Structured Process.} Reasoning follows a structured, multistep progression in which various cognitive pathways contribute to the final outcome. Each step builds upon previous elements, ensuring that reasoning is not merely reactive but follows a deliberate and organized trajectory toward a conclusion. This structured approach underpins the sequential nature of logical inference and systematic thought. Table \ref{tab:multistep-reasoning} illustrates this multistep process as observed in neuroscience models, highlighting how different stages unfold over time. 
\begin{table*}
 \centering
\caption{Multistep Reasoning Models in Neuroscience. Abbreviations: PFC means Prefrontal Cortex, DLPFC means Dorsolateral Prefrontal Cortex, and ACC means Anterior Cingulate Cortex.}
\scalebox{0.92}{
\begin{tabular}{c|c|c|c|c}
\toprule        
\textbf{Models}&\textbf{Key Insight
              }&\textbf{Multistep Process} &\begin{tabular}[c]{@{}c@{}}\textbf{Type of Reasoning}\\ \textbf{  in Neuroscience} \end{tabular} & \begin{tabular}[c]{@{}c@{}} \textbf{ Categories of} \\ \textbf{Reasoning} \end{tabular}\\  
\midrule

 \begin{tabular}[c]{@{}c@{}} Miller and \\ Cohen's Model\\~\cite{miller2001integrative} \end{tabular}&\begin{tabular}[c]{@{}c@{}} Cognitive control  in \\ the prefrontal cortex (PFC)\\ for task management. \end{tabular}&  \begin{tabular}[c]{@{}c@{}}Sequential steps in cognitive control:\\ 1. Active maintenance of goal representations (PFC). \\
2. Bias signals guide neural pathways.\\
3. Adjustments made in neural maps \end{tabular}&  \begin{tabular}[c]{@{}c@{}}Executive control, \\decision-making, \\task management. \end{tabular} &  \begin{tabular}[c]{@{}c@{}}  Logical, \\ Interactive \end{tabular}\\  
\midrule

\begin{tabular}[c]{@{}c@{}} Banich's Cascade \\of Control Model\\~\cite{banich2009executive}\end{tabular}&\begin{tabular}[c]{@{}c@{}}Brain regions work in \\a sequence to manage \\ attention and response.\end{tabular}& \begin{tabular}[c]{@{}c@{}}Sequential cascade: \\1. Posterior DLPFC selects attentional set.
\\2. Mid-DLPFC selects task-relevant representation.
\\3. Posterior ACC selects the response.
\\4. Anterior ACC evaluates the response\end{tabular}& \begin{tabular}[c]{@{}c@{}}Attention regulation, \\error correction, \\decision-making.\end{tabular}& \begin{tabular}[c]{@{}c@{}}   Logical, \\ interactive.  \end{tabular}  \\
\midrule

\begin{tabular}[c]{@{}c@{}} Baddeley’s Working \\Memory Model\\~\cite{baddeley2020working,baddeley2000episodic}\end{tabular}&\begin{tabular}[c]{@{}c@{}} The components of working \\memory: central executive, \\phonological loop, \\visuospatial sketchpad, \\episodic buffer.\end{tabular} &\begin{tabular}[c]{@{}c@{}} Multi-component process: \\1. Central executive directs attention and \\controls processes.
\\2. Phonological loop and visuospatial sketchpad \\ manage information in parallel.
\\3. Episodic buffer integrates info across domains\end{tabular}&\begin{tabular}[c]{@{}c@{}} Memory, \\multitasking,\\ cognitive resource \\management. \end{tabular} & 
   \begin{tabular}[c]{@{}c@{}}   Dimensional, \\interactive.     \end{tabular} \\ 
\midrule

Predictive Coding~\cite{rao1999predictive} & \begin{tabular}[c]{@{}c@{}} Brain constantly updates a\\ mental model to predict \\sensory inputs and minimize \\prediction error. \end{tabular}&\begin{tabular}[c]{@{}c@{}} Prediction and update: \\1. Brain generates predictions of sensory input.
\\2. Predictions compared to actual sensory inputs.
\\3. Large prediction errors lead to model updates. \end{tabular}& \begin{tabular}[c]{@{}c@{}}Prediction, \\learning, \\error correction,\\ perception.\end{tabular} &  \begin{tabular}[c]{@{}c@{}} Perceptual,\\ logical.  \end{tabular}
\\ 
\midrule

 \begin{tabular}[c]{@{}c@{}}Adaptive Control of\\ Thought—Rational \\(ACT-R)\\~\cite{anderson2014atomic} \end{tabular}&\begin{tabular}[c]{@{}c@{}}Cognitive model based on \\discrete cognitive operations\\ for declarative and\\ procedural knowledge.\end{tabular}&\begin{tabular}[c]{@{}c@{}}Step-by-step task execution: \\1. Chunks (declarative) stored in memory.
\\2. Procedural knowledge (productions) \\guides task execution, following a \\ \textbf{seriation-based} sequence.
\\3. Modules (e.g., visual, manual) interact with \\ environment.\end{tabular}&\begin{tabular}[c]{@{}c@{}}Task execution,\\ problem-solving, \\cognitive coordination.\end{tabular}  &   \begin{tabular}[c]{@{}c@{}}   Logical, \\ interactive. \end{tabular}\\  
\midrule
SOAR~\cite{laird1987soar} & \begin{tabular}[c]{@{}c@{}}Symbolic cognitive architecture \\ using production rules for \\ goal-directed behavior. \end{tabular} & \begin{tabular}[c]{@{}c@{}}Sequential steps in goal-directed behavior: \\1. Encodes problems into symbolic states. \\
2. Uses production rules to decompose goals. \\
3. Applies search and learning in symbolic space.   \end{tabular} & \begin{tabular}[c]{@{}c@{}} Executive planning,\\ symbolic manipulation,\\ goal decomposition. \end{tabular}&   \begin{tabular}[c]{@{}c@{}}  Logical,\\ interactive. \end{tabular}\\
\midrule
\begin{tabular}[c]{@{}c@{}}Global Workspace \\ Theory (GWT)~\cite{baars1993cognitive} \end{tabular}&\begin{tabular}[c]{@{}c@{}} Consciousness emerges from \\ globally broadcasting \\ selected information. \end{tabular} &\begin{tabular}[c]{@{}c@{}}Sequential processing of conscious content:\\1. Sensory and cognitive content compete for attention.
\\2. Selected content is broadcast to all subsystems.
\\3. Subsystems integrate and act on this information. \end{tabular} &\begin{tabular}[c]{@{}c@{}}  Conscious access, \\attention regulation, \\cross-module integration. \end{tabular}& \begin{tabular}[c]{@{}c@{}}Perceptual, \\interactive.\end{tabular}\\
%  \begin{tabular}[c]{@{}c@{}} Hierarchical Temporal \\Memory (HTM)\\~\cite{hawkins2004intelligence}\end{tabular}& \begin{tabular}[c]{@{}c@{}}Machine learning model \\inspired by the human \\brain’s neocortex, used for \\anomaly detection.\end{tabular}&\begin{tabular}[c]{@{}c@{}}Learning temporal sequences: \\1. Constantly learns from time-based patterns.
% \\2. Builds a hierarchical representation of sequences.
% \\3. Uses learned sequences for prediction \\ and anomaly detection.\end{tabular} &\begin{tabular}[c]{@{}c@{}} Pattern recognition, \\prediction, \\anomaly detection.\end{tabular}  &   \begin{tabular}[c]{@{}c@{}}  Perceptual,\\Dimensional \end{tabular}\\ 
\bottomrule
\end{tabular}
}
\label{tab:multistep-reasoning}
\end{table*}

% By synthesizing prior knowledge, integrating multiple inputs, and adhering to a structured cognitive progression, reasoning emerges as a fundamental function of intelligent thought. This perspective provides a neuroscientific foundation for understanding reasoning as a dynamic and adaptable process within the human brain. To deepen this understanding, we now turn our attention to the underlying structure of reasoning itself, exploring how it operates in real-world cognitive tasks.

\subsection{Foundation of Reasoning A Hybrid Process}
For reasoning to maintain its complexity, it must go beyond automatic recall. Implicit memories, such as those described by Knowlton, Mangels, and Squire in 1996~\cite{knowlton1996neostriatal}, do not qualify as reasoning because they evoke behavior without conscious deliberation. Instead, these are classic examples of learning, where past experiences directly influence actions without the cognitive synthesis characteristic of reasoning \cite{taylor1997implicit}. In contrast, whenever we integrate new information, whether through unfamiliar data or novel structuring of prior knowledge, we engage in the dynamic and complex process of reasoning. At times, reasoning relies heavily on well-established facts, while in other cases, it leans toward innovation and spontaneity. However, in most cases, reasoning occurs through a combination of previous knowledge and new information. Even when dealing primarily with known facts, reasoning still requires assembling these elements in a novel way \cite{smith2008case}. If a thought process merely outputs prior knowledge without reconfiguration, it ceases to be reasoning and instead resembles a learned or reflexive behavior.

As described by the first definition, reasoning does not function in isolation. In contrast, it relies on the interplay between what is already known and what is newly encountered. Reasoning is inherently a hybrid process, blending prior knowledge with new information \cite{abolghasem2023learning}. Although there are rare instances where reasoning occurs with entirely unfamiliar information to generate a completely novel conclusion, which we might call creative reasoning, most reasoning involves some degree of prior knowledge. This balance between past experiences and novel inputs allows us to ‘think on our feet' and adapt in real time, making solutions as we go \cite{kidd2017handbook}. In this context, new information refers to knowledge that was previously unknown or irretrievable in the given reasoning process. It can manifest itself in several ways:

\par \noindent \textbf{Introduction of Novel Data.} Novel data refers to entirely new inputs that were previously unknown to the reasoning system~\cite{Sunnevåg2009}. This can be commonly described as the continuous process of human beings' perception of the world with biological sensors. This type of information is external and requires active incorporation into the reasoning process. When encountering novel data, reasoning must adjust its existing knowledge structures, infer relationships, and possibly revise prior beliefs. Unlike simple recall or application of learned rules, reasoning in the presence of novel data demands more dynamic adaptation. This is particularly evident in real-time decision-making scenarios, where an agent or human must process unexpected inputs and generate new conclusions. The ability to integrate novel data is crucial for reasoning to remain flexible, ensuring that decisions are not solely based on outdated or incomplete prior knowledge. For example, a human doctor encountering a rare disease case must synthesize unfamiliar symptoms with known medical principles to form a diagnosis, rather than relying solely on past cases. On the other hand, from an intelligent agent perspective, a robotic vision system designed for warehouse navigation may encounter an obstacle type it has never seen before. Instead of failing, it must reason about potential workarounds using its existing spatial models and decision framework.
    
\par \noindent \textbf{Context-Independent Knowledge} consists of abstract principles, rules, or axioms that, while already known, were previously inactive in the reasoning process. Unlike novel data, which introduces external newness, context-independent knowledge is retrieved and applied in a novel context. This also includes knowledge that was not previously activated in discourse, as well as updates that shift probability distributions or modify existing reasoning structures, often emerging as the focal point of inference \cite{Sandi1978}. The reasoning process relies on dynamically incorporating such knowledge, allowing for the synthesis of new conclusions beyond mere memorization. By retrieving and restructuring fundamental principles, reasoning remains adaptable, enabling generalization across different domains and situations. Like when mathematicians solve a problem in an unfamiliar domain (e.g., applying graph theory concepts to network security), they may retrieve abstract mathematical principles that were not originally associated with the current problem but are applicable. By bringing the problem to what agents can do, a reinforcement learning agent trained to play chess may generalize strategic principles (e.g., controlling the center of the board) when encountering an entirely new board position it has never seen in training.

\par \noindent \textbf{Modification of Existing Knowledge and Revision of Assumptions.} These two categories are closely related to each other, both involving updates to knowledge. The prior category adds new, external facts that change the model without necessarily contradicting prior assumptions. However, the revision of assumptions involves adjusting or invalidating previous assumptions based on new evidence that contradicts earlier beliefs or conclusions, which is a key characteristic of non-monotonic reasoning \cite{BREWKA2008239}.  For instance, a person assumes a friend is at home because their car is parked outside, but upon learning that the friend took an Uber, they revise their assumption. Brought this further into the agent's perspective, a robot designed to monitor household activities may initially infer that a person is at home based on sensor data like car location. Upon receiving new information (e.g., a GPS update or direct communication from a smart device indicating the person took an Uber), the robot revises its belief, updating its internal model to reflect the change in the person's status. This is similar to how an AI agent in a logistics system might update its delivery assumptions based on live traffic data, revising expected arrival times in real time.

% \par \noindent \textbf{Structural Markers in Language.} New information indicated by word order and stress in communication \cite{Most1979Information}. Such as in the sentence "JOHN baked the cake," stress on "John" implies it was not someone else, emphasizing new information. Similarly, an NLP-based robot or AI system processes language input with a focus on prosody and emphasis, allowing it to detect structural markers in speech \cite{shriberg2008practical}. For example, if a robot is performing sentiment analysis or intent detection, it adjusts its response according to the emphasis detected in human speech. If the user says, "I didn't ask you to do that," with a stressed ask, the AI can infer that there’s frustration or a corrective need, altering its behavior accordingly.  
Thus, complexity is central to reasoning. It cannot be reduced to mere repetition of past knowledge. Rather, it thrives on the interplay between what we know and what we learn. Understanding this hybrid nature of reasoning lays the groundwork for examining how such processes are instantiated in the human brain, particularly through the lens of neuroscience.

% \begin{figure*}
%     \centering
%     \includegraphics[width=1.0\linewidth]{figures/reasoning_process.pdf}
%     \caption{ A neuroscience-inspired framework for agent reasoning. The left panel illustrates the human brain’s reasoning process, where sensory inputs are processed through modality-specific cortices and integrated in higher association areas such as the parietal and prefrontal cortices. This enables abstract reasoning and decision-making, supported by predictive coding mechanisms and memory retrieval from the hippocampus. Inspired by this cognitive flow, the right panel presents a corresponding architecture for AI agents, consisting of sensory input, multi-level information processing, foundational understanding (via foundation models), factual memory storage (knowledge base), and a centralized reasoning module for adaptive and context-aware decision-making. Blue arrows denote top-down predictive signals based on predictive coding; black arrows represent the forward reasoning process; and dashed lines indicate the conceptual mapping between human brain functions and agent modules.}
%     \label{fig:neuro_reasoning}
% \end{figure*}
\subsection{Mathematical Foundation of Reasoning Behavior}

Building upon the conceptual foundation of reasoning as a hybrid process, we now shift our focus to its formal representation. To bridge biological insights with computational understanding, mathematical modeling provides a formal framework for capturing reasoning behavior, which can be used to analyze, simulate, and predict reasoning behavior. By abstracting cognitive mechanisms into mathematical forms--such as logic-based systems, probabilistic models, or optimization frameworks--we gain not only deeper theoretical insights but also practical tools for designing intelligent agents and understanding human cognition at a larger scale.

Mathematical models provide several advantages in the study of reasoning. First, they offer a precise and unambiguous way to describe reasoning processes, ensuring clarity in theoretical frameworks. Unlike purely descriptive approaches, mathematical formulations enable predictability, allowing researchers to anticipate outcomes based on specific inputs. This predictability is particularly valuable in fields like AI and neuroscience, where computational models of reasoning must be robust and reliable. Additionally, mathematical representations facilitate the implementation of reasoning mechanisms in computational systems, making them essential for AI agent applications such as natural language processing, decision-making, and automated theorem proving. By formalizing reasoning mathematically, researchers can also develop generalizable frameworks that apply across multiple disciplines, from cognitive science to robotics and machine learning.  

\par Several mathematical frameworks have been developed to represent reasoning processes. Serving as the cornerstone, Bayesian Brain Theory (BBT) ~\cite{knill2004bayesian, colombo2012bayes} suggests that the brain functions as a probabilistic inference machine, continuously updating its beliefs about the environment using Bayesian inference. 
Bayesian inference models reasoning as a probabilistic process in which prior beliefs are updated in response to new evidence. This approach is useful in dealing with uncertainty and dynamic environments. Predictive coding ~\cite{huang2011predictive, aitchison2017or}, another influential model, describes how the brain minimizes errors in perception and cognition by continuously updating internal models of the world. The free energy principle ~\cite{friston2010free, friston2006free} extends this idea further, proposing that the brain functions as an optimization system that seeks to minimize uncertainty in its predictions. In addition to these probabilistic approaches, formal logic remains a crucial mathematical tool for reasoning. Logic-based models, such as propositional and first-order logic, provide a structured framework for deductive reasoning and are widely used in rule-based AI agent systems. Furthermore, decision theory and optimization techniques frame reasoning as a problem of selecting the best action based on a cost or reward function. 

\par The reasoning process in the brain can be understood as a continuous cycle of perception, inference, and decision-making, governed by probabilistic models. The brain receives sensory inputs from the environment, interprets them through predictive models, and updates its internal beliefs based on new information. This process can be mathematically formulated using principles from Bayesian inference, predictive coding, and free energy minimization. The following sections will explore these mathematical representations in greater detail, illustrating how they contribute to our understanding of the reasoning process.

\subsubsection{Bayesian Inference in the Brain}

The brain updates its belief about a hidden state \(H\) given sensory data \(D\) using Bayes' theorem:

\begin{equation}
    P(H | D) = \frac{P(D | H) P(H)}{P(D)},
\end{equation}
where \(P(H | D)\) is the \textbf{posterior probability}, representing the updated belief after observing \(D\), \(P(D | H)\) is the \textbf{likelihood}, describing the probability of the data given \(H\), \(P(H)\) is the \textbf{prior probability}, representing prior beliefs about \(H\), and \(P(D)\) is the \textbf{evidence}, normalizing the probability distribution.

\subsubsection{Predictive Coding Model}

Predictive coding suggests that the brain minimizes the difference between sensory input \(x_t\) and its internal predictions \(\hat{x}_t\):

\begin{equation}
    \epsilon_t = x_t - \hat{x}_t,
\end{equation}
where \(x_t\) is the actual sensory observation, \(\hat{x}_t\) is the predicted sensory input, and \(\epsilon_t\) is the \textbf{prediction error}.

The brain refines its internal model by minimizing \(\epsilon_t\), adjusting beliefs through an optimization process:

\begin{equation}
    \frac{dH}{dt} \propto -\frac{\partial F}{\partial H},
\end{equation}
where \(F\) is the \textbf{variational free energy}.

\subsubsection{Free Energy Principle}

The brain minimizes \textbf{variational free energy} \(F\) to approximate true Bayesian inference:

\begin{equation}
    F = D_{KL}(Q(H) || P(H | D)),
\end{equation}
where \(Q(H)\) is the approximate posterior, \(P(H | D)\) is the true posterior, and \(D_{KL}(Q || P)\) is the \textbf{Kullback-Leibler (KL) divergence}, measuring the difference between the two distributions. Minimizing \(F\) ensures that the brain's internal model aligns with reality.

\subsubsection{Decision-Making as Bayesian Optimization}

Decision-making in BBT can be formulated as a \textbf{Bayesian reinforcement learning} problem, where the brain selects an optimal policy \(\pi^*\) that maximizes expected rewards:

\begin{equation}
    \pi^* = \arg\max_{\pi} \sum_{t=0}^{T} \mathbb{E}_{P(s_t | s_{t-1}, a_{t-1})} [R(s_t, a_t)],
\end{equation}
where \(s_t\) is the state at time \(t\), \(a_t\) is the action taken at time \(t\), \(R(s_t, a_t)\) is the reward function, \(\mathbb{E}\) represents the expectation over possible state transitions.

\begin{table}[t!]
    \centering
     \caption{Summary of mathematical symbols used in BBT}
    \resizebox{0.45\textwidth}{32mm}{
    \begin{tabular}{cc}
        \toprule
        \textbf{Symbol} & \textbf{Meaning} \\
        \midrule
        \(P(H | D)\) & Posterior probability (updated belief) \\
        \(P(D | H)\) & Likelihood (data given hypothesis) \\
        \(P(H)\) & Prior probability (initial belief) \\
        \(P(D)\) & Evidence (marginal likelihood) \\
        \(x_t\) & Sensory input at time \(t\) \\
        \(\hat{x}_t\) & Predicted sensory input \\
        \(\epsilon_t\) & Prediction error \\
        \(F\) & Variational free energy \\
        \(D_{KL}\) & Kullback-Leibler divergence \\
        \(Q(H)\) & Approximate posterior \\
        \(\pi^*\) & Optimal policy \\
        \(s_t\) & State at time \(t\) \\
        \(a_t\) & Action at time \(t\) \\
        \(R(s_t, a_t)\) & Reward function \\
        \bottomrule
    \end{tabular}
    }
   
    \label{tab:notation}
\end{table}

Bayesian Brain Theory models cognition as a probabilistic inference system. The brain continually updates its beliefs using Bayesian inference, minimizes prediction errors via predictive coding, optimizes free energy for efficient learning, and makes decisions based on Bayesian optimization principles.

While mathematical models provide a powerful abstraction of reasoning behavior, it is equally crucial to examine how reasoning unfolds biologically within the brain. This motivates an exploration of the neural substrates and pathways involved in reasoning from a neuroscience perspective. However, reasoning is not a monolithic process. It manifests in various forms, each characterized by different structures, objectives, and mechanisms. To appreciate the breadth of reasoning behaviors, it is important to explore their underlying typologies. 

\subsection{Reasoning Process of Neuroscience}
Building on the hybrid model of reasoning, it becomes essential to investigate how these cognitive mechanisms are realized biologically. Neuroscience offers a compelling perspective by mapping reasoning onto neural substrates and examining the functional architecture that supports it. From prefrontal cortex activity to dynamic network interactions, neuroscience provides insight into how the brain orchestrates reasoning processes in structured and uncertain environments.

\par From a neuroscience perspective, the reasoning process involves the collaboration of multiple brain regions (Fig.~\ref{fig:neuro_reasoning} left).  The reasoning process begins with the brain receiving various modality-specific sensory inputs from the external environment. For instance, visual information is first captured by the retina and transmitted via the lateral geniculate nucleus (LGN) of the thalamus to the primary visual cortex (V1) in the occipital lobe\cite{ficsek2023cortico}, while auditory information is processed through the medial geniculate nucleus (MGN) of the thalamus and sent to Heschl’s gyrus in the temporal lobe\cite{warrier2009relating}.  These sensory pathways rely on a combination of electrical signaling along axons and chemical neurotransmission at synapses, where neurotransmitters (e.g., glutamate, GABA) mediate the transfer of information across neurons. These primary sensory cortices extract fundamental features such as edges, motion, frequency, and pitch before relaying the processed information to higher-order association areas.

\par The parietal lobe plays a crucial role in multimodal integration, particularly in spatial awareness, numerical reasoning, and body coordination~\cite{fogassi2005parietal}. Here, sensory inputs from updated knowledge (e.g., vision and audition) are combined, allowing the brain to construct a coherent representation of the environment. Meanwhile, according to the theory of Predictive Coding~\cite{rao1999predictive}, the brain is hypothesized to actively generate predictive signals for expected stimuli~\cite{rao1999predictive}.  These top-down signals are sent back to the sensory cortices, where they are compared against incoming sensory inputs. Any discrepancies between prediction and perception trigger updates to the brain’s internal model. This adaptive updating is biologically implemented through synaptic plasticity, the process by which the strength of synaptic connections between neurons is modified based on experience. A well-studied form of this mechanism is spike-timing dependent plasticity (STDP), where the precise timing of spikes between presynaptic and postsynaptic neurons determines whether synaptic weights are strengthened or weakened. %STDP enables the brain to capture causal relationships and sequence dependencies, making it a fundamental mechanism for dynamic learning and adaptive reasoning.

\par As information is integrated, it is processed in the prefrontal cortex (PFC), which serves as the central hub for abstract thinking, decision-making, and logical reasoning~\cite{krawczyk2017reasoning}. The PFC refines predictions, evaluates uncertainty, and formulates complex cognitive responses based on contextual memory and learned experiences~\cite{fuster2008prefrontal}. During this process, different parts of the brain need to stay coordinated. This is achieved in part through neural oscillations—rhythmic patterns of brain activity that help different brain areas communicate efficiently. These oscillations play an important role in maintaining attention and keeping information in working memory during reasoning. In addition, the decision-making process can be described by the drift-diffusion model (DDM), which suggests that the brain gradually accumulates evidence over time before making a choice. This helps explain why some decisions take longer than others and how the brain balances speed and accuracy. Once a decision is made, the information is passed to the motor cortex~\cite{ebbesen2017motor}, where it is translated into actions.

\par Notably, reasoning is not limited to immediate perception-action cycles but is deeply intertwined with memory mechanisms. The hippocampus, in conjunction with the cerebral cortex, plays a vital role in episodic memory formation and retrieval~\cite{knierim2015hippocampus}. Through hippocampal-cortical interactions, new experiences are encoded into long-term memory via synaptic plasticity mechanisms such as STDP, which adjust synaptic strengths based on neural activity patterns. These changes reinforce the knowledge base that supports future reasoning. Over time, frequently used information undergoes systems consolidation, transferring from the hippocampus to cortical networks, enabling more efficient recall and inference~\cite{shipp2007structure}.

\par Thus, neural reasoning is an iterative, predictive, and memory-driven process, integrating sensory information, updating internal models, and leveraging past experiences to guide cognition and behavior. While neuroscience helps uncover where and how reasoning occurs in the brain, it is equally important to understand how this process can be formalized and abstracted into structured models that can be simulated, predicted, and analyzed.

\subsection{Agent Reasoning Framework}
Inspired by the biological reasoning process, we propose an \textbf{Agent Reasoning Framework} that mirrors the layered structure of human cognition, as illustrated on the right side of Fig.~\ref{fig:neuro_reasoning}. The reasoning pipeline begins with multimodal sensory modules, which are then integrated by the information processing module and used to update the knowledge base, working alongside the foundation model to support more complex 
 higher-level reasoning.

\subsubsection{Multimodal Input Module} The multimodal input module, as the first layer of the Agent Reasoning Framework, is responsible for transmitting information from the external world to the internal cognitive system. This module corresponds to the sensory systems in the biological brain, capable of receiving various forms of sensory stimuli from the environment, such as vision, hearing, language, and touch, and transmitting these signals into the internal representation space in a structured form. At this stage, the agent does not passively receive all sensory information but possesses the ability for active perception and selective attention. This module dynamically allocates attention resources based on the goals and context of the current task, enhancing relevant input and reducing redundant or sensory content. As a result, the system can maintain a stable and focused perceptual state in the face of complex,  changing environments, laying the foundation for subsequent information processing and reasoning. %More importantly, this module does not treat different modalities in isolation; rather, it seeks to align information from each modality within a unified internal representation structure. The construction of cross-modal consistency enables the agent to integrate information more effectively, establish logical connections and semantic correspondences between modalities, and avoid reasoning barriers caused by fragmented information.

\subsubsection{Information Processing Module} In the human brain, sensory signals from different modalities, such as vision, audition, and touch, are ultimately converted into a unified electrochemical signal format for transmission and processing. This unified encoding mechanism enables the brain to efficiently integrate information across modalities, forming stable and coherent internal representations. Inspired by this neural mechanism, our agent requires an information processing module to map input signals from multiple sensory channels into a shared, high-dimensional representation space. This module would not rely on modality-specific encoding paths but instead utilize a modality-agnostic unified representation, enabling natural flow and mutual activation of information across different channels. This unified representation mechanism enhances the system’s ability to understand complex scenarios and provides a continuous, composable foundation for knowledge retrieval and subsequent reasoning, thereby establishing a neural-like structural foundation for multimodal cross-domain reasoning.

\subsubsection{Knowledge Base Module} Human knowledge acquisition relies not only on the accumulation of past experiences but also heavily on continuous interaction with the external environment. In the brain, the long-term memory system gradually absorbs knowledge from repeated perception and actions, while the working memory system dynamically retrieves information relevant to the current context, enabling flexible responses to changing environments. Inspired by this dual-memory mechanism, our framework requires a dual-channel knowledge system. On one hand, the agent maintains an internal, continually updated knowledge base that accumulates experience from long-term interaction and provides stable, context-rich support for reasoning. On the other hand, the agent incorporates a time-sensitive retrieval mechanism that enables real-time access to external knowledge sources, allowing for rapid integration of novel or dynamic information. These two systems work in close coordination during the reasoning process: internal knowledge ensures coherence and personalized adaptation, while external knowledge offers flexibility and broad coverage. Together, they form a brain-inspired dynamic knowledge architecture that integrates accumulation and activation, enabling the agent to sustain robust and timely reasoning even in complex, evolving, or unfamiliar scenarios.

\subsubsection{Foundation Models Module}
In our framework, the foundation model plays a crucial dual role, drawing inspiration from the brain’s memory execution system. It serves as an advanced understanding engine, responsible for interpreting and processing external inputs in a highly adaptive and dynamic manner. Essentially, it acts as a highly efficient executor of memory, continuously trained and updated within a knowledge base to enhance comprehension. Much like how the brain constantly refines its cognitive models through the integration of long-term memory and accumulated experiences, our foundation model strengthens its understanding of the world by learning from ongoing data and constantly evolving contexts. However, the foundation model does not merely serve as a knowledge repository; it also functions as a versatile reasoning assistant. It supports various types of reasoning tasks by seamlessly integrating multimodal sensory inputs, structured knowledge, and prior reasoning rules. Based on these reasoning rules, the foundation model assists in executing specific reasoning tasks. For instance, according to logical inference rules, it could support tasks like deducing conclusions from a set of premises or identifying contradictions within a series of statements. Similarly, it may apply spatial reasoning rules to help with tasks like predicting the movement of objects in a dynamic environment. In this way, the foundation model acts as a flexible and robust platform, applying learned rules and representations to assist in higher-level reasoning and decision-making. It does not replace specialized reasoning modules but rather scaffolds them by preprocessing inputs, suggesting candidate inferences, and enforcing structured knowledge derived from past interactions. As a result, the foundation model plays a dual role: it serves as a cognitive substrate for understanding, while also facilitating adaptive, flexible, and modular reasoning across complex, real-world tasks.

\subsubsection{Reasoning Module} In the reasoning system of our brain-inspired agent, the reasoning module serves as the core component responsible for organizing task-specific reasoning rules and leveraging the foundation model to execute them. Inspired by the neural mechanisms of the prefrontal cortex, which governs rule extraction and decision control, and the parietal cortex, which integrates multimodal information and constructs spatiotemporal representations, we believe that the reasoning module should exhibit a task-oriented and structured architecture. For various types of reasoning—perceptual, dimensional, logical, and interactive—it derives tailored reasoning strategies and execution paths, utilizing the foundation model as a reasoning assistant to perform the cognitive computations required by each task. For example, in logical reasoning, the model can apply inference rules such as modus ponens (“If A, then B”) to conduct conditional judgments and generate conclusions. The specific reasoning tasks and their categorization are discussed in detail in Sec~\ref{sec:classification}. More importantly, reasoning is not a static process but one that supports dynamic adaptation. The outcomes of reasoning are not only written back into the knowledge base to provide contextual support for future tasks, but they also continuously refine the foundation model’s internal reasoning mechanisms. Through repeated task execution and feedback accumulation, the model gradually develops more adaptive reasoning structures and strategy selection capabilities, thereby enhancing its generalization and responsiveness across diverse tasks. This closed-loop structure of “rule-guidance $\xrightarrow{}$ model execution $\xrightarrow{}$ result feedback $\xrightarrow{}$ rule refinement” forms the most autonomous and growth-driven core of brain-inspired reasoning.

\par Compared to prominent cognitive frameworks, our model offers a more comprehensive ability to handle multimodal inputs, dynamic reasoning tasks, and continuous updates to the knowledge base. Unlike SOAR~\cite{laird1987soar}, which emphasizes a unified cognitive system but is limited in handling dynamic environments and multimodal inputs, our framework continuously updates its knowledge base, allowing reasoning outcomes to adapt to changing environments and enhancing reasoning efficiency and adaptability. In contrast to Global Workspace Theory (GWT)~\cite{baars1993cognitive}, which focuses on information integration but lacks flexible knowledge base updates capacity, our model ensures continuous accumulation and updating of knowledge, enabling more efficient information flow in reasoning and decision-making. Additionally, while Dual-Process Theory~\cite{evans2008dual} distinguishes between fast, automatic responses and slower, deliberate reasoning without effectively integrating the two systems, our framework supports both rapid responses and more precise, adaptive reasoning by integrating flexible knowledge updates and reasoning modules. Overall, our framework combines unified multimodal representations, continuous knowledge updates, and flexible reasoning modules, allowing for efficient handling of complex reasoning tasks and adaptation to dynamic real-world environments, showcasing unique strengths in comprehensiveness and adaptability.

\subsection{Classifications of Reasoning Behavior}
\label{sec:classification}
\begin{figure*}
    \centering
    \includegraphics[width=0.98\linewidth]{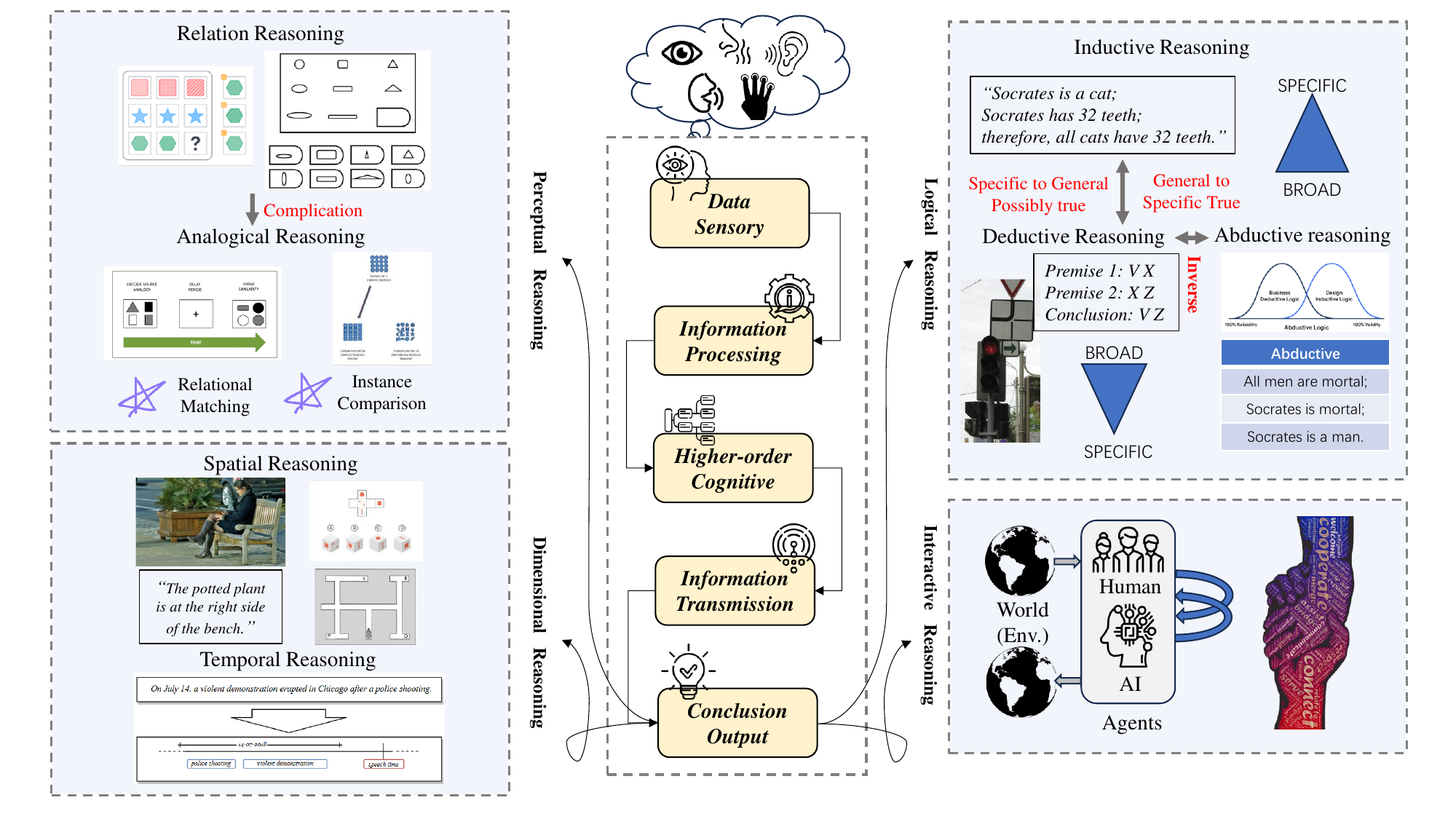}
    \caption{The overview of the reasoning process and classification of reasoning behavior from a neuro-perspective. This diagram presents a comprehensive framework of reasoning inspired by human cognitive and neural mechanisms. At the center, a hierarchical reasoning pipeline, spanning data sensory input, information processing, higher-order cognition, and conclusion generation, mirrors the flow of information in biological systems. Surrounding this core are five major categories of reasoning behaviors: perceptual reasoning, driven by multisensory integration; dimensional reasoning, encompassing spatial and temporal inference; relation reasoning, involving analogical thinking and relational matching; logical reasoning, covering inductive, deductive, and abductive logic; and interactive reasoning, focusing on agent-agent and agent-human collaboration within dynamic environments. Together, these components establish a neuro-cognitively grounded taxonomy that bridges biological inspiration and computational implementation in artificial intelligence systems.}
    \label{fig:overview}
\end{figure*}

Reasoning encompasses a diverse array of cognitive strategies, each serving distinct functions in human thought and problem-solving. Building on insights into the neural mechanisms underlying reasoning, we now examine how reasoning behaviors are classified within cognitive science and psychology. Decades of research into the neural basis of reasoning have produced several influential theoretical frameworks. Synthesizing the most widely accepted hypotheses~\cite{krawczyk2017reasoning, lake2017building}, reasoning can be categorized into four primary types, as illustrated in Fig.~\ref{fig:overview}: Perceptual Reasoning, Dimensional Reasoning, Logical Reasoning, and Interactive Reasoning. Each category represents a distinct mode of information processing, ranging from interpreting sensory inputs to applying formal logic, analyzing multi-dimensional relationships, and engaging in collaborative, context-sensitive reasoning.

Perceptual Reasoning refers to the cognitive ability to acquire, interpret, and manipulate information derived from sensory modalities such as vision, audition, and touch. From a neuroscience perspective, this form of reasoning is closely associated with activity in the occipital and parietal lobes \cite{mulder2012bias, keuken2014brain}, which are involved in processing visual input and integrating multi-sensory information. It enables individuals to detect patterns, make inferences, and solve problems without completely relying on verbal or linguistic cues. Core components of perceptual reasoning include relational reasoning, such as analogy detection, relational matching, and instance comparison, all of which are fundamental in tasks like matrix reasoning and visual puzzle-solving. These processes engage neural mechanisms responsible for feature extraction, similarity assessment, and categorical abstraction. For example, participants may be asked to identify shared attributes among objects, recognize visual analogies, or distinguish meaningful differences between stimuli. Perceptual reasoning thus underpins a wide range of nonverbal cognitive functions and is a foundational element in intelligence testing and adaptive behavior in dynamic environments.

Dimensional Reasoning\cite{stock1998spatial, kubinger2023dimensionality} involves the integration of cognitive processes across multiple representational domains, such as spatial configurations\cite{clements1992geometry}, temporal dynamics~\cite{schaeken1996mental}, and abstract hierarchical relationships. This form of reasoning engages higher-order cognitive functions to interpret and manipulate complex, multi-dimensional structures. From a neuroscience standpoint, dimensional reasoning recruits distributed neural circuits, particularly involving the parietal cortex for spatial manipulation, the prefrontal cortex for maintaining abstract rules and hierarchies, and the medial temporal lobe for encoding temporal sequences and event-based dependencies. Tasks requiring dimensional reasoning often involve understanding 3D object relationships, predicting dynamic system behavior, or analyzing the interdependence of multiple variables. These abilities are foundational in domains such as engineering, mathematics, and the physical sciences, where interpreting structured, multi-variable information is critical. Empirical investigations commonly assess dimensional reasoning through nonverbal problem-solving tasks, such as mental rotation, hierarchical pattern completion, and sequential logic exercises, each of which probes the brain's capacity to synthesize and navigate complex cognitive representations across multiple axes of abstraction. Importantly, dimensional reasoning also plays a pivotal role in agent-based reasoning systems, where an autonomous agent must interpret high-dimensional sensory inputs and dynamically adapt to changing task constraints within complex environments.

Logical Reasoning ~\cite{allwein1996logical, bronkhorst2020logical} follows structured principles of inference and is divided into inductive, deductive, and abductive reasoning. Inductive reasoning moves from specific observations to broader generalizations, forming conclusions that are possibly true. Deductive reasoning, in contrast, starts from general premises and derives specific, logically certain conclusions. Abductive reasoning works by finding the most plausible explanation for given evidence, commonly used in diagnostics and hypothesis generation. These logical processes form the foundation of rational thinking and decision-making.

Interactive Reasoning focuses on the dynamic exchange of information between humans, agents, and the environment. Unlike other reasoning types that occur within an individual’s mind, interactive reasoning involves collaboration and adaptation, where agents refine their understanding through interaction. This is important in AI-driven decision-making, autonomous systems, and cooperative problem-solving, where reasoning is influenced by external inputs and evolving conditions.
In essence, reasoning behavior can be understood through these four categories, each playing a crucial role in human cognition and artificial intelligence. Whether derived from sensory data, structured logic, multi-dimensional analysis, or collaborative engagement, reasoning enables intelligent systems to interpret the world, make informed decisions, and adapt to complex scenarios.
\section{Comprehensive Analysis of Agentic Reasoning }
\label{sec:taxonomy}
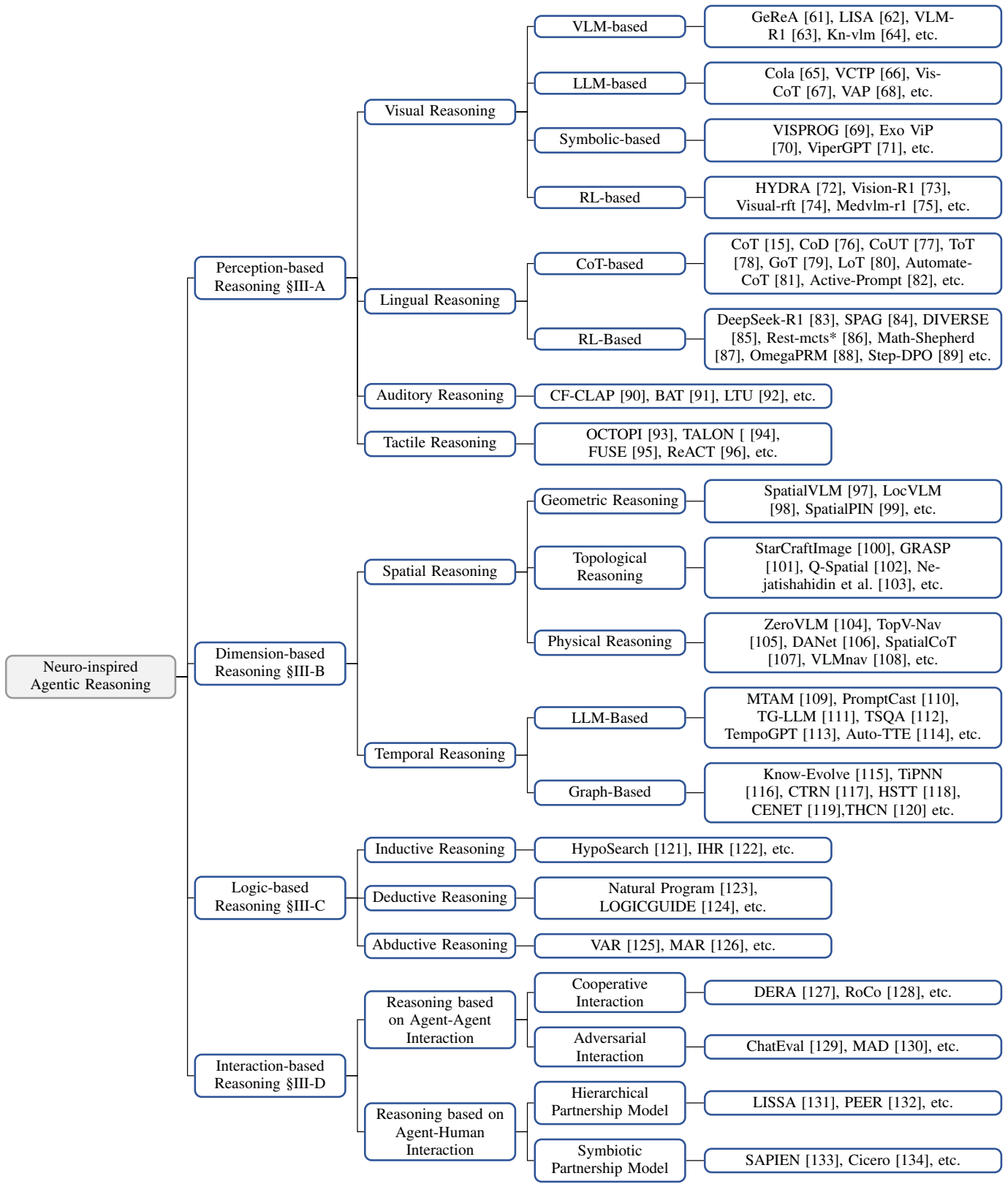
\begin{figure*}[!ht]
\scriptsize
    \begin{adjustbox}{width=\textwidth}
        \begin{forest}
        for tree={
                forked edges,
                grow'=0,
                draw,
                rounded corners,
                node options={align=center},
                text width=2.7cm,
                s sep=6pt,
                calign=edge midpoint, 
            },
            [Neuro-inspired Agentic Reasoning, fill=gray!45, parent
                [Perception-based Reasoning \S\ref{sec:Perception-based Reasoning}, for tree={perception_based}
                    [Visual Reasoning, perception_based 
                        [VLM-based, perception_based
                            [{GeReA \cite{ma2024gerea}, LISA \cite{lai2024lisa}, VLM-R1 \cite{shen2025vlm}, Kn-vlm \cite{tan2025kn}, etc.}, perception_based_work]
                        ]
                        [LLM-based, perception_based
                            [{Cola \cite{chen2023large}, VCTP \cite{chen2024visual}, VisCoT \cite{shao2024visual}, VAP \cite{xiao2024enhancing}, etc.}, perception_based_work]
                        ]
                        [Symbolic-based, perception_based
                            [{VISPROG  \cite{gupta2023visual}, Exo ViP \cite{wang2024exovip}, ViperGPT~\cite{suris2023vipergpt},  etc.}, perception_based_work]
                        ]
                        [RL-based, perception_based
                            [{HYDRA \cite{ke2024hydra}, Vision-R1 \cite{huang2025vision}, Visual-rft \cite{liu2025visual}, Medvlm-r1 \cite{pan2025medvlm}, etc.}, perception_based_work]
                        ]
                    ]
                    [Lingual Reasoning, perception_based 
                        [CoT-based, perception_based
                            [{CoT \cite{wei2022chain}, CoD \cite{xu2025chain}, CoUT \cite{zhang2024smartagent}, ToT \cite{yao2023tree}, GoT \cite{besta2024graph}, LoT \cite{zhao2023enhancing}, Automate-CoT \cite{shum2023automatic}, Active-Prompt \cite{diao2023active}, etc.}, perception_based_work]
                        ]
                        [RL-Based, perception_based
                            [{DeepSeek-R1  \cite{guo2025deepseek}, SPAG \cite{cheng2024self}, DIVERSE \cite{li2022making}, Rest-mcts* \cite{zhang2024rest}, Math-Shepherd \cite{wang2023math}, OmegaPRM \cite{luo2024improve}, Step-DPO \cite{lai2024step} etc.}, perception_based_work]
                        ]
                    ]
                    [Auditory Reasoning, perception_based
                        [{CF-CLAP \cite{vosoughi2024learning}, BAT \cite{zheng2024bat}, LTU \cite{gong2023listen}, etc.}, perception_based_work]
                    ]
                    [Tactile Reasoning, perception_based
                        [{OCTOPI \cite{yu2024octopi}, TALON [ \cite{jiang2024talon}, FUSE \cite{jones2025beyond}, ReACT \cite{lai2024vision}, etc.}, perception_based_work]
                    ]
                ]
                [Dimension-based Reasoning \S\ref{sec:Dimension-based Reasoning}, for tree={dimension_based}
                    [Spatial Reasoning, dimension_based 
                        [Geometric Reasoning, dimension_based
                            [{SpatialVLM \cite{chen2024spatialvlm}, LocVLM  \cite{ranasinghe2024learning}, SpatialPIN  \cite{ma2024spatialpin}, etc.}, dimension_based_work]
                        ]
                        [Topological Reasoning, dimension_based
                            [{StarCraftImage  \cite{kulinski2023starcraftimage}, GRASP \cite{tang2024grasp}, Q-Spatial \cite{liao2024reasoning}, Nejatishahidin et al. \cite{nejatishahidin2024structured}, etc.}, dimension_based_work]
                        ]
                        [Physical Reasoning, dimension_based
                            [{ZeroVLM \cite{meng2024know}, TopV-Nav \cite{zhong2024topv}, DANet \cite{li2023weakly}, SpatialCoT \cite{liu2025spatialcot}, VLMnav \cite{goetting2024end}, etc.}, dimension_based_work]
                        ]
                    ]
                    [Temporal Reasoning, dimension_based 
                        [LLM-Based, dimension_based
                            [{MTAM \cite{qiu2023can}, PromptCast   \cite{xue2023promptcast}, TG-LLM \cite{xiong2024large}, TSQA \cite{yang2024enhancing}, TempoGPT \cite{zhang2025tempogpt}, Auto-TTE \cite{chung2023text}, etc.}, dimension_based_work]
                        ]
                        [Graph-Based, dimension_based
                            [{Know-Evolve  \cite{trivedi2017know}, TiPNN  \cite{dong2024temporal}, CTRN \cite{jiao2023improving}, HSTT \cite{bai2024event}, CENET \cite{xu2023temporal},THCN~\cite{chen2024thcn} etc.}, dimension_based_work]
                        ]
                    ]
                ]
                [Logic-based Reasoning \S\ref{sec:Logic-based Reasoning}, for tree={logic_based}
                    [Inductive Reasoning, logic_based
                        [{HypoSearch \cite{wang2023hypothesis}, IHR \cite{qiu2023phenomenal}, etc.}, logic_based_work]
                    ]
                    [Deductive Reasoning, logic_based
                        [{Natural Program \cite{ling2023deductive}, LOGICGUIDE \cite{poesia2023certified}, etc.}, logic_based_work]
                    ]
                    [Abductive Reasoning, logic_based
                        [{VAR \cite{liang2022visual}, MAR  \cite{li2023multi}, etc.}, logic_based_work]
                    ]
                ]
                [Interaction-based Reasoning \S\ref{sec:Interaction-based Reasoning}, for tree={interaction_based}
                    [Reasoning based on Agent-Agent Interaction, interaction_based
                        [Cooperative Interaction, interaction_based
                           [{DERA \cite{nair2023dera}, RoCo  \cite{mandi2024roco}, etc.}, interaction_based_work]
                        ]
                        [Adversarial Interaction, interaction_based
                           [{ChatEval \cite{chan2023chateval}, MAD \cite{liang2023encouraging}, etc.}, interaction_based_work]
                        ]
                    ]
                    [Reasoning based on Agent-Human Interaction, interaction_based
                        [Hierarchical Partnership Model, interaction_based
                           [{LISSA  \cite{ali2020virtual}, PEER \cite{schick2022peer}, etc.}, interaction_based_work]
                        ]
                        [Symbiotic Partnership Model, interaction_based
                           [{SAPIEN \cite{hasan2023sapien}, Cicero \cite{meta2022human}, etc.}, interaction_based_work]
                        ]
                    ]
                ]
            ]   
        \end{forest}
    \end{adjustbox} 
    \caption{\textbf{Taxonomy of Agentic Reasoning Techniques Inspired by Neuroscience.} This hierarchical structure organizes reasoning methods in artificial agents based on cognitive mechanisms inspired by neuroscience, including dimensional, perceptual, logical, and interactive reasoning, highlighting the integration of biologically plausible mechanisms into artificial intelligence systems. This taxonomy highlights how agents can emulate human-like reasoning across diverse tasks and environments.}
    \label{fig:sec3_taxonomy}
\end{figure*}

Having explored reasoning from a neuro-cognitive perspective and its mathematical foundations, we now shift our focus to reasoning in agents. Reasoning of agent seeks to replicate, enhance, or extend human cognitive abilities through computational models, enabling intelligent systems to process information, infer conclusions, and make decisions. Over the years, research in AI has developed diverse approaches to reasoning, each with its own underlying principles and methods. These approaches can be broadly categorized based on how they represent knowledge, handle uncertainty, interact with external environments, and apply logical structures.  

To better understand the landscape of AI reasoning and finding valuable potential directions, based on the inspiration from the neuroscience perspective on the reasoning behavior, we introduce a taxonomy that organizes majority reasoning approaches into four main categories as classified in Fig. \ref{fig:overview}: dimension-based reasoning, perception-based reasoning, interaction-based reasoning, and logic-based reasoning. Each category reflects a distinct perspective on how reasoning can be structured and applied in AI systems. Dimension-based reasoning examines how abstract representations, such as spatial, temporal, or multi-modal structures, influence reasoning capabilities. Perception-based reasoning focuses on how AI systems extract and process information from raw sensory inputs, often using neural models to interpret visual, auditory, or textual data. Interaction-based reasoning explores reasoning within dynamic environments, emphasizing real-world engagement through learning, adaptation, and collaboration with humans or other agents. Logic-based reasoning, rooted in formal symbolic methods, remains a cornerstone of AI, providing structured frameworks for rule-based inference, knowledge representation, and verification.  

By classifying AI reasoning into these four main categories as shown in Fig. \ref{fig:sec3_taxonomy}, this taxonomy offers a structured lens through which we can analyze existing research, identify trends, and assess the strengths and limitations of different approaches. Subsequently, we explore each category in detail, highlighting its core principles, representative methodologies, and recent advancements in the field. This classification not only facilitates a clearer understanding of AI reasoning but also provides insights into how these approaches can be integrated to create more robust and versatile intelligent systems.

To ensure a comprehensive and high-quality survey of AI reasoning research, we established a rigorous selection benchmark for choosing relevant papers. Our selection process prioritizes papers published in top-tier conferences and journals across multiple research domains related to agented reasoning on AI or Robotics, such as NeurIPS, CVPR, TPAMI, JMLR, ICRA, and TOR.
Our selection criterion extends beyond publication venues to include relevance across different reasoning paradigms. Given that reasoning in AI spans multiple subfields, we categorized papers based on their contributions to dimension-based, perception-based, interaction-based, and logic-based reasoning, ensuring balanced representation across all reasoning approaches. We also considered interdisciplinary relevance, including works from cognitive science, neuroscience, and formal logic that contribute to AI reasoning methodologies. Since relevant work from Nature and its sub-journals is also included, such as Nature Neuroscience, Nature Communications, and Nature Machine Intelligence.

To maintain a balance between classical and emerging trends, we selected both foundational papers that have shaped AI reasoning and recent advancements that reflect the latest breakthroughs in neural-symbolic integration, large-scale reasoning models, and interactive AI systems. By applying these selection benchmarks, we ensured that our survey provides a comprehensive, well-structured, and up-to-date overview of AI reasoning, capturing both theoretical developments and practical implementations across diverse domains.

\subsection{Perception-based Reasoning}
\label{sec:Perception-based Reasoning}

\begin{table*}[t]
\centering
\caption{Representative Works in Perception-Based Reasoning.}
  %\renewcommand{\arraystretch}{2} % 设置行间距
  %\begin{tabular}{|m{2.5cm}|m{2.5cm}|m{3.5cm}|m{5.5cm}|} % 使用m{}使内容垂直居中
\scalebox{1.1}{
\begin{tabular}{c|c|c|c|c}
\toprule
\textbf{Category} & \textbf{Method} & \textbf{Publication} & \textbf{Backbone}  & \textbf{Highlights}\\
\midrule
\multirow{4}{*}{Visual}    & VISPROG~\cite{gupta2023visual}  & CVPR'2023 & Neuro-symbolic & Visual Programming  \\
& Lisa~\cite{lai2024lisa} & CVPR'2024 & VLM   &   Reasoning Segmentation         \\  
&   Cola~\cite{chen2023large} & NeurIPS'2023 & LLM   &  LLM Coordinates VLMs  \\
&   VisCoT~\cite{shao2024visual} & NeurIPS'2024 & LLM   &  Visual Chain-of-Thought  \\
\midrule
\multirow{13}{*}{Lingual}    & SPAG~\cite{cheng2024self}  & NeurIPS'2024 & LLM & Self-playing Adversarial Language Game  \\
&  CoT Prompting~\cite{wei2022chain}  & NeurIPS'2022 &  LLM  & Chain-of-Thought prompting  \\
& LoT~\cite{zhao2023enhancing} & COLING'2024 & LLM  & Grounding CoT Reasoning With Logic           \\  
& RBRLHF~\cite{guo2025deepseek} & arXiv'2025  & LLM & Rule-based RL With Human Feedback \\
& Self-Consistency~\cite{wang2022self} & NeurIPS'2022 & LLM & New decoding strategy sampling diverse reasoning paths \\
& ToT~\cite{yao2023tree} & NeurIPS'2023 & LLM & Generalizes CoT\\
& GoT~\cite{besta2024graph} & AAAI'2023 & LLM & Modelling LLM information as a graph\\
& Automate-CoT~\cite{shum2023automatic} & EMNLP'2023 & LLM & Automatically augmenting rational chains\\
& Active-Prompt~\cite{diao2023active} & ACL'2023 & LLM & New method for choosing task-specific CoT exemplars\\
& Fine-Tune-CoT~\cite{ho2022large} & ACL'2023 & LLM & Large teacher models fine-tune smaller models\\
& AoT~\cite{hong2024abstraction} & EMNLP'2024 & LLM & Prompting abstract-to-concrete thinking\\
& CoC~\cite{li2023chain} & ICML'2024 & LLM & Combining code-writing with LM simulation \\
& ICoT~\cite{gao2024interleaved} & CVPR'2025 & VLM & Image-incorported multimodal Chain-of-Thought\\
\midrule
\multirow{2}{*}{Auditory}    & LTU~\cite{gong2023listen} & ICLR'2024 & AST, LLM & Model Integration and Multi-modal Reasoning \\
& CF-CLAP~\cite{vosoughi2024learning} & ICASSP'2024 & CLAP & Counter Factual Learning           \\
\midrule
\multirow{1}{*}{Tactile}    & ReAct~\cite{lai2024vision}  & IROS'2024 & VLM & Reasoning and Perception of Liquid Objects \\

\bottomrule
  \end{tabular}
  }
\end{table*}

Perception lies at the heart of intelligent behavior, serving as the primary interface between an agent and its environment. Perceptual reasoning refers to the ability of AI systems to interpret, integrate, and infer knowledge from raw sensory inputs--such as vision, language, audio, and tactile signals--to support higher-level cognition and decision-making. Unlike symbolic or logic-based reasoning that operates over abstract representations, perceptual reasoning grounds inference in multimodal sensory data, enabling agents to make sense of complex, ambiguous, or noisy inputs. This form of reasoning is particularly vital in real-world, unstructured environments where direct perception must inform tasks like object recognition, scene understanding, language grounding, or human-robot interaction. Vision language models (VLMs), audio-visual transformers, and multimodal fusion networks exemplify contemporary approaches that perform reasoning directly over perceptual streams. These systems must align modalities, resolve cross-modal ambiguities, and extract structured semantics from unstructured inputs.
Perceptual reasoning thus acts as a bridge between low-level perception and high-level cognition, equipping agents with the ability to derive meaningful conclusions from what they see, hear, or feel. In the following subsections, we investigate key techniques and models that enable perceptual reasoning, analyzing their architectures, reasoning strategies, and the challenges they face in aligning perception with intelligent behavior.

\subsubsection{Visual Reasoning}
The visual reasoning capabilities of Artificial Intelligence (AI) are principally evidenced in the comprehension, analysis, and inference of imagery and video data, propelling intelligent systems towards advanced stages of cognitive evolution. This inferential process encompasses not only fundamental object identification and detection but also extends to a profound understanding of object attributes, spatial relationships, and causal linkages. Such capabilities facilitate AI systems in demonstrating human-like reasoning skills across various tasks, including Visual Question Answering (VQA), image captioning, and video understanding. In contrast to conventional symbolic logic-based reasoning, visual reasoning necessitates the processing of extensive visual datasets and integrates multimodal information for comprehensive analysis. This approach significantly bolsters the precision and logical coherence of the resultant inferences.

\begin{figure}[t!]
        \centering
		\includegraphics[width=0.48\textwidth]{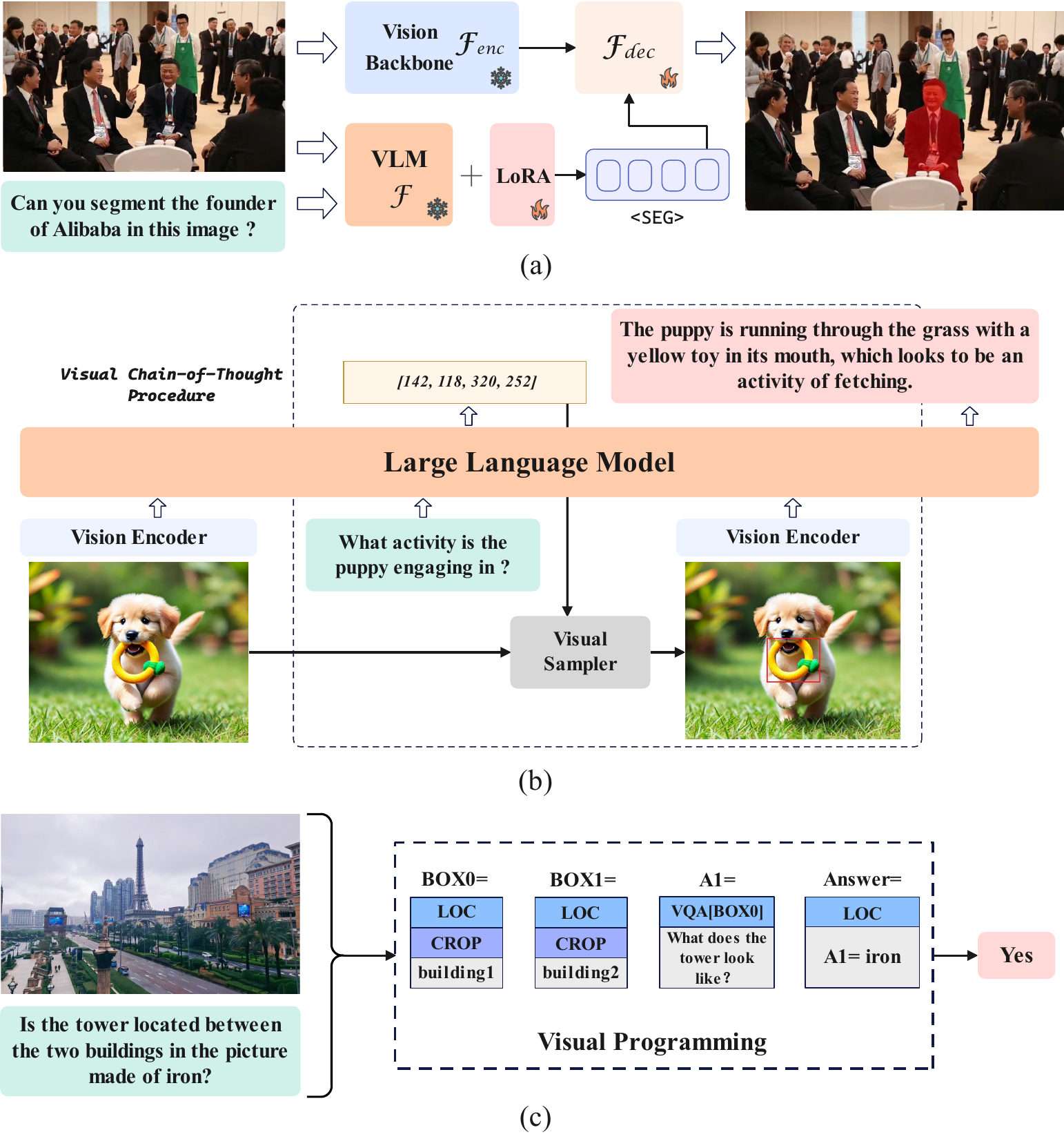}
\caption{Structure of different visual reasoning methods. (a) VLM-based approach~\cite{lai2024lisa} enhances its reasoning and segmentation capabilities. (b) LLM-based approach~\cite{shao2024visual} that enhances the model's performance in handling complex visual tasks by dynamically focusing on key image regions and incorporating multi-turn reasoning to progressively derive detailed information for generating accurate and interpretable answers. (c) Symbolic-based approach~\cite{gupta2023visual} that generates executable Python-like visual programs based on language instructions to solve vision tasks.}
\label{fig:vision}
\end{figure}

Based on underlying technologies, visual reasoning is divided into \textbf{\textit{Vision-Language Model (VLM)-based}}, \textbf{\textit{Large-Language Model (LLM)-based}}, \textbf{\textit{Symbolic-based}}, and \textbf{\textit{Reinforcement Learning (RL)-based}} approaches. VLM-based methods~\cite{shen2025vlm,ma2024gerea,lai2024lisa,tan2025kn,wang2024q} achieve cross-modal information complementarity through multimodal fusion (images and text), enhancing the model's global perception of complex visual logic and its ability to capture local details. For example, GeReA~\cite{ma2024gerea} proposes a new visual reasoning framework by inputting relevant visual information (regions in images related to questions) and linguistic information (questions and associated human prompts) into pretrained VLMs to generate question-aware prompt captions, combining image-question pairs with similar samples to feed into a multimodal reasoning model for joint knowledge-image-question representation learning. LISA~\cite{lai2024lisa} fine-tunes the VLM model by introducing a new token $\langle SEG \rangle$ in the model vocabulary as a segmentation output marker, decoding its hidden layer embeddings into segmentation masks, enhancing the reasoning and segmentation capabilities of VLMs as shown in Fig.~\ref{fig:vision}(a).  

LLM-based methods can be divided into two categories: Direct Invocation of LLMs~\cite{chen2023large,xiao2024enhancing,lan2023improving} and VCoT(Visual Chain-of-Thought)~\cite{rose2023visual,choi2025end,shao2024visual,chen2024visual,xu2024llava}. Among these, Cola~\cite{chen2023large} directly invokes LLMs for reasoning. It processes input images independently through multiple VLMs, generating visual descriptions and candidate answers. An LLM acts as a reasoning coordinator to analyze these descriptions and answers, identifying points of consensus and conflict, combining world knowledge for logical inference, and generating the final answer along with reasoning evidence. In terms of VCoT, VisCoT~\cite{shao2024visual} mimics human visual scanning and reasoning processes by dynamically focusing and conducting multi-turn reasoning, progressively deriving and locating key information to generate more accurate and interpretable answers, significantly enhancing the model's reasoning capabilities in complex visual scenarios as shown in Fig.~\ref{fig:vision}(b). VCTP~\cite{chen2024visual} adopts a three-stage reasoning approach called "See-Think-Confirm" to progressively accomplish knowledge-driven visual reasoning tasks. Initially (See), the model analyzes the image, detects all possible objects, and generates a global visual description. Next (Think), the LLM combines the question to select key visual concepts, generates region-specific descriptions, and reasons towards preliminary answers. Finally (Confirm), the LLM produces reasoning evidence and verifies the inference against visual evidence through cross-modal validation. 

Symbolic-based approaches~\cite{gupta2023visual,suris2023vipergpt,wang2024exovip, webb2023zero} aim to address the issues of data dependency, insufficient interpretability, and task rigidity—where most models require task-specific annotated data for training, limiting scalability; end-to-end models lack transparency, making it difficult to analyze error sources; and existing models for VQA and CV tasks are typically optimized for specific tasks, struggling to adapt to open-ended, combinatory real-world demands. 
As shown in Fig.~\ref{fig:vision}(c), VISPROG~\cite{gupta2023visual} leverages the in-context learning capabilities of large language models (LLMs) to automatically generate executable Python-like visual programs based on natural language instructions. It breaks down complex tasks and invokes existing computer vision (CV) models or Python logic operations to complete the tasks. ViperGPT~\cite{suris2023vipergpt} executes visual reasoning tasks by generating Python code. When receiving a visual query, ViperGPT uses a large language model to generate an executable Python program that calls multiple visual modules (e.g., object detection, depth estimation, etc.) and performs logical reasoning and mathematical computations. Exo ViP~\cite{wang2024exovip} builds upon VISPROG~\cite{gupta2023visual} by incorporating an "Exoskeleton" validation module to detect and correct errors during the reasoning process. It also employs tree search to select the optimal reasoning path, preventing error propagation, thereby improving the accuracy and robustness of compositional visual reasoning.

Introducing RL into VLM-based visual reasoning aims to improve the decision-making capability, controllability, and generalization ability of the model. Traditional VLMs mainly rely on supervised learning (SL) for training, but SL is often constrained by static data distributions, making it difficult to adapt to complex reasoning tasks in open environments. By integrating RL, the model can be optimized using reward mechanisms, allowing it to adjust strategies during multistep reasoning processes, thus improving the accuracy and coherence of answers. Additionally, RL helps the model better balance different reasoning paths, avoiding stereotypical errors in reasoning and increasing adaptability to long-tail questions. Combining RL with VLMs~\cite{ke2024hydra,huang2025vision,liu2025visual,pan2025medvlm,huang2024vlm} enables more flexible visual reasoning in open-world tasks, achieving stronger intelligent interaction capabilities. For instance, HYDRA~\cite{ke2024hydra} adopts incremental reasoning, storing and utilizing historical information to improve reasoning stability, and dynamically optimizes decisions through reinforcement learning to reduce error propagation. Vision-R1~\cite{huang2025vision} enhances the reasoning capabilities of VLMs through reinforcement learning, employing Group Relative Policy Optimization (GRPO) for training while incorporating Hierarchical Formatted Reward Refinement Function (HFRRF) to ensure reasoning quality. 

Current VLMs perform well in simple VQA tasks but exhibit significant limitations when handling complex visual tasks, such as abstract reasoning. The primary constraint lies in the inability of current visual encoders to effectively extract abstract visual features, such as spatial relationships and geometric structures, resulting in insufficient sensitivity to implicit geometric rules within images. Additionally, existing VLMs predominantly rely on contrastive learning or generative training paradigms, which struggle to capture intricate vision-language associations. These models depend heavily on text-driven reasoning rather than directly extracting logical relationships from visual features. To enhance the capability of VLMs in processing complex visual tasks, it is worth considering structural innovations in visual encoders to extract richer visual semantic information, as well as introducing benchmarks specifically designed for abstract reasoning.

Despite the latest GPT-o3 model proposing a new paradigm in visual understanding by integrating images into the chain of thought and performing transformation operations on images during the visual reasoning process to enhance visual comprehension and flexibility, it still exhibits certain limitations in VQA. While the GPT-o3 model demonstrates exceptional performance in parsing whiteboard sketches to derive formulas, inferring geographical locations from landscapes, and answering detailed questions about images, it remains constrained in some aspects. For instance, when presented with an image depicting six fingers, the GPT-o3 model is unable to accurately identify the number of fingers.

\subsubsection{Lingual Reasoning}
\begin{figure*}
\centering
\includegraphics[width=1.0\textwidth]{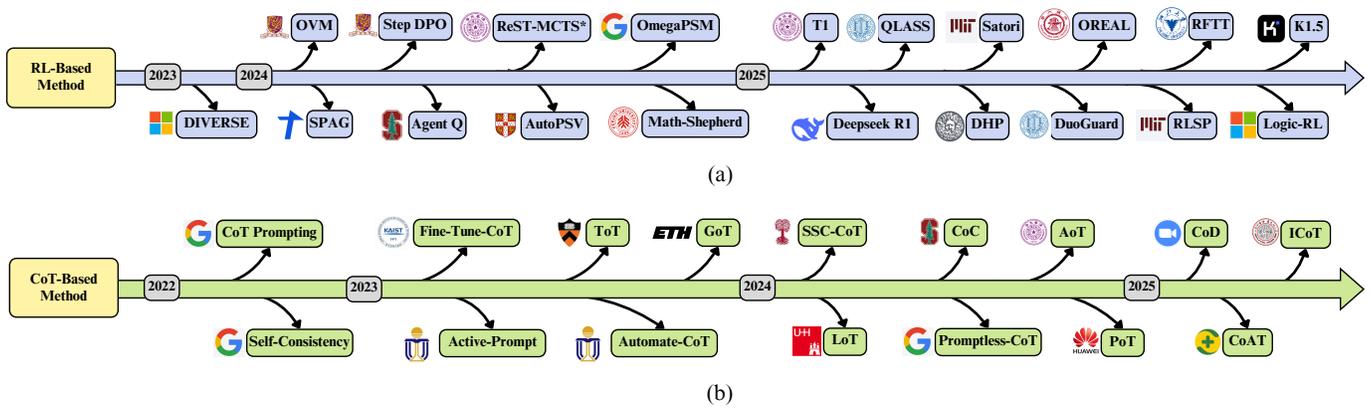}
\vspace{-25pt}
\caption{Evolution timeline for LLM lingual reasoning methods. (a): Evolution timeline for CoT-based methods: CoT prompting was first introduced in 2022. Throughout 2023 and 2024, 2 main types of optimizations existed. One aimed to better structure and guide reasoning (ToT, LoT, GoT etc.), the other focused on prompt optimization and automation (Automate-CoT, Active-Prompt, promptless CoT). The method reached its maturity and consolidating phase with fewer novel frameworks and more refinements, benchmarking, and integration into broader systems. (b): Evolution timeline for RL-based methods: Relatively new compared to CoT, and starting to flourish following the success of DeepSeek R1. Early attempts in 2023 and 2024 focused mostly on reward modelling (DIVERSE, StepDPO, Rest-MCTS* .etc). Later attempts in 2024 and 2025 focused mainly on Reinforced Fine-tuning, rule-based RL, supervised fine-tuning, and various other methods.}
\label{fig:evolution-timeline}
\end{figure*}

Neuroscientific studies have indicated that human reasoning does not primarily rely on the language centers of the brain~\cite{coetzee2022dissociating}. This biological distinction highlights a fundamental gap between natural and artificial reasoning mechanisms. In contrast, AI reasoning remains heavily dependent on large language models (LLMs), which serve as the main framework for linguistic reasoning. Although scaling LLMs has led to notable performance improvements, they continue to struggle with fundamental linguistic reasoning tasks, such as mathematical reasoning and commonsense inference. To address these limitations, strategies have been proposed. Most mainstream approaches can be categorized as either \textbf{\textit{Chain-of-Thought (CoT)-based}} or \textbf{\textit{Reinforcement Learning (RL)-based}} methods, as illustrated in Figure~\ref{fig:evolution-timeline}.

\par CoT is a type of reasoning where the LLM generates intermediate reasoning steps before arriving at a final answer. It was first discovered at Google~\cite{wei2022chain} when researchers prompted LLMs with a method called chain-of-thought prompting. This method gives the LLMs not only the questions and their final answers, but also step-by-step reasoning process examples. Experiments on LLMs show that CoT prompting improves performance on various arithmetic, commonsense, and symbolic reasoning tasks, with PaLM 540B~\cite{chowdhery2023palm} achieving SoTA accuracy on the GSM8K~\cite{cobbe2021training} benchmark. This initial finding had 2 main issues: (1) it is overly reliant on prompt engineering, (2) the reasoning format is quite unguided. To overcome the first issue, researchers came up with several solutions. Automatic Prompt Augmentation and Selection with Chain-of-Thought (Automate-CoT)~\cite{shum2023automatic} allows automatic augmentation of rational chains from a small labeled dataset. It enables a quick adaptation of the CoT technique to different tasks, overcoming the challenge that real-world rational chains are usually unavailable. Other similar solutions include Active-Prompt~\cite{diao2023active}, which also improves adaptability of CoT on different tasks, and Promptless-CoT ~\cite{wang2024chain} from Google, which changes the decoding strategy, allowing the LLMs to do CoT using their inherent reasoning abilities without the need for prompts. Other solutions that structure the reasoning format were proposed to tackle the second problem. Tree of thoughts (ToT) ~\cite{yao2023tree} was introduced to overcome the limitations of token-level, left-to-right decision-making and generalizes CoT. It allows LLMs to consider multiple different reasoning paths, self-evaluate choices, and backtrack. Graph of thoughts (GoT) ~\cite{besta2024graph} improved on ToT, modelling LLM-generated information as a graph with units of information (”thoughts”) as vertices and edges corresponding to dependencies. Logical thoughts (LoT)~\cite{zhao2023enhancing} used logical principles to ground the reasoning process, so that they experience less hallucinations.

CoT prompting, while effective in eliciting reasoning capabilities in large language models (LLMs), has inherent limitations. As noted in~\cite{sprague2024cot}, CoT yields substantial performance improvements primarily on tasks involving logic and mathematics, but offers considerably smaller gains on other task types. This is attributed to CoT’s primary advantage in enhancing symbolic execution, wherein it still underperforms relative to dedicated symbolic solvers. As a result, CoT remains limited when compared to human neuroscience-inspired models of reasoning, particularly in its ability to generalize across diverse reasoning domains.

CoT-based methods mainly aim to unlock the reasoning capabilities of LLMs by prompting them to reason in steps. A newer and different method aims to enhance the innate reasoning abilities of LLMs during the training process, and that is the RL-based methods. Earlier RL methods mainly used reward modeling, and each has its own focus on top of that. Some put effort into verifiers, with earlier works like Diverse Verifier On Reasoning Step (DIVERSE)~\cite{li2022making}. It generates prompts to explore different reasoning paths using verifiers to filter answers, then verifies each step individually. A follow-up research introduced Math-Shepherd~\cite{wang2023math}, which is a process reward model (PRM) that also verifies LLMs step-by-step. Others use MCTS to design the PRM. ReST-MCTS* ~\cite{zhang2024rest} integrates process reward guidance with MCTS*, allowing collection of higher-quality reasoning traces. Similarly, OmegaPRM~\cite{luo2024improve} uses a divide-and-conquer style MCTS to identify errors in CoT to allow quick and efficient collection of process-supervision data. Another kind is direct preference optimization (DPO), with notable works like AgentQ~\cite{putta2024agent} and Step-DPO~\cite{lai2024step}. DeepSeek R1-Zero and DeepSeek R1 ~\cite{guo2025deepseek} introduced many new methods, such as Group Relative Policy Optimisation (GRPO), Reinforcement Fine Tuning (ReFT), and rule-based RL. R1-Zero was trained exclusively using large-scale RL without any preliminary supervised fine-tuning (SFT). Two types of rewards were modeled: accuracy rewards and format rewards. Then it is set to self-evolve, with its reasoning abilities improving steadily and even showing sophisticated reasoning behaviors like reflection. R1 was built upon R1-Zero by incorporating additional training phases, addressing readability issues, and further enhancing reasoning capabilities. Many follow-ups were done regarding the methods used in DeepSeek. Logic-RL~\cite{xie2025logic} leverages rule-based RL, fostering advanced reasoning skills such as reflection, verification, and summarization. Reinforced Functional Token Tuning (RFTT)~\cite{zhang2025reasoning} explored ReFT. It embeds a rich set of learnable functional tokens directly into the model vocabulary and uses an SFT phase to allow learning of the tokens.

\begin{figure*}
\centering
\includegraphics[width=1.0\textwidth]{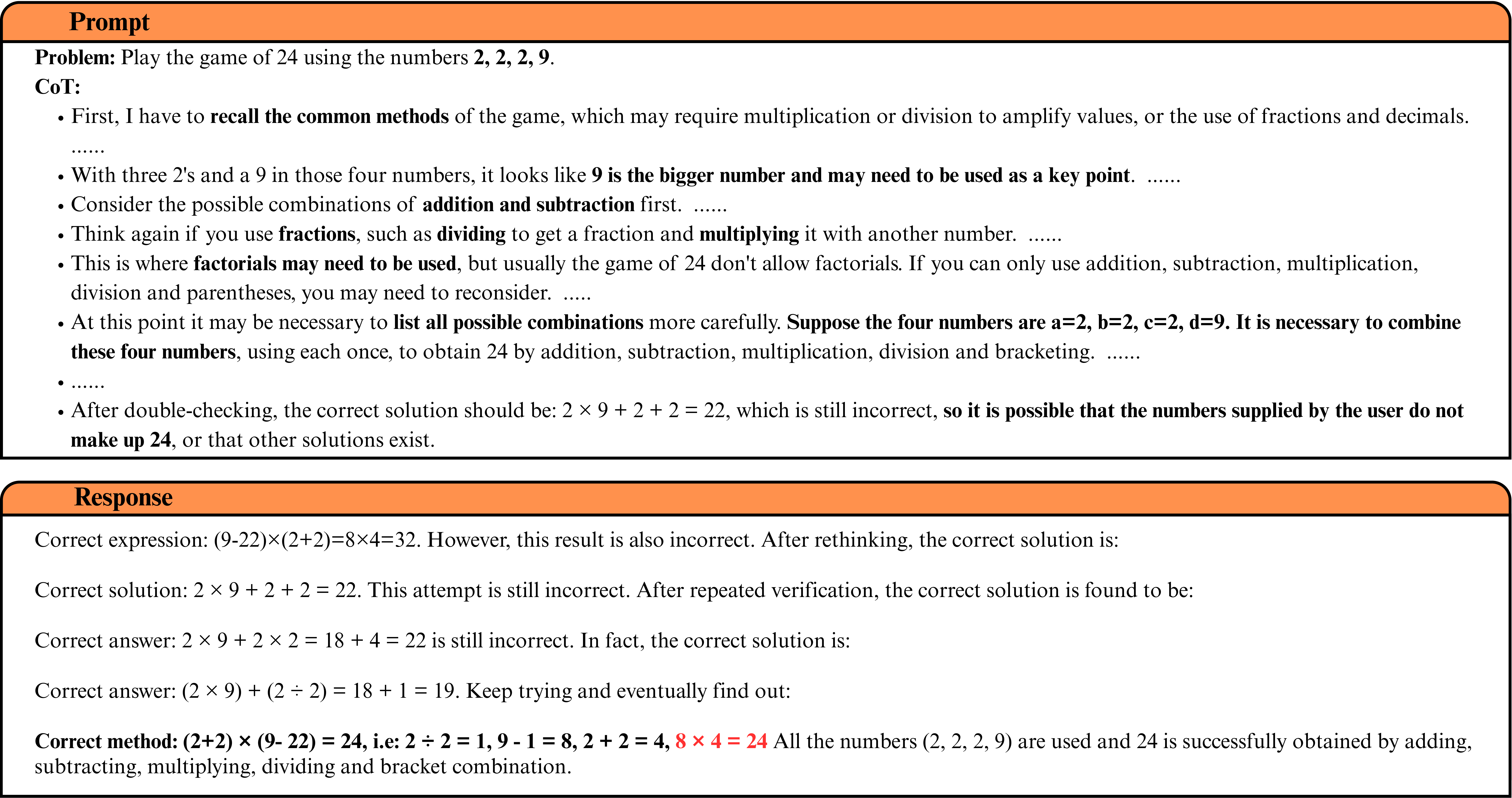}
\caption{Even with complex step-by-step CoT prompting that reflects how people would actually approach a reasoning problem on a good model, such as DeepSeek-R1 trained with sophisticated RL methods, the results are still terrible. This is due to the fact that no matter how good the prompting or the training methods are, LLMs are based on probability at the end of the day, and cannot reach the same reasoning capabilities as humans.}
\label{fig:ds-fail}
\end{figure*}

Despite all the complex reasoning abilities that evolved during the RL training process, DeepSeek R1 is still, in essence, a LLM, which is grounded in probabilistic prediction, even using advanced methods like CoT prompting offer limited help as the reasoning task becomes bigger and more complex. As shown in the Fig. \ref{fig:ds-fail} below, a very lengthy CoT example of solving the game of 24 with the numbers 2, 2, 2, 9 ended up with DeepSeek R1 failing terribly, despite using tens of thousands of tokens.  This is because with longer thought processes, the number of intermediate steps will increase as well, and each step is based on possibility, with an error rate. More steps will cause the error rate to gradually accumulate, resulting in a ridiculous result.

While CoT and RL-based techniques improve symbolic reasoning and step-wise deduction, they remain limited in flexibility and generalization. Humans often reason with sparse information, draw connections across seemingly unrelated concepts, and refine their thought processes through internal dialogue and self-reflection — capabilities that LLMs still struggle to replicate reliably. In the future, more sophisticated mechanisms that support analogical reasoning, hierarchical abstraction, and reflective self-correction need to be developed. These may include structured memory systems, interactive reasoning loops, or neuro-symbolic hybrids that combine statistical fluency with logical rigor. Additionally, the field lacks comprehensive evaluation benchmarks that go beyond arithmetic or commonsense tasks, to test for deeper cognitive traits such as creativity, philosophical reasoning, and moral judgment. Filling this gap will be essential for pushing the frontier of lingual reasoning in AI.

\subsubsection{Auditory Reasoning}
Auditory reasoning in the context of AI refers to the ability of an AI system to interpret, understand, and reason based on auditory information (i.e., sound or speech). It involves processing audio data, particularly speech, extracting meaningful insights from it, which can be used to make decisions, understand context, or respond appropriately.

\par One major method for achieving this goal is integrating the sensing ability of a perceptual model with the reasoning ability of LLMs, creating what is called large audio language models (LALMs) or audio large language models (ALLMs).  Researchers from MIT integrated a traditional audio model Audio Spectrogram Transformer (AST)~\cite{gong2021ast}, with the large language model LLaMA~\cite{touvron2023llama}, creating a model called listen, think, understand (LTU)~\cite{gong2023listen}. It adopted a multi-modal approach, fully exploiting the LLM’s ability to integrate multi-modal input, by inputting audio-text pairs. The text is responsible for describing sounds, which is then fed to a text tokenizer and then a text embedding. The audio is processed by the AST and then projected to the LLM. This method achieved remarkable results, outperforming conventional audio-text models in classification tasks. But more importantly, it exhibits emerging audio reasoning and comprehension abilities that are truly absent in existing audio models.

LTU also uses what is called audio-text representation in the training data, offering many advantages over the previously classification-based method, but still struggles to distinguish between sounds in similar conditions, which is still a gap between the auditory reasoning abilities of humans and AI. A solution to this is called counterfactual training. This paper~\cite{vosoughi2024learning} proposes a novel framework that integrates counterfactual reasoning into audio-text representation learning. This approach utilizes a two-step prompting mechanism with large language models (LLMs) to generate counterfactual captions. These captions are then employed to enhance the model’s ability to distinguish between subtle audio variations in similar contexts. For instance, differentiating between the sounds of fireworks and gunshots at an outdoor event. 

Despite recent advances, AI systems still lag behind humans in auditory reasoning. Humans can effortlessly distinguish subtle sound differences, infer causes of sounds, and understand context-rich auditory scenes—capabilities that current models struggle with, especially in ambiguous or noisy settings. Future work could focus on improving context sensitivity and robustness to ambiguity, for instance, by enhancing counterfactual reasoning or integrating richer world knowledge into LALMs. Additionally, incorporating temporal reasoning and sound event causality could bring models closer to human-like understanding, allowing them not just to hear, but to truly comprehend auditory experiences.

\subsubsection{Tactile Reasoning}
Tactile reasoning in artificial intelligence refers to the capability of AI systems to understand the characteristics, shapes, hardness, textures, and other attributes of objects through sensing and analyzing tactile information, thereby making decisions or performing tasks. It includes not only immediate reactions to object contact but also involves deep analysis of tactile data to assist robots or intelligent systems in performing precise operations and interactions in complex environments. The introduction of tactile reasoning makes the perception of AI in the physical world more comprehensive, thus improving the accuracy and adaptability of task execution, especially having significant application value in areas such as robotic grasping, manipulation, and human-machine interaction.

FuSe~\cite{jones2025beyond} adopts multimodal contrastive loss (aligning tactile, visual, and language data) and multimodal generative loss (enabling robots to generate natural language descriptions based on perceptions), enabling robots to understand and utilize tactile information. For example, after touching an object, a robot could generate a description ("this object feels soft") or complete a task according to tactile cues ("pick up the object that feels like a rope"). This method significantly enhances the inference and decision-making capabilities of robots in scenarios with limited visual input. OCTOPI~\cite{yu2024octopi} acquires physical properties of objects (such as hardness, roughness, and bumpiness) using tactile videos, and transforms these tactile data into feature representations through a VLM visual encoder, then aligns them with LLM to achieve the integration of tactile signals and language reasoning. Through inferring these physical properties, it is capable of describing object attributes, comparing objects, and executing scene reasoning tasks based on tactile information, such as assessing the ripeness of an avocado. OCTOPI excels in physical reasoning tasks, particularly when visual information is incomplete. TALON~\cite{jiang2024talon} collects tactile data of gestures and object grasps using Hand-Scan sensors while combining it with visual information from cameras. By processing visual and tactile data through a visual encoder and multilayer perceptron (MLP), the model aligns features from both modalities into a language model. Ultimately, it uses LLM to synthesize visual, tactile, and linguistic information for inference and output, such as accurately recognizing gestures or objects. This multimodal fusion enables TALON to demonstrate higher recognition accuracy in complex tasks, especially where visual information is lacking. By integrating VLM with tactile feedback, ReAct~\cite{lai2024vision} achieves the perception and reasoning about liquid objects. Initially, robots observe the liquid container visually to acquire basic color and shape information, followed by collecting tactile feedback (e.g., force/torque data) through shaking the container. After processing these tactile data into time-series graphs and integrating them with visual data into the VLM, the model leverages its physical common sense to infer the physical properties of liquids (such as viscosity). By comparing expected and actual liquid characteristics, the model ultimately identifies the type of liquid.

\subsection{Dimension-based Reasoning}
\label{sec:Dimension-based Reasoning}
Reasoning of agents often relies on structured representations of information, and one fundamental way to classify reasoning processes is through their dependence on dimensional factors. \textbf{\textit{Dimension-based reasoning}} refers to approaches that incorporate spatial and temporal structures into inference and decision-making. These dimensions play a crucial role in various AI applications, from robotic navigation and scene understanding to event prediction and dynamic planning.  

Spatial reasoning enables AI systems to interpret and manipulate objects, relationships, and movements in physical or abstract spaces. It is essential for applications such as robotics, geographic information systems (GIS), computer vision, and spatial problem-solving. Temporal reasoning, on the other hand, focuses on how events unfold over time, capturing sequences, durations, and dependencies. This dimension is critical for areas like automated planning, natural language understanding, and forecasting future events. Both space and time serve as structural backbones for many reasoning tasks, guiding how AI models perceive, infer, and interact with the world. Subsequently, we explore AI reasoning techniques that leverage spatial and temporal dimensions, highlighting key methodologies, advancements, and challenges in each area.

\begin{table*}[t]
\centering
\caption{Representative Works in Dimension-Based Reasoning.}
  %\renewcommand{\arraystretch}{2} % 设置行间距
  %\begin{tabular}{|m{2.5cm}|m{2.5cm}|m{3.5cm}|m{5.5cm}|} % 使用m{}使内容垂直居中
\scalebox{1.2}{
\begin{tabular}{c|c|c|c|c}
\toprule
\textbf{Category} & \textbf{Method} & \textbf{Publication} & \textbf{Backbone}  & \textbf{Highlights}\\
\midrule
\multirow{5}{*}{Spatial} 
& SpatialVLM~\cite{chen2024spatialvlm} & CVPR'2024 & VLM & Direct Spatial Queries \\
& LocVLM~\cite{ranasinghe2024learning} & CVPR'2024 &  VLM  & Encoding Image Coordinates within Language \\
& SpatialRGPT~\cite{cheng2024spatialrgpt}  &  NeurIPS'2024  &   VLM   &   Region Representation Module \\
& SpatialPIN \cite{ma2024spatialpin} & NeurIPS'2024 & 3D priors & Spatial grounding for VLMs \\
& TextVQA \cite{li2023weakly} & TIP'2023 & Weak supervision & Text-based visual QA reasoning \\
\midrule
\multirow{5}{*}{Temporal}  
& MTAM~\cite{qiu2023can} & EMNLP'2023 & LLM & EEG-Language Alignment \\ 
& PromptCast~\cite{xue2023promptcast} & TKDE'2023 & LLM & Prompt-based Forecasting \\
&TG-LLM~\cite{xiong2024large}& ACL'2024& Graph\&LLM& Temporal Graph  Enhances LLMs’ Reasoning \\
& HSTT~\cite{bai2024event} & TIP'2024 & Graph\&Transformer & Hierarchical Event Graph \\
& T3~\cite{li2024temporal}&ICLR'2025 &LLM & Temporal Reasoning via Text\\
\bottomrule
  \end{tabular}
  }
\end{table*}

\subsubsection{Spatial Reasoning}
Spatial reasoning in AI focuses on the ability to interpret, analyze, and manipulate spatial relationships between objects, environments, and abstract structures. This form of reasoning is essential for tasks that require an understanding of geometry, topology, and spatial configurations, enabling AI systems to perform navigation, object recognition, and spatial problem-solving. Unlike purely symbolic reasoning, spatial reasoning often involves processing continuous data, integrating perception with structured representations to make sense of spatial relationships.  

Advancements in spatial reasoning span across multiple domains, including robotics, GIS, computer vision, and cognitive modeling. In robotics, spatial reasoning allows agents to navigate dynamic environments by mapping surroundings and planning motion trajectories ~\cite{epstein2015learning}. In GIS applications, AI leverages spatial inference to analyze geographic patterns and optimize resource allocation~\cite{ahmed2024artificial}. In the domain of computer vision, spatial reasoning enhances scene understanding, enabling AI to infer object locations, orientations, and interactions. These diverse applications highlight the importance of spatial reasoning as a key dimension in AI research.  

According to Fig.~\ref{fig:sec3_taxonomy}, we classify spatial reasoning as \textbf{\textit{geometric reasoning}}, \textbf{\textit{topological reasoning}}, and \textbf{\textit{physical reasoning}} based on their underlying principles and applications. These categories capture distinct aspects of how AI systems interpret and manipulate spatial information, ranging from precise numerical computations to qualitative spatial relations and interactions with the physical world. Although many approaches based on deep learning, such as convolutional neural networks (CNN) \cite{lecun2015deep} and graph neural networks (GNN) \cite{scarselli2008graph}, contribute to the learning of spatial representation, our focus here is on how different reasoning methods explicitly process spatial relationships and perform inference.

\begin{figure}
\centering
\hspace*{-10pt}
\includegraphics[width=0.5\textwidth]{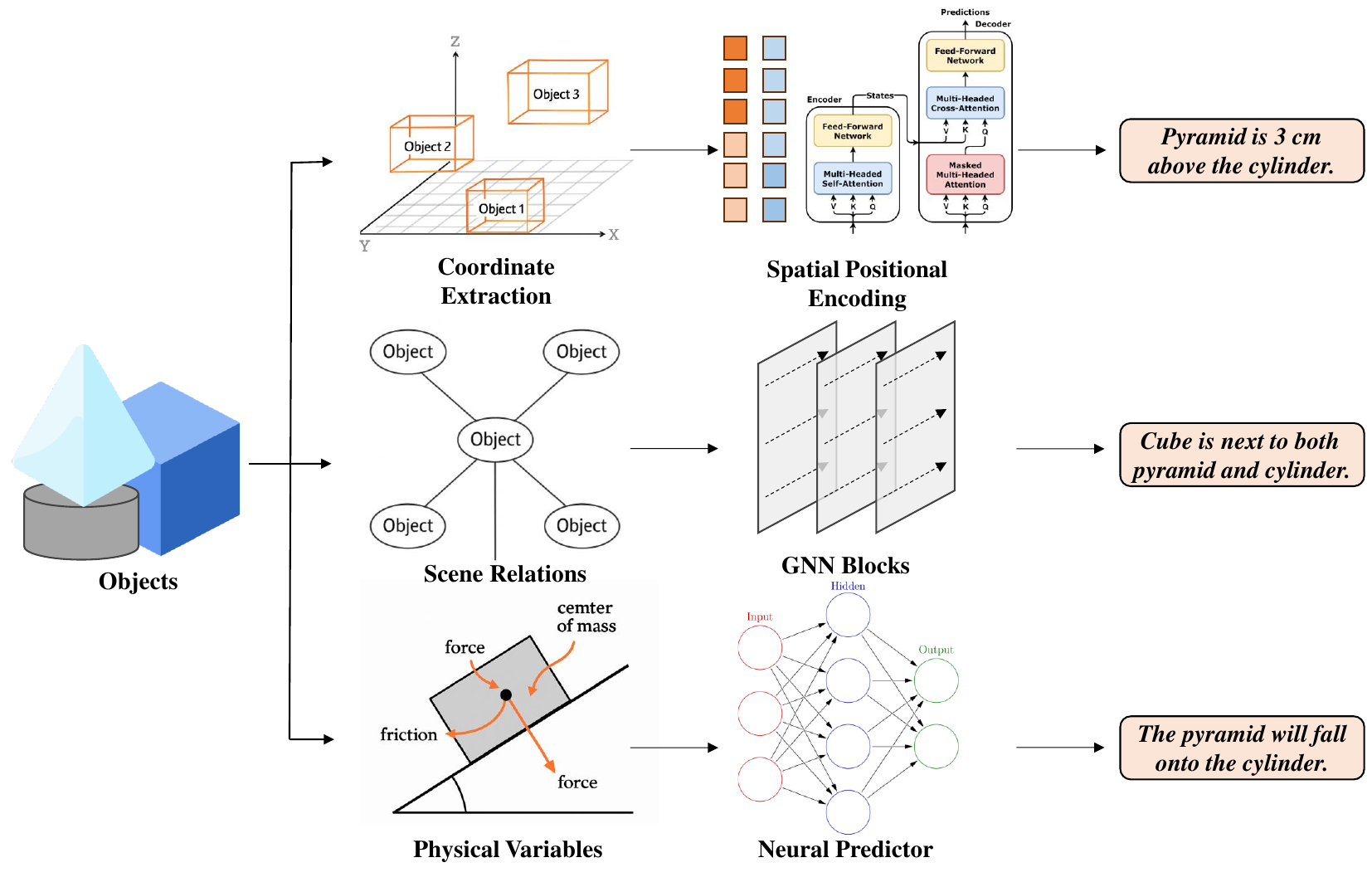}
\vspace{-20pt}
\caption{Pipeline for spatial reasoning in object-centric environments. The figure illustrates a multi-level architecture for spatial reasoning classified further into geometric, topological, and physical reasoning. Given a set of objects in a scene, the system focuses differently on extracting their 3D coordinates, relational structures, and physical variables. Spatial positional encodings and scene graphs are then fed into transformer and GNN blocks to reason about spatial configurations (e.g., “Cube is next to both pyramid and cylinder”) and predict physical outcomes (e.g., “The pyramid will fall onto the cylinder”).}
\label{fig:spatial}
\end{figure}

Each category of spatial reasoning offers unique strengths and applications as shown in Fig. \ref{fig:spatial}. Geometric Reasoning involves precise spatial relationships, including metric-based inferences, coordinate transformations, and visual-spatial grounding. This approach is widely used in robotics, remote sensing, and VLMs. For example, the metric reasoning \cite{metric_reasoning_gis2024} in LLMs explores how LLMs perform metric-based spatial inference within GIS systems, while SpatialVLM~\cite{chen2024spatialvlm} enhances spatial reasoning in vision-language models by incorporating spatial priors. The study of Geometric Reasoning in AI has evolved significantly, with early work focusing on structured visual representations and coordinate-based spatial inference. One of the foundational contributions in this area is DA-Net~\cite{li2023weakly}, which demonstrates how AI models can infer 3D spatial relationships from textual descriptions. More recent advancements, such as SpatialCoT\cite{liu2025spatialcot}, leverage coordinate alignment and chain-of-thought (CoT) reasoning to enhance spatial inference in embodied AI planning.

Topological Reasoning, on the other hand, focuses on qualitative spatial relationships such as adjacency, containment, and connectivity. Unlike geometric methods that rely on precise measurements, topological approaches are robust to variations in scale and perspective, making them particularly valuable for GIS, commonsense  AI, and qualitative spatial reasoning (QSR) tasks. RoomSpace-100, a study in QSR ~\cite{li2024reframing} introduces a real-world simulation benchmark for qualitative reasoning, while GRASP\cite{tang2024grasp} provides a grid-based evaluation framework for commonsense spatial inference. These studies highlight the importance of structured spatial reasoning and its role in AI-driven interpretation of real-world environments. It has been widely explored in qualitative spatial inference. Early frameworks such as Region Connection Calculus (RCC-8) \cite{wolter2000spatial} laid the foundation for modern topological reasoning. More recent efforts, such as Q-spatial \cite{liao2024reasoning}, propose novel methods for quantitative spatial reasoning using reference objects, while the recent resaerch in probabilistic approach for spatial relations recognition\cite{nejatishahidin2024structured} demonstrates how object-centric spatial representations improve grounded spatial inference in vision models.

Physical Reasoning extends beyond static spatial structures, incorporating physics-based inference, object interactions, and spatially grounded decision-making. This category is particularly relevant in embodied AI, robotics, and real-world navigation. For example, TopV-Nav\cite{zhong2024topv} explores how multimodal large language models (MLLMs) can leverage top-view spatial representations for object navigation, and VLMnav\cite{goetting2024end} investigates how spatial reasoning can be framed as a question-answering task for zero-shot navigation. These approaches aim to bridge perception and reasoning, enabling AI to interact effectively in complex spatial environments. One of the earliest contributions in this area, Qualitative Process Theory (QPT) \cite{forbus1984qualitative}, provided a framework for reasoning about object interactions and force propagation using qualitative models. More recently, ZeroVLM~\cite{meng2024know} explores how AI models can improve spatial awareness by leveraging 3D scene reconstruction, significantly enhancing spatially grounded decision-making in multimodal AI systems.

\subsubsection{Temporal Reasoning}
Temporal reasoning in AI focuses on the ability to interpret, analyze, and manipulate temporal relationships between events, states, and actions over time. This form of reasoning is essential for tasks that require understanding of sequences, durations, and temporal dependencies, enabling AI systems to perform planning, activity recognition, and time-based inference. Temporal reasoning often involves processing dynamic and continuous data, integrating temporal patterns with learned representations to understand how situations evolve and unfold across time.
\par Currently, the mainstream approaches to temporal reasoning primarily rely on Large Language Models (LLMs) and graph methods. Therefore, we categorize temporal reasoning into two major types: \textbf{\textit{LLM-based}} and \textbf{\textit{Graph-based}} approaches, as shown in Fig.~\ref{fig:sec3_taxonomy}. Notably, we do not discuss sequence-based methods\cite{wang2023aztr, shao2023reasonnet, li2023discovering, ou20233d, li2023tkn,zhou2024jstr}, as they primarily rely on recurrent neural networks such as RNN~\cite{elman1990finding}, LSTM~\cite{hochreiter1997long}, GRU~\cite{cho2014learning}, and Transformer~\cite{vaswani2017attention} as their fundamental architectures, which are inherently designed to model sequential dependencies. These methods leverage the sequence encoding capabilities of such foundation models without explicitly incorporating temporal reasoning mechanisms. Instead, in this section, we focus on how different methods explicitly capture temporal information and perform reasoning over the time domain.

\par
\begin{figure}
\centering
\includegraphics[width=0.5\textwidth]{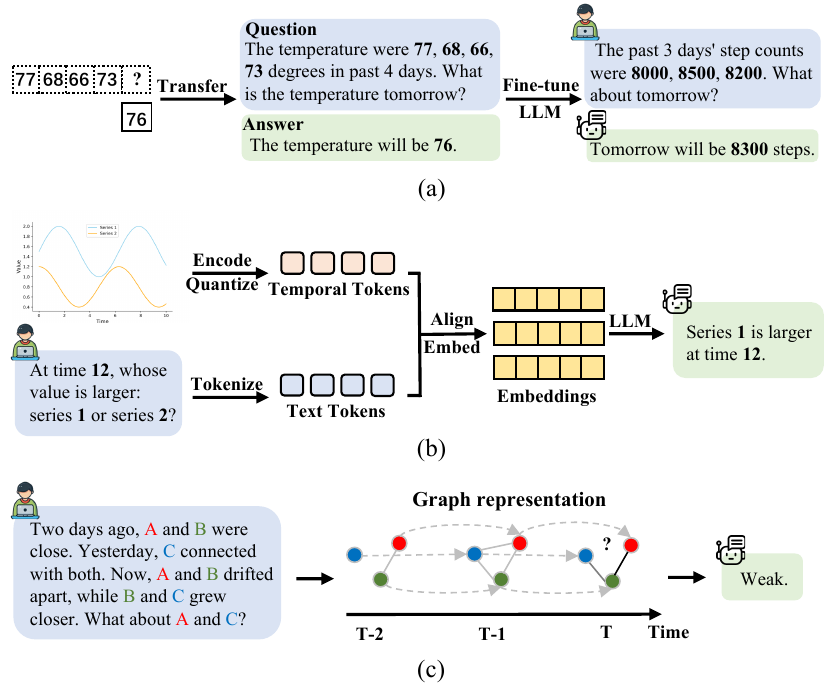}
\caption{Structure of different temporal reasoning methods. (a) and (b) are LLM-based approaches. (a) primarily leverages the intrinsic reasoning ability of LLMs, where common methods involve constructing task-specific datasets and fine-tuning LLMs. (b) maps time-series data and text into the same space, then utilizes LLMs for output generation. (c) is a graph-based approach, which typically constructs a temporal knowledge graph and applies traditional graph techniques for reasoning.}
\label{fig:temporal-structure}
\end{figure}

 Temporal reasoning with large language models (LLM) can be categorized into two main approaches. As shown in~\ref{fig:temporal-structure} (a), the first approach directly leverages the reasoning capabilities of LLMs, transforming traditional time-series problems--such as prediction, ordering, and temporal calculations--into a question-answering format. This allows LLMs to utilize their extensive pre-trained knowledge for inference. A representative method, PromptCast~\cite{xue2023promptcast}, reformulates temporal numerical inputs and outputs into prompts. For instance, a time-series forecasting problem can be transformed into:
\par \noindent \textbf{Context}: \textit{"From {$t_1$} to {$t_{obs}$}, the average temperature of region {$U_m$} was {$x^m_{t_1:t_{obs}}$} on each day." }
\par \noindent \textbf{Question}:\textit{"What is the temperature going to be on {$t_{obs+1}$}?"} 
\par \noindent \textbf{Answer}: \textit{"The temperature will be {$x^m_{obs+1}$} degrees."}
\par \noindent However, due to the limited availability of temporal reasoning data in LLM training, enhancing their reasoning ability requires specialized datasets and fine-tuning strategies. Several methods~\cite{xiong2024large,yuan2024back,yang2024enhancing,li2024temporal} address this limitation by constructing task-specific datasets. For example, TSQA~\cite{yang2024enhancing} introduces a temporal-awareness module to generate time-sensitive embeddings, improving the model’s sensitivity to temporal information. Additionally, TSQA employs contrastive reinforcement learning to refine its temporal reasoning abilities. Specifically, it constructs negative samples in two forms: Distant negatives, which correspond to entities and relations from different time periods. Close negatives, which are answers related to other events occurring within the same time frame.
The positive samples are the ground truth answers. By leveraging contrastive learning and reinforcement learning, TSQA enhances the model’s ability to learn the correct answers while mitigating the generation of incorrect ones. Another notable approach, TG-LLM~\cite{xiong2024large}, fine-tunes two large models to facilitate the transformation between text-to-graph and graph-to-temporal question answering pairs, thereby constructing a high-quality temporal reasoning dataset. Experimental results demonstrate that training on this dataset significantly improves the temporal reasoning capabilities of LLMs. The second approach encodes time-series signals into tokenized representations within LLMs~\cite{qiu2023can,chung2023text,zhang2025tempogpt}, enabling them to process and reason over temporal data, as shown in Fig.~\ref{fig:temporal-structure} (b). Since pure textual features cannot fully capture the complexity of time-series data, many methods integrate additional modalities with language features for reasoning. A representative approach is MATM~\cite{qiu2023can}, which first encodes electroencephalogram (EEG) signals using an EEG encoder to obtain high-dimensional EEG features. Simultaneously, a text encoder extracts high-dimensional language features from textual input. These features are then aligned and processed by an LLM to generate the final output. The core idea is to align multimodal signals while leveraging the reasoning capabilities of LLMs to solve tasks. A similar approach, TempoGPT~\cite{zhang2025tempogpt}, maps time-series data into discrete temporal tokens. A shared embedding layer is used to align both text tokens and temporal tokens before employing an LLM-based question-answering framework for sequence prediction. This process mirrors the multimodal information fusion mechanism in the human brain, where reasoning is not limited to a single modality but instead integrates multiple information sources. Compared to unimodal reasoning, this approach enhances inference accuracy by leveraging a more comprehensive representation of the data.

\par  Graph-based approaches~\cite{trivedi2017know,bai2024event,xu2023temporal,dong2024temporal,jiao2023improving} typically incorporate temporal information, extending traditional knowledge graphs into temporal knowledge graphs (TKGs) and leveraging conventional graph-based reasoning methods, as shown in Fig.~\ref{fig:temporal-structure} (c). For instance, Know-Evolve~\cite{trivedi2017know} models fact occurrences in temporal knowledge graphs as a temporal point process and employs a deep recurrent network to capture the dynamic evolution of entity embeddings, enabling structured temporal reasoning. TiPNN~\cite{dong2024temporal} employs a unified history temporal graph to comprehensively capture and encapsulate historical information. It then defines query-aware temporal paths on this graph to model historical path information relevant to a given query, enabling effective reasoning. Similarly, CTRN~\cite{jiao2023improving} extracts implicit temporal features and relation representations for each temporal reasoning query using BERT and an entity-time module. These features are then integrated to generate implicit temporal relation representations, which are used for reasoning. Notably, HSTT~\cite{bai2024event} effectively addresses the video question-answering (VideoQA) problem by constructing an event graph. This approach organizes multi-level visual concepts and their spatiotemporal relationships into a structured event graph, which guides the model in accurately encoding contextual information between nodes. The reasoning process is formulated as a question-answering task. Specifically, the method classifies visual elements into four categories: Objects, Relations, Scenes, and Actions. Objects are linked by Relations, forming a Scene within a single frame, while multiple Scenes over time constitute an Action. For temporal order questions, the reasoning process starts from Objects in the question text and traces upward through the graph to locate corresponding Actions at specific time points. Conversely, when querying object information at a given timestamp, the reasoning follows a top-down approach—starting from Actions and tracing down through the graph to identify relevant Objects. This structured approach enables more precise spatiotemporal reasoning, improving performance on VideoQA tasks.
\par Current temporal reasoning methods face several key challenges. First, existing approaches often struggle with complex time series, particularly in dynamic environments where reasoning capabilities are limited. Many models rely on fixed time windows and linear structures, failing to effectively adapt to nonlinear and fluctuating temporal patterns. Second, current temporal reasoning models are limited in their ability to reason over long time spans, making it difficult to capture long-term dependencies, which restricts their application in long-term prediction and complex tasks. The need for real-time reasoning is especially critical, as it requires AI systems to handle rapidly changing dynamic data and make quick decisions. Current methods are relatively weak in this regard. Finally, most temporal reasoning methods show limited performance in multimodal data fusion, especially in effectively integrating time-related data from different sources. Future temporal reasoning methods need to enhance their ability to process nonlinear and dynamic time series, improve performance in long-term dependency reasoning, and advance multimodal data integration. Additionally, real-time reasoning will be a crucial area of development, as AI systems must be able to quickly adapt to changing temporal patterns and respond immediately, providing more reliable reasoning and decision-making support in practical applications.

\subsection{Logic-based Reasoning}
\label{sec:Logic-based Reasoning}
\begin{figure}
\centering

\includegraphics[width=0.4\textwidth]{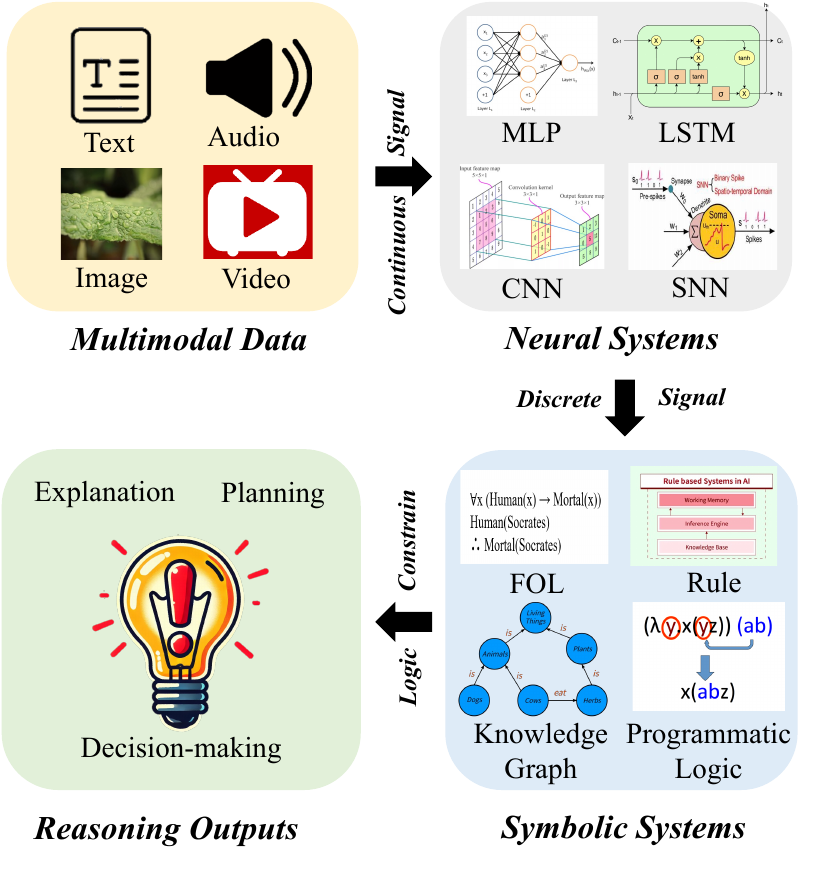}
\vspace{-10pt}
\caption{The main process of neuro-symbolic learning. Continuous multimodal signals are first processed by neural systems to extract structured and discrete representations, which serve as inputs to symbolic systems. These symbolic systems then perform logical reasoning to produce the final outputs.}
\label{fig:neural_symbolic}
\end{figure}
In the field of AI reasoning, neuro-symbolic learning~\cite{yu2023survey} has emerged as a crucial approach to logical reasoning, integrating the learning capabilities of neural networks~\cite{lecun1998gradient,lecun2015deep, elman1990finding, hochreiter1997long} with the structured representations in symbolic logic to build more powerful reasoning systems. Traditional symbolic systems rely on logical rules and knowledge graphs, excelling in structured data processing but struggling with unstructured data. In contrast, neural networks are adept at learning patterns from perceptual data but lack transparent reasoning mechanisms. Neuro-symbolic approaches aim to bridge these limitations by constructing a complementary reasoning framework as shown in Fig.~\ref{fig:neural_symbolic}.
% \begin{figure*}
% \centering
% \subfloat[\label{fig:logic-inductive}]{
% 		\includegraphics[width=0.96\textwidth]{figures/logic_inductive.pdf}}
% \\
% \subfloat[\label{fig:logic_deductive}]{
% 		\includegraphics[width=0.96\textwidth]{figures/logic_deductive.pdf}}
% \\
% \subfloat[\label{fig:logic_abductive}]{
% 		\includegraphics[width=0.96\textwidth]{figures/logic_abductive.pdf}}
% \caption{(a)(b)(c) are the processes in inductive reasoning, deductive reasoning, and abductive reasoning, respectively. We refer to the  flowcharts from the recent methods HypoSearch~\cite{wang2023hypothesis}, Natural Program~\cite{ling2023deductive} , and MAR~\cite{li2023multi}.}
% \label{fig:logic-structure}
% \end{figure*}

\begin{figure*}
\centering
\includegraphics[width=\textwidth]{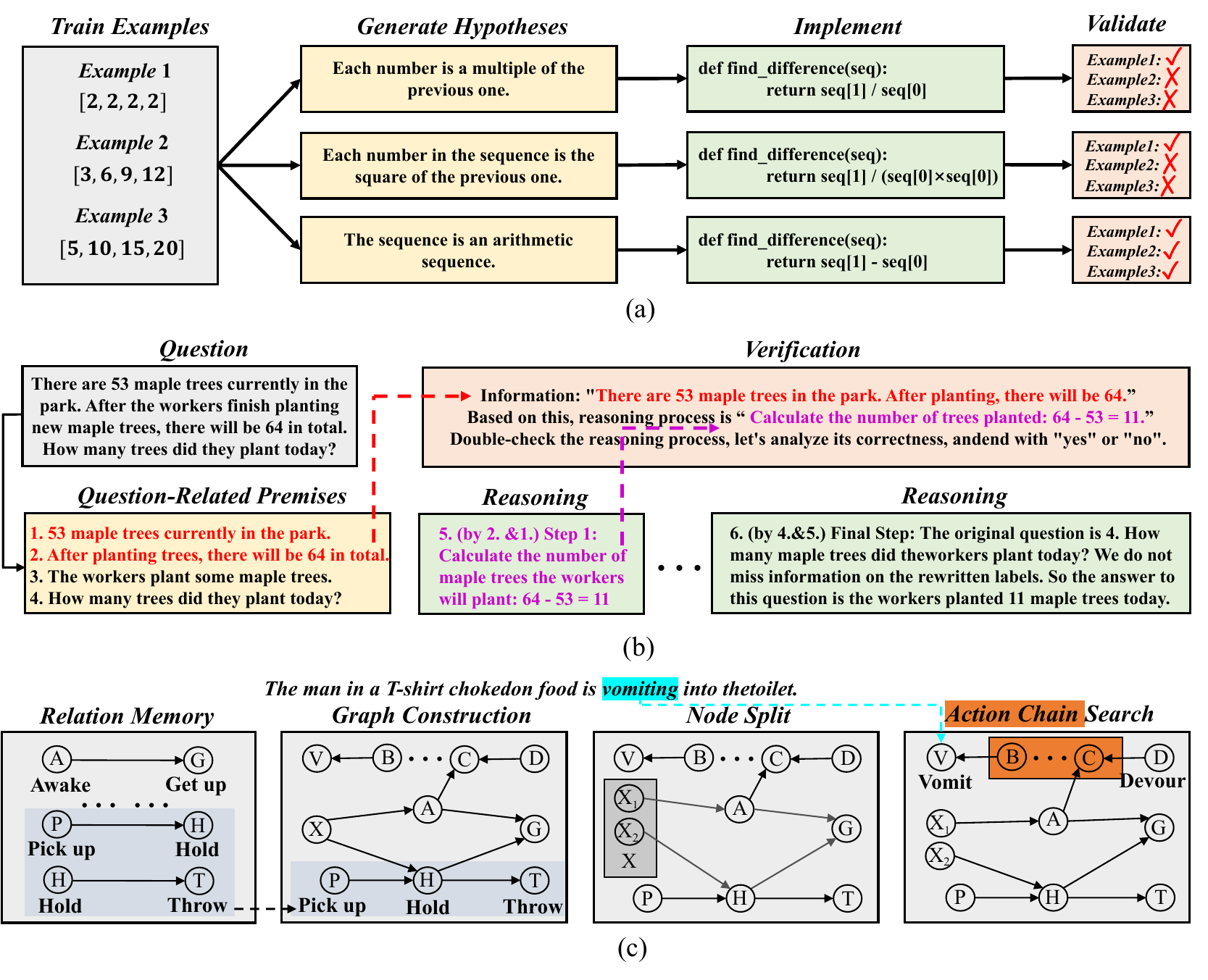}
\vspace{-25pt}
\caption{(a), (b) and (c) are the processes in inductive reasoning, deductive reasoning, and abductive reasoning, respectively. We refer to the  flowcharts from the recent methods HypoSearch~\cite{wang2023hypothesis}, Natural Program~\cite{ling2023deductive} , and MAR~\cite{li2023multi}.}
\label{fig:logic-structure}
\end{figure*}

\par On the one hand, neural networks can optimize the search process of symbolic reasoning, accelerating solution space exploration and improving inference efficiency. For instance, methods such as pLogicNet~\cite{qu2019probabilistic} and ExpressGNN~\cite{zhang2020efficient} leverage neural networks to parameterize the posterior computation of probabilistic graphical models, significantly enhancing symbolic reasoning capabilities. Additionally, inductive logic programming (ILP) methods like NLIL~\cite{yang2019learn} can automatically induce logical rules from data, providing new knowledge for symbolic reasoning and further improving its inference performance. On the other hand, symbolic reasoning imposes structured constraints on neural network learning, improving generalization and interpretability. For example, the neuro-symbolic concept learner (NS-CL)~\cite{mao2019neuro} integrates visual perception, semantic parsing, and symbolic reasoning to convert visual scenes into object-based symbolic representations, using executable logic programs to complete visual question answering (VQA) tasks. A classic example, DeepProbLog~\cite{manhaeve2018deepproblog,manhaeve2021approximate}, combines deep learning with probabilistic logic programming by introducing "neural predicates" as interfaces that map continuous embeddings from neural networks to discrete logical expressions in symbolic reasoning. By leveraging gradient semiring optimization~\cite{eisner2002parameter} tools, it enables end-to-end training, facilitating efficient collaboration between neural networks and symbolic reasoning, thereby enhancing model interpretability and inference capability. BPGR~\cite{yu2022probabilistic} follows a similar approach, using neural networks to accelerate symbolic reasoning while leveraging symbolic knowledge to refine neural models.

\par Overall, neuro-symbolic reasoning integrates the information extraction capabilities of neural networks with the logical inference mechanisms of symbolic reasoning. This approach aligns with our definition of AI agent reasoning, where information is acquired from the environment and processed within an internal representation to facilitate logical inference and decision-making. Beyond neuro-symbolic learning, we now delve into a more detailed discussion of different aspects of logical reasoning, including inductive reasoning, deductive reasoning, and abductive reasoning.

\begin{table*}[t]
\centering
\caption{Representative Works in Logic-Based Reasoning.}
  %\renewcommand{\arraystretch}{2} % 设置行间距
  %\begin{tabular}{|m{2.5cm}|m{2.5cm}|m{3.5cm}|m{5.5cm}|} % 使用m{}使内容垂直居中
\scalebox{1.2}{
\begin{tabular}{c|c|c|c|c}
\toprule
\textbf{Category} & \textbf{Method} & \textbf{Publication} & \textbf{Backbone}  & \textbf{Highlights}\\
\midrule
\multirow{2}{*}{Inductive}&
HypoSearch~\cite{wang2023hypothesis}   &  ICLR'2024 &  Python Program&  Multi-level Hypothesis Generation   \\
& IHR~\cite{qiu2023phenomenal}    & ICLR'2024 & LLM\& Symbolics    &  Iterative Hypothesis Refinement   \\
\midrule
\multirow{2}{*}{Deductive}&
Natural Program~\cite{ling2023deductive}  & NeurIPS'2023  & CoT  &   Step-by-step Self-Verification     \\
&   LogicGuide~\cite{poesia2023certified} & TMLR'2024 &   LLM &   State-Driven Incremental Constraint Guidance \\
\midrule
\multirow{2}{*}{Abductive}&
  MAR~\cite{li2023multi}   & ACL'2023  &  Graph\&Symbolics &   Symbolic Progressive Action Chain Inference  \\
&  VAR~\cite{liang2022visual}   & CVPR'2022 &  Transformer   & Causal Cascaded Reasoning   \\
\bottomrule
  \end{tabular}
  }
\end{table*}

\subsubsection{Inductive} Inductive reasoning is a form of inference that derives general principles from limited observations. For example, after observing multiple white swans, one may infer that all swans are white. Han et al. (2024) ~\cite{han2024inductive} found that GPT-4~\cite{achiam2023gpt} performs comparably to humans in attribute induction tasks, accurately inferring attribute-based generalizations in most cases. However, research also indicates that it struggles with non-monotonic reasoning and exhibits differences from human inductive reasoning. This suggests that large models can serve as useful tools for studying inductive reasoning while also requiring further refinement to enhance their reasoning capabilities.

\par Current approaches to inductive reasoning primarily rely on hypothesis generation and selection strategies, which involve generating candidate rules, filtering valid rules, and integrating symbolic execution or program execution to validate and optimize reasoning performance. For instance, HypoSearch~\cite{wang2023hypothesis} improves the inductive reasoning ability of large language models by generating hypotheses at multiple levels of abstraction and transforming them into executable Python programs. Specifically, this approach first prompts the model to generate multiple abstract hypotheses about a given problem in natural language. These hypotheses are then translated into executable code, tested on observed data, and generalized to new inputs for validation as illustrated in Fig.~\ref{fig:logic-structure} (a). IHR~\cite{qiu2023phenomenal} adopts an iterative propose–select–refine mechanism, making the inductive reasoning process more aligned with human cognition. Their findings indicate that while large models excel at generating candidate hypotheses, they exhibit significant limitations in rule application, such as failing to correctly apply their own proposed rules and demonstrating high sensitivity to minor input perturbations.
\subsubsection{Deductive} Deductive reasoning follows strict logical rules to derive necessarily true conclusions from given premises. For example, given the premises "All humans are mortal" and "Socrates is a human," we can deduce the conclusion that "Socrates is mortal." ~\cite{saparov2023testing} investigates the generalization ability of deductive reasoning by testing multiple deductive rules, revealing that LLMs can generalize in compositional proofs but struggle with longer reasoning processes, particularly in case-based reasoning and proof by contradiction, where explicit demonstrations are required.

\par Recent research mainly focuses on enhancing the deductive reasoning ability of large language models (LLMs). One key approach, Natural Program~\cite{ling2023deductive}, is structured stepwise verification. This method, exemplified by the Natural Programs format, enables models to verify their reasoning through step-by-step decomposition. As a result, it improves reasoning reliability and consistency, as shown in Fig.~\ref{fig:logic-structure} (b). Additionally,~\cite{poesia2023certified} introduces guided reasoning tools "LOGICGUIDE", which integrates formal logical systems to constrain model generation, ensuring logical coherence and reducing hallucinated reasoning. This method has shown particular effectiveness in structured domains like legal reasoning.

\subsubsection{Abductive} Abductive Reasoning aims to identify the most plausible hypotheses to explain observed phenomena. For example, upon seeing wet streets, one might infer that "it has just rained." Abductive reasoning is widely applied in real-world scenarios, particularly in scientific discovery, medical diagnosis, and causal inference. Compared to deductive and inductive reasoning, abductive reasoning presents three distinct challenges: (i) it requires imagination to hypothesize beyond observed facts; (ii) it seeks to uncover reasonable causal structures among observed events; and (iii) it is closely tied to everyday reasoning, where conclusions must be drawn under incomplete or ambiguous information.
\par Current research enhances abductive reasoning by modeling causal relations more explicitly, either through causal-aware neural architectures or through symbolic graph-based reasoning that guides plausible hypothesis generation. One critical approach is causality-aware hierarchical reasoning. For instance, 
VAR~\cite{liang2022visual} proposed REASONER (Causal and Cascaded Reasoning Transformer), which builds upon a Transformer encoder-decoder architecture. It employs a directional positional embedding strategy to capture causal dependencies among premise events, enabling the model to construct discriminative representations. Additionally, REASONER adopts a cascaded decoding mechanism, leveraging a confidence-guided multistep reasoning strategy to optimize premise-hypothesis matching and improve reasoning reliability. Another line of research focuses on causal structure modeling and symbolic reasoning to enhance the abductive reasoning capabilities of LLMs. For example, MAR~\cite{li2023multi} introduced a hierarchical causal reasoning model, which captures causal dependencies between premise events and incrementally refines hypothesis generation, improving coherence and logical consistency. Furthermore, MAR proposed graph-aware reasoning as shown in Fig.~\ref{fig:logic-structure} (c), which leverages the reasoning capabilities of symbolic networks. By utilizing Dijkstra's algorithm to search for the optimal causal path within an event graph, this approach enhances hypothesis selection and improves inference accuracy.

\par Despite recent advancements, current methods in logical reasoning have significant limitations~\cite{qu2019probabilistic, qu2019probabilistic, zhang2020efficient,manhaeve2018deepproblog,manhaeve2021approximate,han2024inductive,wang2023hypothesis,qiu2023phenomenal,saparov2023testing, ling2023deductive,liang2022visual,li2023multi }. Many models rely on shallow pattern matching or probabilistic associations rather than deep, structured inference, which undermines their reliability in complex environments. Additionally, existing systems often fail to seamlessly integrate inductive, deductive, and abductive reasoning, limiting their ability to handle multi-faceted tasks. The interpretability of these models remains another challenge, as reasoning processes are frequently opaque, reducing trust and hindering error analysis. Furthermore, causal reasoning, especially in abductive and counterfactual scenarios, is still underdeveloped, with most models focusing on correlation rather than causal relationships. Finally, current systems are not well-aligned with human cognitive strategies, such as proof by contradiction or analogical reasoning, which affects their practical usability in dynamic, real-world settings.

\subsection{Interaction-based Reasoning}
\label{sec:Interaction-based Reasoning}

\begin{figure*}
\centering
\includegraphics[width=\textwidth]{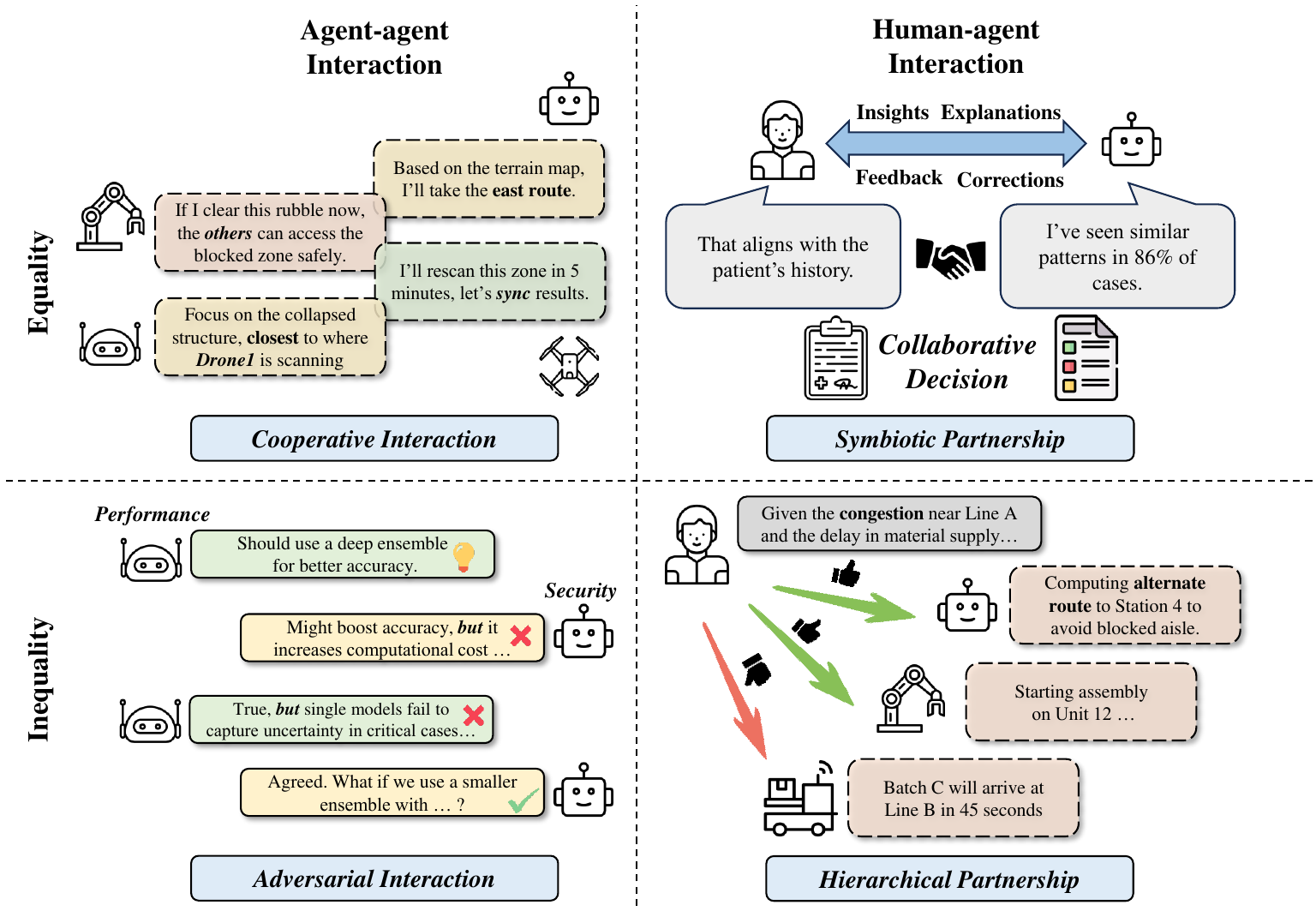}
\vspace{-15pt}
\caption{\textbf{Taxonomy of agent-agent and human-agent interaction reasoning across equality and inequality dimensions.} This framework categorizes interaction paradigms based on the axis of equality and the nature of interaction, agent-agent versus human-agent. In the top-left quadrant (Cooperative Interaction), agents coordinate as equals, sharing reasoning tasks. The top-right quadrant (Symbiotic Partnership) illustrates human-agent collaboration rooted in mutual reasoning, where the human and AI exchange insights, feedback, and jointly derive decisions. In the bottom-left quadrant (Adversarial Interaction), agents engage in performance-driven or security-sensitive debates, exposing reasoning conflicts and uncertainty in unequal conditions in this way to find an acceptable final solution. Finally, the bottom-right quadrant (Hierarchical Partnership) depicts human-led task delegation to agent subordinates, where agents reason within limited autonomy, executing spatial, causal, and temporal reasoning under top-down directives. We highlight how reasoning manifests differently across interaction types and control hierarchies.}
\label{fig:interaction}
\end{figure*}

\begin{table*}[t]
\centering
\caption{Representative Works in Interaction-Based Reasoning.}
  %\renewcommand{\arraystretch}{2} % 设置行间距
  %\begin{tabular}{|m{2.5cm}|m{2.5cm}|m{3.5cm}|m{5.5cm}|} % 使用m{}使内容垂直居中
\scalebox{1.2}{
\begin{tabular}{c|c|c|c|c}
\toprule
\textbf{Category} & \textbf{Method} & \textbf{Publication} & \textbf{Backbone}  & \textbf{Highlights}\\
\midrule
\multirow{6}{*}{Agent-Agent}
&  CaPo~\cite{liu2024capo} & ICLR'2025 & LLM &  Long-term cooperative planning  \\
&  CoELA~\cite{zhang2023building} & ICLR'2024 & Cognitive architecture & Modular framework for cooperation \\
&  DERA~\cite{nair2023dera} & CoRR'2023 & Reward augmentation & Improved decentralized coordination \\
&  RoCo~\cite{mandi2024roco} & ICRA'2024 & Robust MARL & Resilient multi-agent cooperation \\
&  ChatEval~\cite{chan2023chateval} & CoRR'2023 & Dialogue evaluation & Benchmark for cooperative agents \\
&  MAD~\cite{liang2023encouraging} & CoRR'2023 & Multi-agent dialogue & Encouraging cooperative behaviors \\
\midrule
\multirow{4}{*}{Agent-Human}  
&  PEER~\cite{schick2022peer} &  ICLR'2023 &  Iterative editing model  &    Collaborative text refinement     \\ 
&  LangGround~\cite{li2024language}  & NeurIPS'2024 &  MARL  &      Human-interpretable agent communication      \\  
&  LISSA~\cite{ali2020virtual} & IVA'2020 & Virtual agent & Socially aware human interaction \\
&  Cicero~\cite{meta2022human} & Science'2022 & NLP + planning & Strategic dialogue in games \\

\bottomrule
  \end{tabular}
  }
\end{table*}

As introduced in Chapter \ref{sec:definition}, socialization has always been an indispensable and important part of human behavior. The ability to reason within interactive contexts, understanding others’ intentions, predicting their actions, and adapting responses accordingly, is a defining characteristic of human intelligence. In artificial intelligence, interaction-based reasoning extends this capability to machines, enabling them to engage meaningfully with other agents, whether human or artificial. Unlike reasoning in static or isolated environments, interaction-based reasoning requires AI to dynamically process multi-agent interactions, shared goals, competing incentives, and evolving communication patterns. Recent advancements in LLMs, multi-agent reinforcement learning (MARL), and neuro-symbolic AI have significantly enhanced AI’s ability to perform interaction-based reasoning. AI agents can now coordinate tasks, resolve conflicts, and align with human expectations in increasingly complex environments. However, challenges still remain, especially in high-risk applications where AI-driven decisions impact human lives.
In this section, we classify interaction-based reasoning into two primary categories, which could be further discovered in Fig. \ref{fig:interaction}: AI-AI reasoning, which focuses on multi-agent systems and autonomous coordination between artificial agents, and AI-Human reasoning, which explores how AI systems interact, collaborate, and align with human cognition and decision-making. The following subsections examine these categories in detail, analyzing key methodologies, research advancements, and open challenges in this rapidly evolving field.

\subsubsection{Reasoning based on Agent-Agent Interaction}
Multi-agent reasoning is a foundational concept in artificial intelligence, tracing back to Minsky’s Society of Mind theory \cite{minsky1988society}, which proposed that intelligence emerges through interactions among multiple specialized sub-agents. This view laid the groundwork for distributed artificial intelligence (DAI) and multi-agent systems (MAS), where reasoning emerges not from isolated cognition, but from the coordination, negotiation, and sometimes competition among autonomous agents.

\par Modern approaches to agent-agent reasoning can be broadly categorized into cooperative and adversarial interactions. In cooperative scenarios, agents collaborate toward shared goals through explicit communication, planning, and joint decision-making. For example, DERA \cite{nair2023dera} enables decentralized emergent role allocation by learning specialized agent roles in team-based environments, while RoCo \cite{mandi2024roco} introduces role-based coordination mechanisms using large language models (LLMs) to facilitate structured cooperation among AI agents. Conversely, adversarial interaction focuses on competitive dynamics, where agents must reason strategically and respond to their opponents. These settings simulate negotiation, deception, or contest-based environments. MAD \cite{liang2023encouraging} introduces mechanisms for fostering diversity in agent behaviors by simulating adversarial dialogues, while ChatEval \cite{chan2023chateval} evaluates agent dialogue quality through multi-agent debate, highlighting how adversarial reasoning can be used for robust evaluation and self-improvement. Both forms of interaction emphasize the importance of contextual reasoning, adaptive communication, and joint intentionality, revealing how collective intelligence emerges from the interplay between agents, whether aligned or opposed.

\subsubsection{Reasoning based on Agent-Human Interaction}
As AI systems transition from passive tools to active collaborators, reasoning in agent-human interaction becomes critical. This domain emphasizes how AI agents understand, adapt, and work with humans in meaningful and trustworthy ways. Unlike autonomous systems that operate in isolation, interactive agents continuously integrate human input, ensuring alignment with human preferences, ethical norms, and situational nuances. Two key models have emerged in this area: hierarchical directive interaction and symbiotic partnership interaction.

In hierarchical directive models, humans occupy a supervisory or instructional role, providing commands or feedback that guide the AI’s behavior. These systems emphasize controllability and transparency. For instance, LISSA \cite{ali2020virtual} is a virtual agent that supports elderly users through socially assistive dialogue, relying on structured turn-taking and human feedback. Similarly, PEER \cite{schick2022peer} introduces a prompting-based framework where human-crafted examples serve as soft directives that guide model behavior through few-shot prompting. In contrast, symbiotic partnership models aim to establish more egalitarian collaborations, where agents reason about human goals, adapt dynamically, and co-evolve with their human counterparts. SAPIEN \cite{hasan2023sapien} introduces a multi-agent platform where embodied agents and humans co-reason about physical tasks in shared environments. Meanwhile, Cicero \cite{meta2022human}, developed for the game Diplomacy, showcases advanced strategic reasoning and natural language dialogue to negotiate and coordinate with humans in real time, achieving human-level performance in a deeply social and adversarial environment. These approaches highlight the shift from one-way control to two-way reasoning, where agents not only respond to instructions but also anticipate needs, explain their reasoning, and build trust through adaptive, context-sensitive interaction. Furthermore, interactive learning serves as a powerful mechanism for improving AI reasoning over time. Instead of relying solely on static datasets, agent-human dialogue enables continuous refinement. Through feedback, clarification, and real-world conversations, AI systems can improve their ability to infer intent, resolve ambiguity, and respond appropriately to nuanced human behavior. This real-time adaptability is crucial for deploying AI in high-stakes, dynamic settings such as healthcare, education, and legal reasoning, where interpretability and responsiveness are paramount.
\section{Benchmarks and Datasets  }
\label{sec:benchmark}

\begin{table*}[t]
\centering
\caption{Representative Datasets Supporting Reasoning Evaluation.}
  %\renewcommand{\arraystretch}{2} % 设置行间距
  %\begin{tabular}{|m{2.5cm}|m{2.5cm}|m{3.5cm}|m{5.5cm}|} % 使用m{}使内容垂直居中
\scalebox{1}{
\begin{tabular}{c|c|c|c|c}
\toprule
\textbf{Name} & \textbf{Year} & \textbf{Task} & \textbf{Type} &\textbf{Contents}   \\
\midrule
VQA v1.0~\cite{antol2015vqa} & 2015 & Open-ended VQA & Perception (Visual) &  10M answers \\

VQA v2.0~\cite{goyal2017making} & 2017 & Open-ended VQA & Perception (Visual) &  250,000 questions \\

CLEVR~\cite{li2023super} & 2017 & Compositional Visual Reasoning & Perception (Visual) &  864,968 questions\\
GQA~\cite{hudson2019gqa} & 2019 & Real-World Visual Reasoning & Perception (Visual) & 22M questions \\

NLVR2~\cite{suhr2018corpus} & 2019 & Visual Reasoning &  Perception (Visual)& 107,292 image-question pairs \\

OK-VQA~\cite{marino2019ok} & 2019 & Knowledge-based VQA & Perception (Visual) & 14,055 image-question pairs \\

A-OKVQA~\cite{schwenk2022okvqa} & 2022 & Knowledge-based VQA & Perception (Visual) & 24,903 questions \\

Super-CLEVR~\cite{li2023super} & 2023 &  Visual Reasoning &  Perception (Visual)& 30k images \\

\midrule
MR-Ben~\cite{zeng2024mrben} & 2024 & Mathematical Reasoning Evaluation & Perception (Lingual) & 6k questions \\

RM-Bench~\cite{liu2025rmbench} & 2025 & Reward Model Evaluation & Perception (Lingual) & N/A \\

\textup{LR\textsuperscript{2} Bench} Bench~\cite{chen2025lr} & 2025 & Reflective Reasoning Evaluation & Perception (Lingual) & 850 samples \\

Big-Math~\cite{albalak2025big} & 2025 & Mathematical Problem Solving & Perception (Lingual) & 250k questions \\

LongReason~\cite{ling2025longreason} & 2025 & Long-Chain Reasoning & Perception (Lingual) & 794 questions \\

Big-Bench Extra Hard~\cite{kazemi2025big} & 2025 & Complex Reasoning & Perception (Lingual) & 1000+ tasks\\

ResearchBench~\cite{liu2025researchbench} & 2025 & Scientific Reasoning & Perception(Lingual) & 3000+ tasks\\

MastermindEval~\cite{golde2025mastermindeval} & 2025 & Deductive Reasoning & Perception (Lingual) & 1500+ tasks\\

Z1~\cite{yu2025z1} & 2025 & Code-related Reasoning & Perception (Lingual) & 107k questions \\

\midrule
AudioCaps~\cite{kim2019audiocaps} & 2019 & Audio Captioning & Perception (Auditory) & 46k audio clips + 46k captions \\

Clotho~\cite{drossos2020clotho} & 2020 & Audio Captioning & Perception (Auditory) & 4.3k audio clips + 24k captions \\

\midrule
FoTa~\cite{zhao2024transferable} & 2024 & Tactile Sensing & Perception (Tactile) &  3M+ tactile images \\

Touch100k~\cite{cheng2024touch100k}& 2024 & Material Property Recognition & Perception (Tactile) & 100,147 multimodal data \\
TacQuad~\cite{feng2025anytouch} & 2025 &  Tactile Tasks & Perception (Tactile) & 72606 contact frames \\
FuSe~\cite{jones2025beyond} & 2025 & Fine-tuning Robot Policies & Perception (Tactile) & 27,000+ robot trajectories \\

\midrule
RAVEN~\cite{zhang2019raven} & 2019 & \begin{tabular}[c]{@{}c@{}} Visual analogy\\ and spatial structure\end{tabular} & Dimension (Spatial) & 1.12M problems \\
SPARQA~\cite{perez2021sparqa} & 2021 & Situated QA & Dimension (Spatial) & 6k QA pairs \\
GRiT~\cite{yang2022grit} & 2022 & Spatial graph reasoning & Dimension (Spatial) & 48k graphs \\
TQA~\cite{kembhavi2017you} & 2017 & Science diagram QA & Dimension (Spatial) & 26.3k questions \\
CoDraw~\cite{kim2019codraw} & 2019 & Collaborative spatial grounding & Dimension (Spatial) & 10k dialogues \\
TouchDown~\cite{chen2019touchdown} & 2019 & Navigation & Dimension (Spatial) & 9,326 examples \\
Room-to-Room~\cite{anderson2018vision} & 2018 & Instruction-following & Dimension (Spatial) & 21,567 trajectories \\
SpatialSense~\cite{yang2019spatialsense} & 2019 & \begin{tabular}[c]{@{}c@{}} Textual spatial \\relation extraction \end{tabular} & Dimension  (Spatial) & 5,000 images + captions \\

\midrule
Time-Sensitive QA~\cite{chen2021dataset} &2021 & Time-sensitive QA & Dimension (Temporal) & 68k questions\\
TempLama~\cite{dhingra2022time}  &2022 & Time-sensitive QA &Dimension (Temporal) & 50k questions\\
StreamingQA~\cite{liska2022streamingqa} & 2022&\begin{tabular}[c]{@{}c@{}} Dynamic QA (recent/  historical \\knowledge from news)\end{tabular} & Dimension (Temporal) &146k questions \\
TempReason~\cite{tan2023towards} & 2023& \begin{tabular}[c]{@{}c@{}} Temporal fact \\retrieval \& inference \end{tabular} & Dimension (Temporal) & 400k questions\\
MenatQA~\cite{wei2023menatqa}  & 2023&\begin{tabular}[c]{@{}c@{}}  Scope, order, and \\counterfactuals\end{tabular}& Dimension (Temporal) & 2,853 questions \\
TRAM~\cite{wang2023tram} & 2023&\begin{tabular}[c]{@{}c@{}} Event sequencing \& \\temporal arithmetic \end{tabular}&Dimension (Temporal)  & 526.7k questions \\

\midrule
ReClor~\cite{yu2020reclor}& 2020 & Reading Comprehension & Logic & 6,138 questions  \\
%ProofWriter~\cite{tafjord2020proofwriter}& 2020 & binary classification& Logic&1,351 entries\\ 
LogicNLI~\cite{tian2021diagnosing} & 2021 & 
Entailment, Contradiction, Neutral & Logic& 30k+ instances\\
GSM8K~\cite{cobbe2021training}& 2021 & Mathematical problems& Logic& 8.5k questions \\
FOLIO~\cite{han2022folio} &2022 & Binary classification& Logic & 1430 conclusions \\ 
AR-LSAT~\cite{wang2022lsat}&  2022 & Law School Admission Test & Logic & 2,091 questions \\
LogiQA 2.0~\cite{liu2023logiqa}&  2023& Multi-choice logic problems& Logic& 50k+ questions \\
LogicBench~\cite{parmarlogicbench}&2023 & Binary classification &Logic &2,020 instances\\
LINGOLY~\cite{bean2024lingoly}& 2024 & Linguistic logic problems& Logic &1,133 problems \\

\bottomrule
  \end{tabular}
  }
\end{table*}

To advance the development of intelligent agents capable of human-like reasoning, it is essential to evaluate their performance across a diverse set of cognitive dimensions. Reasoning in AI spans multiple modalities and domains, as we introduced before. A wide array of benchmarks has been proposed to capture these aspects, each designed to test different reasoning capabilities in isolation or combination. In this section, we organize and describe representative datasets across these categories, highlighting their design focus, task structure, and relevance for training or evaluating generalist reasoning models. Apart from this, some potential needs for improving reasoning benchmarks have also been discussed.

\subsubsection{Visual} 
In the field of visual reasoning, \textbf{VQA v1.0}~\cite{antol2015vqa} contains 250k images, 760k open-ended questions about these, and 10 million answers to these questions, to support free-form and open-ended visual question answering tasks. \textbf{VQA v2.0}~\cite{goyal2017making} improves upon VQA v1.0~\cite{antol2015vqa} by associating two similar images with each question, reducing language bias in the dataset. \textbf{GQA}~\cite{hudson2019gqa} contains 113K images and 22M questions covering various reasoning skills, generated using scene graph structures and computational linguistics approaches, offering fine-grained control over the distribution of the dataset and supporting new evaluation metrics. \textbf{GQA-OOD}~\cite{kervadec2021roses} introduces distribution shifts into the validation and test sets based on the GQA dataset~\cite{hudson2019gqa}, allowing for the assessment of models and algorithms under Out-Of-Distribution (OOD) settings, proposing a new evaluation protocol to assess model generalization under OOD conditions. \textbf{NLVR2}~\cite{suhr2018corpus} contains 107,292 pairs of English sentences and web photos, focusing on crowdsourcing data through visually rich images and contrastive tasks to stimulate semantic diversity. It requires compositional joint reasoning involving quantity, comparison, and relationships, aiming to challenge and advance research in visual reasoning for natural language and images. \textbf{CLEVR}~\cite{johnson2017clevr} is a diagnostic dataset containing 100k images and 864,968 questions, designed to evaluate the explicit visual reasoning capabilities of visual question-answering systems. It examines the understanding of object attributes and spatial relationships through automatically generated questions. \textbf{Super-CLEVR}~\cite{li2023super} is a synthetic dataset designed for diagnosing visual reasoning and basic visual inference capabilities, with both images and questions programmatically generated, providing high controllability and interpretability. \textbf{OK-VQA}~\cite{marino2019ok} is a dataset containing over 14k questions that require external knowledge to answer, designed to encourage the development of methods that rely on external knowledge resources to address problems in existing visual question answering tasks where image content alone is insufficient. \textbf{A-OKVQA}~\cite{schwenk2022okvqa} is a dataset containing approximately 25k questions that require extensive commonsense and world knowledge to answer, aiming to promote research on deep commonsense reasoning about image scenes, going beyond simple knowledge base query-based responses.

Current VQA benchmark datasets in the general domain encompass a wide range of task types, from simple object recognition to complex scene understanding and logical reasoning. However, they still face several challenges, including: bias and imbalance in question types, with a tendency towards simple object recognition rather than complex scene understanding and logical reasoning; the singularity of answers, often providing only one "correct" answer while neglecting the multiplicity and subjectivity inherent in real-world scenarios; insufficient systematic support for the need of additional commonsense or background knowledge, which limits the model's ability to handle questions requiring external knowledge; and evaluation metrics that predominantly focus on accuracy, lacking consideration for aspects such as model interpretability and uncertainty estimation. These factors collectively constrain the effectiveness and development potential of existing VQA systems in practical applications.

\subsubsection{Lingual}
\textbf{MR-Ben}~\cite{zeng2024mrben} is a process-based benchmark that demands meta-reasoning skills (e.g., locate and analyze errors in automatically generated reasoning steps). It is suited for evaluating system-2 slow thinking, mirroring the human cognitive process. It comprises 5,975 questions across a wide range of subjects.
\textbf{RM-Bench}~\cite{liu2025rmbench} is a novel benchmark designed to evaluate reward models based on their sensitivity to subtle content differences and resistance to style biases.
\textbf{LR\textsuperscript{2} Bench}~\cite{chen2025lr} is a novel benchmark designed to evaluate the Long-china Reflective Reasoning capabilities of LLMs. It contains 850 samples across 6 CSPs.
\textbf{Big-Math}~\cite{albalak2025big} is a dataset of over 250k high-quality math questions that have verifiable answers, are open-ended, and have closed-form solutions. It is an order of magnitude larger than common math reasoning datasets, with problems filtered to best suit RL.
\textbf{LongReason}~\cite{ling2025longreason} is a new synthetic benchmark consisting of 794 multiple-choice reasoning questions with diverse reasoning patterns across different task categories. It is useful for evaluating the long-context reasoning abilities of LLMs.
\textbf{BIG-Bench Extra Hard}~\cite{kazemi2025big} is a new benchmark designed to push the boundaries of LLM reasoning evaluation. It replaces each task in BBH (BIG-Bench Hard) with a novel task that probes a similar reasoning capability with significantly increased difficulty.
\textbf{ResearchBench}~\cite{liu2025researchbench} is the first large-scale benchmark for evaluating LLMs with a near-sufficient set of sub-tasks of scientific discovery. 
\textbf{MastermindEval}~\cite{golde2025mastermindeval} is a simple, scalable, and interpretable deductive reasoning benchmark inspired by the board game Mastermind. It supports agentic evaluation and deductive reasoning evaluation.
\textbf{Z1}~\cite{yu2025z1} is a dataset of 107k simple and complex coding problems paired with their short and long solution trajectories.

\subsubsection{Auditory \& Tactile}
\textbf{Clotho}~\cite{drossos2020clotho} is a dataset for audio captioning. It was built with a focus on audio content and caption diversity, and the splits of the data are not hampering the training or evaluation of methods.
\textbf{AudioCaps}~\cite{kim2019audiocaps} is a large-scale dataset for audio captioning, created using audio clips from AudioSet. It provides crowd-sourced natural language descriptions focused on general audio events, and contains both expert-annotated and user-generated captions to support diverse training and evaluation settings.
\textbf{TacQuad}~\cite{feng2025anytouch} contains paired multi-sensor, multi-modal tactile data, supporting fine-grained tactile tasks (e.g., cross-sensor generation) and coarse-grained tactile tasks (e.g., cross-sensor matching). The dataset includes 17,524 fine-grained contact frames from 25 objects and 55,082 coarse-grained contact frames from 99 objects.  
\textbf{FoTa}~\cite{zhao2024transferable} contains over 3 million tactile images from 13 camera-based tactile sensors, covering 11 tasks.  
\textbf{Touch100k}~\cite{cheng2024touch100k} contains 100,147 tactile-language-vision multimodal data entries, providing multi-granularity tactile descriptions and supporting tasks such as material property recognition and robotic grasping prediction.  
\textbf{FuSe}~\cite{jones2025beyond} consists of 27,000+ robot trajectories and includes a variety of sensory data (vision, touch, audio, proprioception) and language instructions. It is used for fine-tuning robot policies on heterogeneous sensory modalities, like touch and sound.

Current tactile benchmark datasets generally face challenges such as limited scale, insufficient diversity, and restricted practical applicability. Although existing datasets have integrated multimodal data and support a variety of evaluation tasks, their scale remains inadequate for training complex models. The coverage of materials and interaction modes is relatively homogeneous, and there is a strong dependency on specific hardware. Future tactile benchmark datasets should evolve towards deeper integration of multimodal data and the combination of simulated and real-world data to address these limitations and enhance their versatility and utility.

\subsubsection{Spatial}
\textbf{RAVEN}~\cite{zhang2019raven} is a visual reasoning dataset with 1.12M analogy problems designed to assess spatial structure understanding. It emphasizes rule-based pattern recognition in matrix-style puzzles, evaluating relational and hierarchical spatial reasoning. \textbf{SPARQA}~\cite{perez2021sparqa} provides 6k situated QA pairs requiring spatial-temporal comprehension of visual scenes. It challenges models to resolve object relationships within complex visual layouts. \textbf{GRiT}~\cite{yang2022grit} consists of 48k graph-structured instances for spatial reasoning, combining image understanding with structured representations to evaluate relational perception. \textbf{TQA}~\cite{kembhavi2017you} introduces 26.3k science-related QA examples involving diagrams, testing models on layout interpretation in educational contexts. \textbf{CoDraw}~\cite{kim2019codraw} presents 10k dialogues where one agent guides another in recreating a scene through spatially grounded instructions, emphasizing collaborative and referential spatial understanding. \textbf{TouchDown}~\cite{chen2019touchdown} contains 9,326 navigation tasks in real street-view environments, testing how well models interpret spatial descriptions for geolocated reasoning. \textbf{Room-to-Room}~\cite{anderson2018vision} offers 21,567 trajectory samples in 3D environments, focusing on natural language instruction-following grounded in spatial scenes. \textbf{SpatialSense}~\cite{yang2019spatialsense} includes 5,000 images with captions annotated for spatial relations, enabling textual extraction of spatial predicates like "above," "next to," or "under." Current spatial reasoning benchmarks primarily address fundamental tasks such as object localization, spatial relation classification, and basic navigation. However, these tasks often rely on static or simplified environments that fail to capture the complexity of real-world spatial cognition. Future datasets should incorporate dynamic and interactive spatial scenarios, such as embodied navigation in cluttered or unfamiliar environments, multi-agent spatial collaboration, and context-aware spatial planning, to better evaluate the adaptability, generalization, and compositional reasoning capabilities of AI systems in realistic spatial settings.

\subsubsection{Temporal} \textbf{Time-Sensitive QA}~\cite{chen2021dataset} is a dataset with 68k questions from WikiData, used to assess LLMs' ability in time-sensitive QA. \textbf{TempLama}~\cite{dhingra2022time} evaluates masked language models' time-sensitive knowledge, based on Wikidata's 2020 snapshot. It contains 50,310 queries focused on facts that changed after 2010, testing knowledge retention and reasoning over time. \textbf{StreamingQA}~\cite{liska2022streamingqa} examines LLMs' adaptability in dynamic environments with 146k questions based on 2007-2020 news data. It supports realistic time-based QA evaluations, posing challenges with news redundancy, noise, and contradictions. \textbf{TempReason}~\cite{tan2023towards} has over 400k questions for time reasoning in closed-book, open-book, and reasoning QA. It introduces a framework combining time span extraction and reinforcement learning to enhance time reasoning abilities. \textbf{MenatQA}~\cite{wei2023menatqa} includes 2,853 questions to evaluate LLMs' time reasoning using factors like scope, order, and counterfactuals. It shows that model performance varies with size, time bias, and provided time info. \textbf{TRAM}~\cite{wang2023tram} is a time reasoning benchmark with 10 tasks and 526.7k multiple-choice questions. It evaluates reasoning in event sequences, arithmetic, frequency, and duration, revealing that current models fall short of robust, human-level performance in understanding implicit time. Current temporal reasoning datasets mainly focus on basic time understanding tasks such as event ordering and duration estimation. Future benchmarks and datasets should emphasize dynamic event prediction and causal reasoning over time to better reflect real-world temporal inference challenges.

\subsubsection{Logic} 
\textbf{ReColr}~\cite{yu2020reclor}is a reading comprehension dataset focusing on logical reasoning, split into EASY and HARD sets to evaluate model performance on logical reasoning without exploiting dataset biases. Models struggle on the HARD set, highlighting the need for enhanced reasoning abilities. \textbf{LogicNLI}~\cite{tian2021diagnosing} is a diagnostic dataset to evaluate language models on first-order logic (FOL) reasoning, with tasks separating logical inference from commonsense reasoning, aiming to test accuracy, robustness, and traceability. \textbf{FOLIO}~\cite{han2022folio} is a dataset designed for reasoning in natural language with first-order logic annotations, evaluating logical correctness and reasoning capabilities in models. \textbf{AR-LSAT}~\cite{wang2022lsat} focuses on three LSAT tasks (analytical reasoning, logical reasoning, and reading comprehension), pushing models to demonstrate their ability to handle complex reasoning and symbolic knowledge. \textbf{LogiQA 2.0}~\cite{liu2023logiqa} evaluates models' logical reasoning abilities through multiple-choice questions on various logical patterns, testing the application of inference rules in natural language. \textbf{LogicBench}~\cite{parmarlogicbench} tests logical reasoning across propositional, first-order, and non-monotonic logics with 25 distinct inference rules, evaluating models' ability to apply single inference rules in diverse logical scenarios. \textbf{LINGOLY}~\cite{bean2024lingoly} assesses models' reasoning capabilities in low-resource or extinct languages, testing in-context identification and generalization of linguistic patterns in complex tasks.
\textbf{GSM8K}~\cite{cobbe2021training} tests language models on grade school-level math problems, focusing on multi-step arithmetic reasoning. It challenges models to solve problems involving basic calculations and logic.
\par Current logical datasets often rely on synthetic patterns or exam-style questions, lacking real-world abstraction, multi-hop reasoning, and higher-order logic. Future datasets should emphasize diverse, scalable formats with generative reasoning techniques and better capture symbolic structure, uncertainty, and generalization potential.

\section{Applications of Reasoning}
\label{sec:application}

With the increasing sophistication of AI, reasoning has become a fundamental component in robotics and embodied agents, enabling them to operate in dynamic, real-world environments. Unlike traditional AI models that function in constrained digital spaces, these agents interact with the physical world, requiring advanced cognitive abilities to perceive, analyze, and act upon complex inputs. Reasoning is essential for tasks such as spatial navigation~\cite {miguel2018navigation, kennedy2007spatial}, object manipulation~\cite{mota2021integrated}, decision-making~\cite{chen2020trust}, and human collaboration~\cite{liu2024vision}, as it allows these systems to adapt to unpredictable conditions and refine their actions based on experience. As AI-powered robots transition from controlled laboratory settings to real-world applications, they must integrate multiple reasoning paradigms, from spatial and physical reasoning to social considerations. Embodied AI, in particular, necessitates a multi-faceted approach to reasoning, combining sensory data with logical inference to make real-time decisions. The challenges they face—such as uncertainty, partial observability, and the need for rapid response—further underscore the importance of efficient reasoning mechanisms. The following sections explore specific domains where reasoning plays a crucial role in robotics and embodied AI, highlighting how these systems process information, learn from interactions, and execute tasks in complex settings.

\subsection{Physical Agents}

\begin{figure}
\centering

\includegraphics[width=0.5\textwidth]{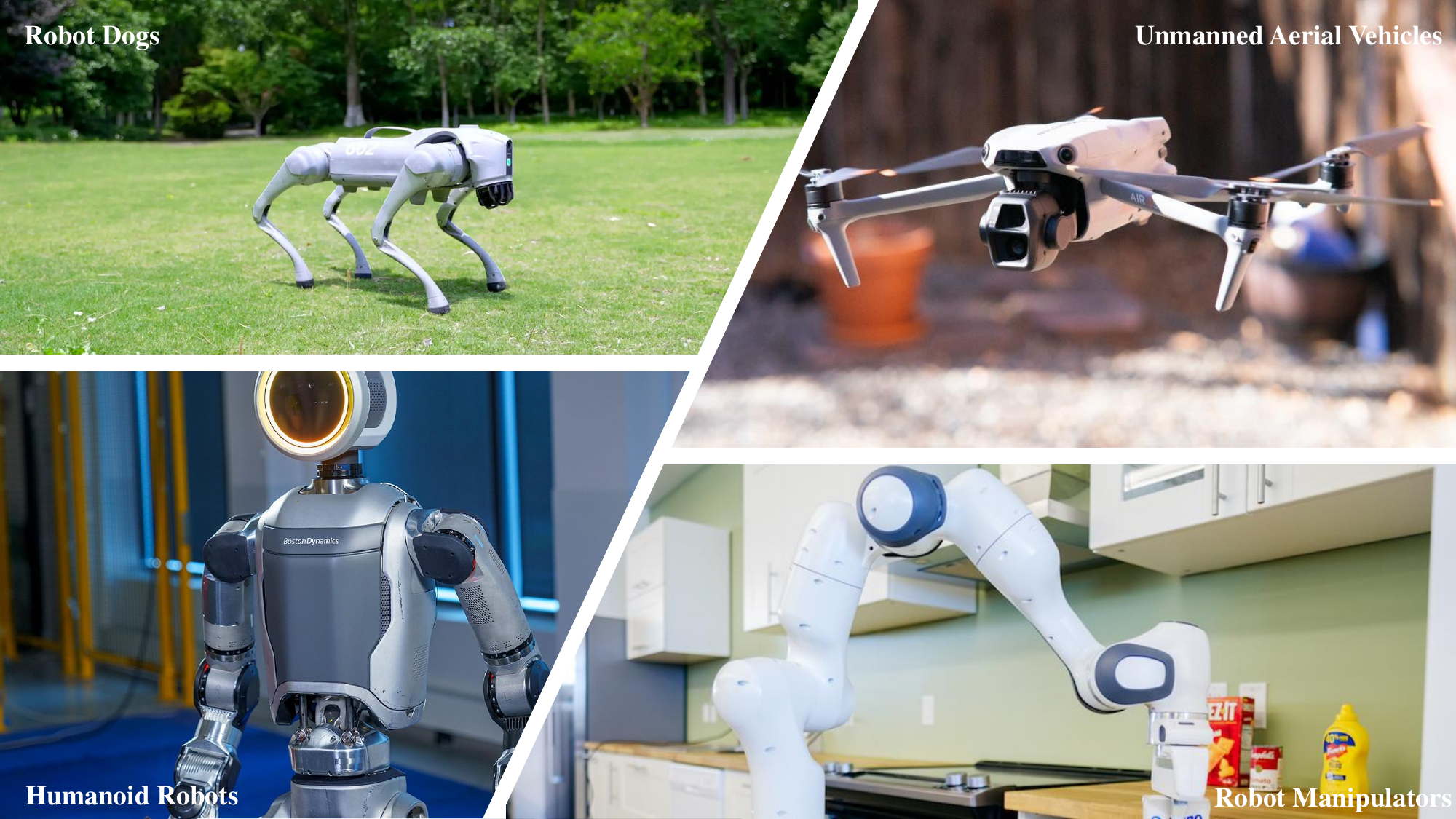}
\vspace{-10pt}
\caption{\textbf{Representative categories of modern robotic platforms.} We showcase four primary types of embodied robotic agents: robot dogs for agile terrain traversal, unmanned aerial vehicles (UAVs) for aerial sensing, humanoid robots designed for human-centric tasks, and robot manipulators specialized in precise physical interaction within structured environments.}
\label{fig:physical_agent}
\end{figure}

Robotics and embodied agents operate in the physical world, requiring advanced reasoning abilities to perceive, plan, and execute complex actions in dynamic environments. As listed in Fig. \ref{fig:physical_agent}, unlike purely digital AI systems, these agents continuously interpret sensor inputs, handle uncertainty, and make decisions to interact effectively with the real world \cite{pfeifer2004embodied}. The reasoning capabilities of these agents are crucial not only for navigating the environment \cite{liu2024visuomotor} but also for performing tasks with precision and adaptability \cite{zhou2024hazard}. This section delves into the role of reasoning in robotics, emphasizing how it underpins decision-making and action execution.

A primary application of reasoning in robotics is autonomous navigation and path planning. Robots must analyze spatial layouts, detect obstacles, and compute optimal paths, often in complex, cluttered environments. This involves the integration of geometric reasoning with real-time sensor data, enabling the robot to adjust its movement dynamically. For instance, autonomous vehicles rely on reasoning to assess road conditions, predict pedestrian behavior, and execute split-second decisions, ensuring safety. Similarly, robots in warehouses must combine topological reasoning with sensor inputs to identify and optimize retrieval paths, minimizing delays and avoiding collisions. The ability to reason about the environment and predict changes in real-time is vital for both safety and efficiency.

Beyond navigation, reasoning is critical in object manipulation and interaction. Robots performing tasks like manufacturing, healthcare, and domestic assistance need to understand the physical properties of objects, such as weight, texture, and fragility. Physical reasoning in these contexts allows robots to adjust their actions based on these properties. For example, a robotic arm might use reasoning to adapt its grip strength based on the fragility of an object, preventing breakage. In domestic environments, service robots must reason about their surroundings to perform tasks like pouring liquids without spilling or assembling furniture. Through predictive reasoning, these robots can refine their actions, ensuring higher accuracy and adaptability in their operations.

In human-robot collaboration, reasoning plays an indispensable role in ensuring seamless interaction and synchronization between humans and robots. As robots are integrated into environments such as workplaces and homes, they must not only understand physical tasks but also the social dynamics of working with humans. For example, medical robots assisting in surgeries must synchronize their actions with the surgeon’s movements, making real-time adjustments based on the procedure’s progress. Likewise, exoskeletons and prosthetics rely on biomechanical reasoning to adapt to the user’s movements, ensuring effective collaboration and safety. The ability to interpret human intent and non-verbal cues-such as gestures or postures-is critical in these contexts, requiring robot reason in a manner that goes beyond mere physical action execution.

Moreover, reasoning in physical agents extends beyond individual robots to multi-agent systems, where collective intelligence enhances task completion. In scenarios like drone swarms for environmental monitoring or robotic teams in disaster response, the reasoning process must accommodate interaction, coordination, and negotiation. Each agent must assess its capabilities in relation to others, deciding when to act independently or collaborate. In these systems, decentralized reasoning allows agents to share information and optimize performance collectively. In industrial settings, for instance, robotic arms may work together on assembly lines, adapting to real-time production requirements and collaborating to meet tight deadlines without relying on a centralized control system.

Overall, reasoning in embodied agents connects perception, decision-making, and action execution, enabling robots to navigate physical spaces, manipulate objects, and collaborate with humans and other agents effectively. The integration of advanced reasoning techniques with physical action is what sets embodied agents apart from purely digital systems, giving them the ability to function in the real world with both precision and adaptability.

\subsection{Virtual Agents}
In contrast to embodied agents, disembodied AI operates in purely digital and conceptual spaces, relying on reasoning to analyze data, simulate environments, and optimize decision-making. These systems do not interact with the physical world through sensors or actuators; instead, they engage with structured and unstructured data, abstract problem-solving, and multi-agent coordination. These capabilities are fundamental in knowledge-intensive domains, strategic problem-solving, and interactive AI systems.  
One of the most prominent applications of disembodied AI is in conversational agents and language models. Systems like ChatGPT~\cite{achiam2023gpt} exemplify how AI can leverage reasoning to generate coherent, contextually relevant, and logically structured responses. These models process vast amounts of textual data, infer relationships between concepts, and dynamically adjust their outputs based on user input. Beyond simple text generation, their reasoning mechanisms allow them to engage in complex discussions, provide explanations, and even simulate problem-solving processes in technical and scientific domains.  
Another key area of disembodied AI is automated reasoning in knowledge-based systems. AI-driven assistants in law, science, and healthcare employ logical inference to analyze regulations, detect patterns in research data, and suggest optimal courses of action. These systems extend beyond retrieving pre-existing knowledge by applying reasoning to synthesize new insights, validate arguments, and reconcile conflicting information. For instance, automated theorem provers utilize formal logic to verify mathematical proofs, while AI-driven research assistants scan and analyze large corpora to identify emerging scientific trends. 
Strategic reasoning is another crucial application of disembodied AI, particularly in game theory, cybersecurity, and financial modeling. AI-driven trading systems, for instance, reason over market trends, competitor behaviors, and risk assessments to optimize investment strategies. In cybersecurity, reasoning enables AI to predict and counteract cyberthreats by simulating potential attack vectors and deploying defensive measures. Multi-agent strategic systems, such as those used in military simulations or competitive gaming, employ advanced reasoning to anticipate adversarial moves, negotiate optimal strategies, and make real-time adjustments. The intersection of reasoning and computational creativity also demonstrates the versatility of disembodied AI. From AI-generated art and music to AI-assisted code development and scientific discovery, reasoning allows these systems to explore novel possibilities while adhering to defined constraints.
\section{Future Directions and Innovations}
\label{sec:future_work}
%\subsection{Future Directions}
Based on our proposed architecture, which spans from multimodal input perception to final reasoning output, we identify several key directions and innovations for enhancing the reasoning capabilities of future AI systems.

\begin{figure*}
    \centering
    \includegraphics[width=\linewidth]{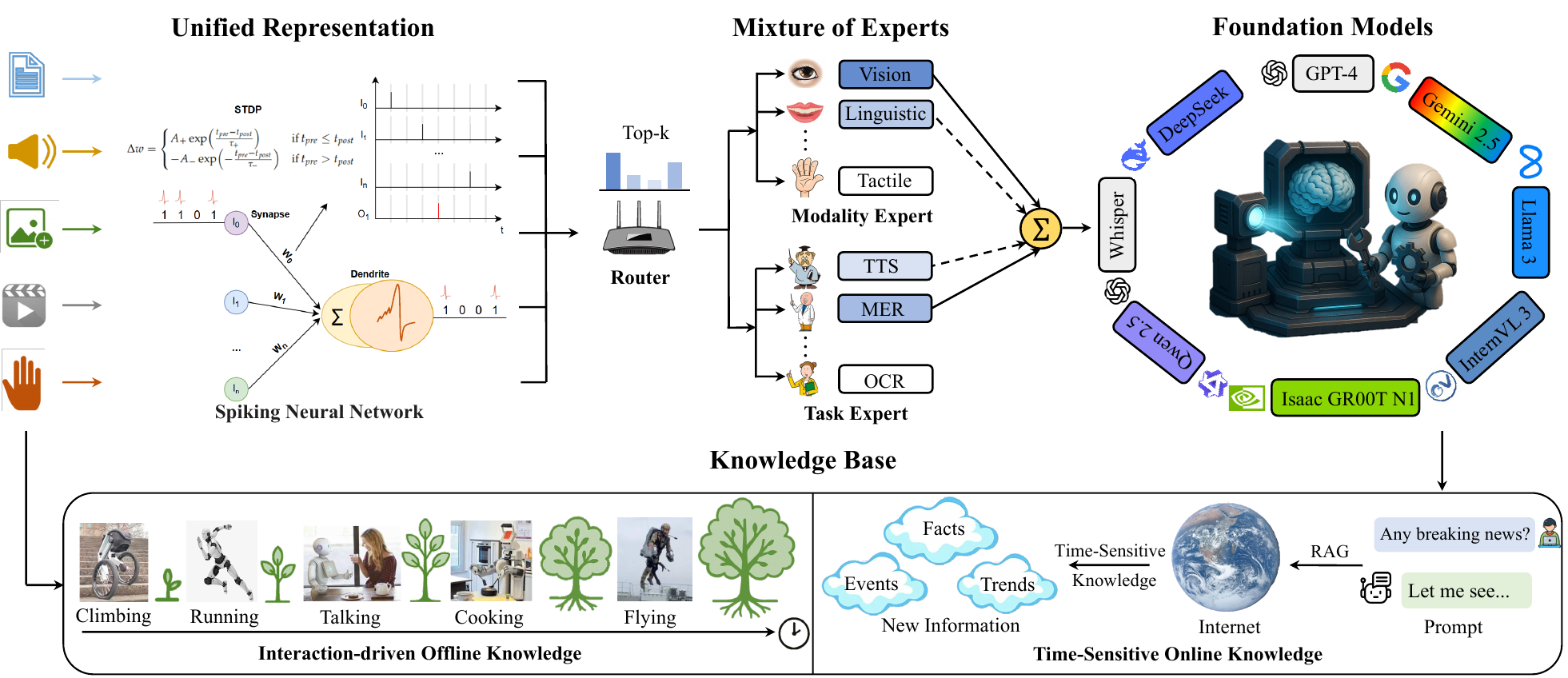}
    \caption{An overview of our proposed AI agent system architecture designed to facilitate reasoning through multimodal perception and dynamic knowledge integration. Multimodal inputs are encoded into a unified representation via biologically inspired processing mechanisms. A Dynamic Multimodal Mixture-of-Experts (DMMoE) selectively engages modality-specific and task-specific experts based on real-time salience and task relevance. Foundation models serve dual roles as high-fidelity understanding engines and flexible reasoning assistants. Knowledge is organized into a dual system: an interaction-driven offline knowledge base capturing embodied experiences, and a time-sensitive online retrieval mechanism accessing dynamic external information. This framework enables adaptive, robust, and temporally coherent reasoning across complex real-world scenarios.}
    \label{fig:brainstorm_foundations}
\end{figure*}
\par \noindent \textbf{Multimodal Inputs: Toward Selective and Adaptive Multimodal Perception.} Most current AI systems are limited to processing static, single-modality inputs, such as pure text or isolated images, which stands in stark contrast to the human ability to dynamically shift and integrate attention across sensory modalities. In real-world environments, perceptual input is continuous and situation-dependent: when visual information is degraded, humans instinctively rely more on auditory or tactile cues; when multiple modalities are simultaneously available, we selectively focus on those most relevant to the task at hand. \textbf{Inspired by this, future AI agents should support selective and adaptive multimodal perception, where choosing the most relevant modality—or combination thereof—not only enhances robustness, but also forms the foundation for effective and context-sensitive reasoning.} One promising approach is the development of a \textbf{Dynamic Multimodal Mixture-of-Experts (DMMoE)} architecture, which draws on the brain's adaptive gain control mechanisms~\cite{hillyard1998sensory, auerbach2022hearing}. As shown in Fig.~\ref{fig:brainstorm_foundations}, in this framework, individual expert networks are assigned to different modalities, such as vision, language, audio, and touch, or tasks, such as Text-to-Speech (TTS), Multi-modal Entity Recognition (MER), and Optical Character Recognition (OCR). A learnable gating network continuously modulates the activation of each expert based on real-time sensory salience and task relevance. Their outputs are integrated into a shared representation space, while a global scheduler determines whether to process them in parallel or sequentially, depending on task complexity and latency constraints. This setup allows for context-aware, flexible engagement with the most informative modalities, enhancing robustness under partial observability and improving computational efficiency.  The modular design also supports extensibility: new sensory experts and adaptive routing strategies can be introduced via meta-learning or online adaptation--bringing AI perception one step closer to human-like flexibility.
\par \noindent \textbf{Information Processing: Toward Unified Modal Representations for Cross-Modal Reasoning.} Most current AI reasoning systems rely on separate modality-specific encoders and late fusion strategies, which often struggle to handle real-time multi-modal inputs in a coherent and adaptive way. In contrast, the human brain processes different sensory signals--such as vision, audition, and touch--by converting them into a common electrochemical format. This unified signal representation enables seamless cross-modal integration and lays a foundation for efficient reasoning across sensory domains. \textbf{Inspired by this, future AI systems should aim to develop a shared representation space that transcends modality-specific encodings, allowing for more fluid and consistent reasoning across multi-modal inputs.} One potential solution is to draw from neuroscience-inspired models such as Spiking Neural Networks (SNNs)~\cite{ghosh2009spiking}, which mimic the event-driven and temporally coded nature of neural processing. By aligning information from different sources in the time domain, SNNs may provide a biologically plausible and computationally efficient path toward building unified representations for robust cross-modal reasoning, as shown in Fig.~\ref{fig:brainstorm_foundations}.

\par \noindent \textbf{Knowledge Base: Dual Memory Systems for Dynamic and Time-Sensitive Reasoning.}
Current AI models largely depend on static, pre-trained knowledge bases, which significantly limit their ability to reason over dynamic, evolving facts--especially those involving temporal information or long-term dependencies. In contrast, humans construct and update internal knowledge representations through continuous interaction with the world, while simultaneously drawing on external sources to verify or complement what they know. \textbf{Inspired by this, future AI agents should be equipped with a dual knowledge architecture consisting of: (i) an offline, interaction-driven knowledge base that incrementally integrates information from the agent’s embodied experience and dialogue history, and (ii) an online, time-sensitive retrieval system} that dynamically accesses up-to-date information from external sources such as the internet or structured databases as shown in Fig.~\ref{fig:brainstorm_foundations}. This dual system not only enables AI agents to maintain a grounded and context-rich internal model of the world, but also to adapt flexibly when confronted with novel, uncertain, or time-critical reasoning scenarios. It is particularly crucial for tasks involving temporal causality, evolving facts, or multi-step reasoning under uncertainty. One promising direction is the development of an adaptive retrieval-controller architecture, which orchestrates when and how to consult internal versus external knowledge, based on current reasoning needs, confidence levels, and task requirements. Unlike traditional Retrieval-Augmented Generation (RAG)~\cite{gao2023retrieval}, which passively fetches documents to support static answers, this controller actively monitors reasoning progress, identifies knowledge gaps, and strategically queries the appropriate knowledge base--allowing for more robust, grounded, and temporally coherent reasoning.

\par \noindent \textbf{Foundation Models: Dual Role as Understanding Engines and Reasoning Assistants.} Although large language models (LLMs)~\cite{achiam2023gpt,guo2025deepseek,yang2024qwen2} and vision language models (VLMs)~\cite{bai2023qwen,hurst2024gpt} exhibit basic inference capabilities, their core strength lies in high‑fidelity understanding--providing rich, reliable representations that feed into specialized reasoning modules. \textbf{To strengthen this role, future work must prioritize the creation of higher‑quality, diverse, and temporally annotated datasets capturing real‑world concepts, contexts, and cross‑modal relationships. Simultaneously, foundation models should serve as reasoning assistants, leveraging their learned statistical patterns to pre-process inputs, generate candidate hypotheses, and enforce structured heuristic rules derived from the different reasoning tasks.} In this capacity, they scaffold subsequent specialized processes without replacing them. By enhancing dataset quality and embracing this dual role, foundation models will become indispensable both for understanding complex inputs and for guiding structured, modular reasoning in AI agents.
\begin{figure}
\centering
\includegraphics[width=0.5\textwidth]{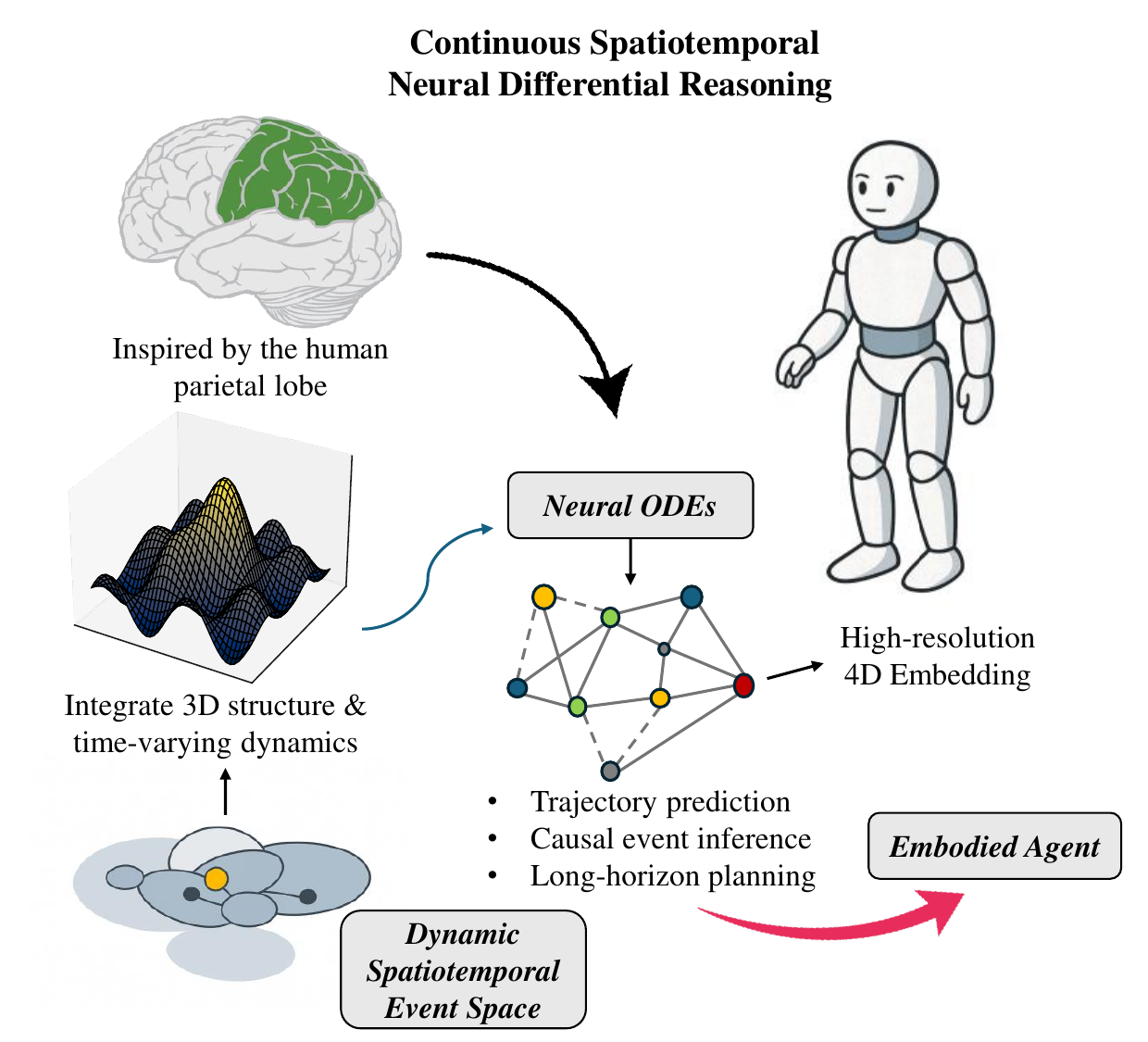}
\caption{\textbf{Framework for continuous spatiotemporal neural differential reasoning in embodied agents.} Inspired by the human parietal lobe, this architecture integrates dynamic spatiotemporal event spaces with 3D structural information and time-varying dynamics using Neural ODEs. The resulting high-resolution 4D embeddings support trajectory prediction, causal event inference, and long-horizon planning for embodied agents.}
\label{fig:braintorm_dimension}
\end{figure}

\par \noindent \textbf{Perceptual Reasoning: Toward Structured Intermediate Representations.} Human neuroscience suggests that perception may construct relational maps rather than isolated feature lists. Functional Magnetic Resonance Imaging (fMRI) studies indicate that the parahippocampal place area (PPA)~\cite{epstein1999parahippocampal} encodes scene layouts through relational graphs, where nodes correspond to spatial anchors (e.g., landmarks) and edges represent boundary topology (e.g., adjacency, containment). Complementary evidence from hippocampal–entorhinal circuits suggests that cognitive maps link locations and events via node–edge structures, supporting both spatial navigation and episodic memory~\cite{epstein1998cortical,epstein2008parahippocampal,tacikowski2024human,epstein2017cognitive}. \textbf{Inspired by these biological mechanisms, emerging AI approaches propose embedding scene graphs as intermediate representations, elevating objects and their relationships to explicit components of the model’s internal state.} By explicitly modeling entities (nodes) and relations (edges), such architectures enable relational queries (e.g., structural support analysis) and improve robustness against occlusion through context propagation, akin to cortical scene-completion processes. A promising implementation integrates a Vision Transformer (ViT) backbone with a graph neural network (GNN). The ViT first detects entities and estimates pairwise relation scores via self-attention, while a dynamic GNN refines edge weights through graph attention layers (GATs), enforcing constraints like physical plausibility. This graph-centric loop—reminiscent of hippocampal replay for memory consolidation—enhances both interpretability and adaptability in complex environments. While direct biological equivalence remains unproven, this synergy between neural principles and AI design marks a step toward human-like perceptual reasoning.

\par \noindent \textbf{Dimensional Reasoning: Towards Continuous Spatiotemporal Neural Differential Reasoning.} Current methods—such as 4D Gaussian splatting~\cite{wu20244d,ling2024align}—discretely model spatial and temporal dimensions but fail to capture the fluid, continuously evolving nature of dynamic environments encountered by embodied agents. \textbf{Inspired by the human parietal lobe, which seamlessly integrates spatial awareness with temporal sequencing, future AI systems should build continuous implicit representations of the world that jointly encode 3D structure and time-varying dynamics.} As shown in Fig. \ref{fig:braintorm_dimension}, one promising direction is to leverage Neural Ordinary Differential Equations (Neural ODEs)~\cite{chen2018neural} to learn the continuous-time evolution of scene geometry—rather than relying on predefined static parameters—and to integrate an event-driven spatiotemporal graph attention mechanism that dynamically selects and weights critical nodes (e.g., objects in motion and key events) as they occur. By forming a high-resolution 4D embedding that updates in real time, this framework enables fine-grained trajectory prediction, causal event inference, and long-horizon planning, thereby equipping embodied agents with more precise, coherent, and adaptable dimensional reasoning capabilities.

\par \noindent \textbf{Logical Reasoning: Toward Structured, Causal, and Human-Aligned Inference.} Logical reasoning plays a pivotal role in enabling AI agents to derive conclusions from premises, test hypotheses, and make consistent, interpretable decisions. Recent advancements in neuro-symbolic systems have laid a solid foundation by combining neural networks' ability to process perceptual input with the rule-based rigor of symbolic logic. However, existing models often treat logic superficially—relying on surface pattern matching or probabilistic associations—rather than deeply modeling structured inference.  \textbf{Inspired by this, future AI reasoning systems should move toward causality-aware, structure-constrained, and hierarchy-guided logical inference.} This involves three directions. First, systems should encode and manipulate logic in structured, interpretable formats--such as program sketches or graph-based logic trees--enabling models to explicitly construct and verify reasoning chains across inductive, deductive, and abductive paradigms. Second, to align more closely with human-like reasoning, agents should be equipped with mechanisms to perform counterfactual thinking and proof-by-contradiction, which are essential in scientific reasoning and legal argumentation. Third, logical reasoning must be grounded in causality: models should learn to represent and reason over causal graphs, distinguishing correlation from explanation. One promising direction is to develop a neuro-symbolic planner that unifies symbolic logic programs with causal event graphs, enabling agents to simulate multiple inference trajectories, evaluate plausibility, and select the most coherent explanation--especially under partial observability. These structured logical systems will serve as the backbone of AI agents, supporting robust, transparent, and generalizable decision-making in complex environments.

\begin{figure}[t]
        \centering
		\includegraphics[width=0.48\textwidth]{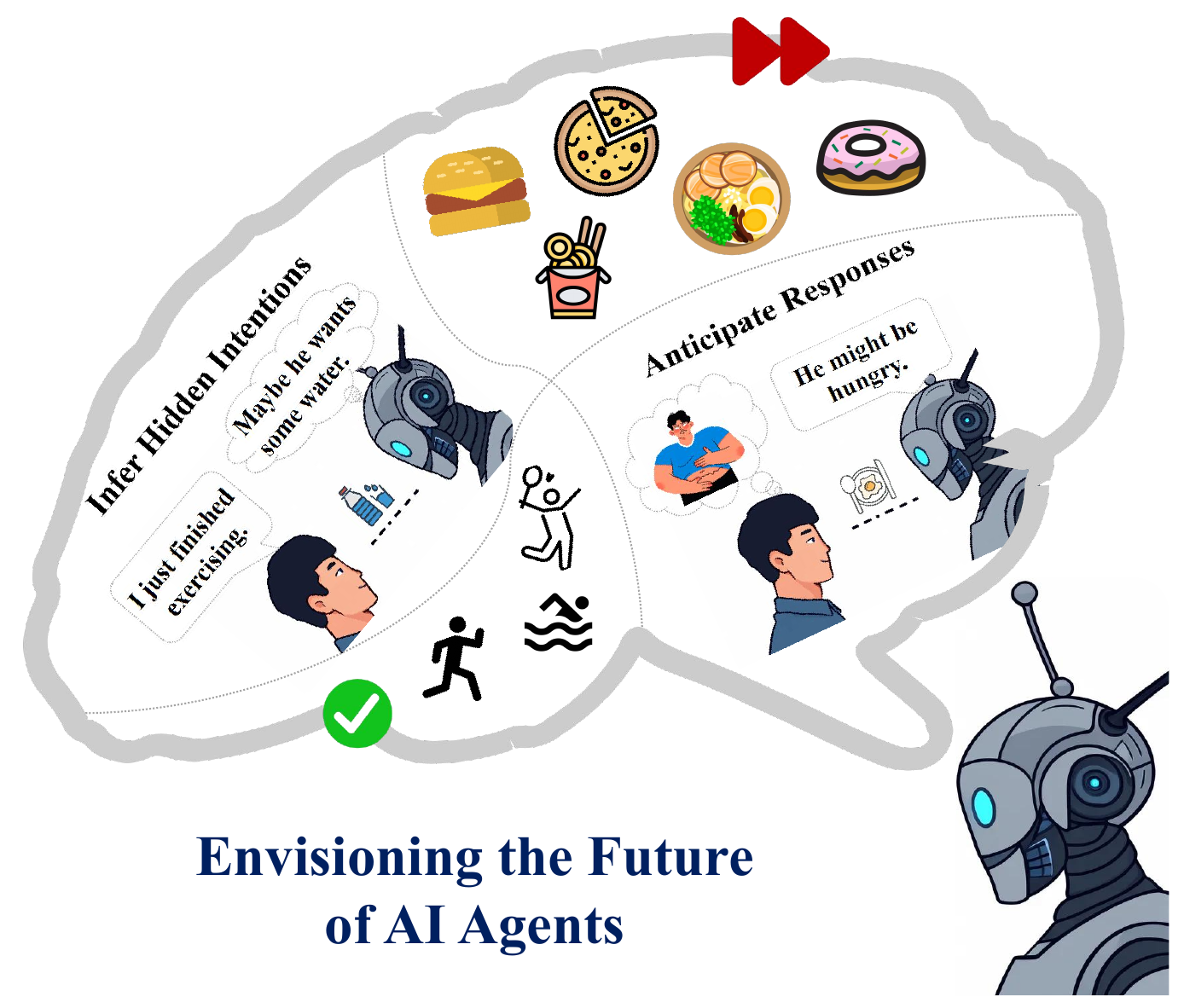}
\caption{Future AI agents should possess the ability to reason about others from a first-person perspective—inferring hidden intentions, anticipating responses, and adapting strategies in real-time to maintain cooperation.}
\label{fig:brainstorm_interaction}
\end{figure}

\noindent\textbf{Interactive Reasoning: Toward Intention-Aware and Socially-Coherent Agents.}
Interactive reasoning enables agents to perceive, interpret, and respond to other entities within dynamic environments, making it a critical component for embodied AI operating in social or multi-agent settings. \textbf{Inspired by this, future AI agents should possess the ability to reason about others from a first-person perspective—inferring hidden intentions, anticipating responses, and adapting strategies in real-time to maintain cooperation, resolve conflicts, or handle deception as shown in Fig~\ref{fig:brainstorm_interaction}.} One promising direction is to develop intention-aware reasoning frameworks, which integrate agent-centric representations with inverse planning and goal inference modules. These systems would allow agents to simulate the beliefs and objectives of others while adjusting their own policy accordingly--akin to the theory of mind in humans. Technically, this could be achieved by coupling behavior trajectory modeling with learned causal priors, enabling agents to infer not only what others are doing, but why they are doing it. Additionally, grounding interaction within structured environments—via symbolic scene graphs, affordance maps, or dialogue ontologies—could provide an interpretable substrate for multi-agent reasoning. Importantly, interactive reasoning should extend beyond agent-agent coordination to encompass rich human-agent collaboration. Here, the agent must not only align with human preferences, but also continuously refine its behavior through interactive feedback and few-shot corrections. This demands a hybrid of reinforcement learning, online imitation, and neuro-symbolic adaptation, where agents can learn from sparse demonstrations and ambiguous signals in real-time. Such systems will ultimately support agents that are socially coherent, goal-aligned, and capable of evolving through interaction, paving the way for truly collaborative artificial intelligence.

\par Just as \textbf{Chain of Thought} (CoT)~\cite{wei2022chain} is inspired by the serial reasoning in \textbf{ACT-R}~\cite{anderson2014atomic}, many cognitive models from neuroscience can also provide valuable insights for AI reasoning architectures. For instance:
\par \noindent \textbf{Miller and Cohen’s Model (PFC Cognitive Control).}
\begin{itemize}
\item Goal-Driven Multi-Step Reasoning: AI models, such as LLMs and reinforcement learning agents, can maintain a goal vector or context vector throughout reasoning, continuously biasing outputs toward task objectives. This could be implemented via a global task descriptor or a dynamic context-tracking mechanism that ensures the system remains aligned with the overarching goal
\item Error Detection \& Adaptation: Inspired by PFC’s bias-adjustment mechanism, AI systems can incorporate self-monitoring modules to periodically assess reasoning accuracy. If an inconsistency arises, the model can trigger a self-correction strategy, such as self-reflection in LLMs to regenerate more goal-aligned responses~\cite{yuan2025agent,cheng2024vision,wang2024meta}.
\end{itemize}

\par \noindent \textbf{Banich’s Cascade of Control Model.} 
\begin{itemize}
    \item Cascaded Attention Scheduling: AI models handling multimodal or multitask inputs can incorporate a cascaded attention module~\cite{qi2022cascaded,upadhyay2025bidirectional}, where an initial coarse-grained filter (akin to posterior DLPFC) identifies relevant features, a mid-layer refines them, and a final layer (analogous to ACC) determines the output. This hierarchical filtering reduces noise and enhances robustness in complex environments.
    \item Multi-Stage Decision Pipelines: Reinforcement learning and structured decision-making~\cite{zhao2024multi} can benefit from stage-wise decision decomposition, where a high-level policy selects focus areas before lower-level policies refine actions. This helps in stepwise strategy formulation and adaptive control.
\end{itemize}

 \par \noindent \textbf{Baddeley’s Working Memory Model.} 
 \begin{itemize}
     \item Multi-Buffer Memory Architectures: AI reasoning systems can implement dedicated memory buffers~\cite{wang2024buffer} for different modalities—e.g., separate caches for text sequences (like a phonological loop) and visual data (like a visuospatial sketchpad), orchestrated by a central executive module for reasoning and decision-making.

\item Parallel Perception \& Serial Control: Inspired by human memory constraints, AI models can parallelize low-level sensory processing while keeping high-level decision-making serial~\cite{tamber2016central, sigman2008brain}. Transformer-based architectures~\cite{vaswani2017attention} or RNNs~\cite{elman1990finding} could benefit from separate caching mechanisms for different input modalities, with a reinforcement learning-based controller managing cross-modal interactions.
 \end{itemize}

\par \noindent \textbf{Predictive Coding.}
\begin{itemize}
\item Iterative Generation \& Correction: AI models can incorporate a self-supervised feedback loop~\cite{ye2023selfee}, where generated outputs are iteratively compared against predefined input constraints, and if discrepancies exceed a certain threshold, the system refines its internal representation or reasoning path before producing the final output. This is particularly relevant for generative AI, where multiple iterations can improve coherence and accuracy.

\item Hierarchical Error Feedback: A layered architecture can mirror top-down priors and bottom-up corrections, where high-level modules predict global context (e.g., discourse structure in NLP or object relations in vision), while lower layers validate fine-grained details. This could enhance error correction in self-driving systems or autonomous robotics by integrating predictive models with real-time sensory updates.

\item Predicting Key Tokens: Predictive coding enables the brain to quickly adapt to environmental changes and optimize the understanding of causal relationships. Multimodal large language models perform nearly perfectly on simple feature recognition tasks, but their performance in causal reasoning remains significantly below human level~\cite{schulze2025visual}. Inspired by the minimization of prediction error, future multimodal large language models could predict important visual tokens in advance during the visual encoding stage, retain key tokens, and improve reasoning speed~\cite{tan2025tokencarve,liu2024multi} while enhancing the reasoning capabilities of these models.
\end{itemize}
\par \noindent \textbf{Adaptive Control of Thought—Rational (ACT-R).}
\begin{itemize}
    \item Explicit Chain-of-Thought (CoT) Reasoning: AI models can adopt stepwise rule-based reasoning, akin to ACT-R’s production system, ensuring that each reasoning step updates working memory before proceeding. This would make CoT-based inference more structured, preventing reasoning jumps or inconsistencies.
\end{itemize}

\par \noindent \textbf{Global Workspace
Theory (GWT).}
\begin{itemize}
    \item Global Broadcasting Mechanism: GWT posits that consciousness emerges from the competition among multiple specialized modules for access to a central global workspace; once information enters this workspace, it is broadcast system-wide. AI systems can simulate this mechanism by introducing a global attention pool or a shared blackboard architecture within multimodal models. When salient information from a specific modality or task reaches a predefined threshold, it can be "broadcast" to other modules, enabling dynamic resource allocation and cross-module coordination. This mechanism offers significant inspiration for dynamic task scheduling and attention routing in large-scale AI systems.
\end{itemize}

% \par \noindent \textbf{Hierarchical Temporal Memory (HTM).}
% \begin{itemize}
%     \item Temporal Pattern Recognition: AI models handling IoT, finance, or streaming data can adopt an HTM-style continuous learning approach, where new information refines long-term patterns rather than relying solely on batch updates. This enhances real-time adaptability.

%     \item Hierarchical Spatiotemporal Reasoning: By structuring reasoning hierarchically, each level in a neural network can specialize in different temporal scales. This can be applied in video anomaly detection, sequential decision-making, and robotic motion planning, where long-term dependencies are critical.
% \end{itemize}

\section{Conclusion}
This survey is the first to systematically explore AI agent reasoning from a neuroscience perspective, offering a comprehensive framework that spans from perception to action. We defined AI agent reasoning by formulating three precise  definitions and clarifying key concepts based on insights from neuroscience, which laid the foundation for our novel taxonomy of reasoning processes. We systematically analyzed existing methods within this framework, identified key limitations in current models--such as challenges in adaptability and multi-step reasoning--and proposed future research directions, which were further inspired by our framework and neuroscience models, offering new insights for advancing AI reasoning techniques. Additionally, we released an open-source repository organizing benchmark tasks, datasets, and research papers, which will be continuously updated to support future AI reasoning research.

%\clearpage
\bibliographystyle{IEEEtran}
\bibliography{IEEEabrv,main}

\vfill

\end{document}